\documentclass[useAMS,usenatbib,usegraphicx]{mn2e}
\usepackage{aasmacros,amssymb,amsmath}

\newcommand{\Msol}{$M_{\odot}$}
\newcommand{\hh}{$h^{-1}$}
\newcommand{\kms}{km s$^{-1}$}
\newcommand{\be}{\begin{equation}}
\newcommand{\ee}{\end{equation}}

\title[Satellites are the main drivers of environmental effects]{zCOSMOS 20k\thanks{Based on observations obtained at the European Southern Observatory (ESO) Very Large Telescope (VLT), Paranal, Chile, as part of the Large Program 175.A-0839 (the zCOSMOS Spectroscopic Redshift Survey)}: Satellite galaxies are the main drivers of environmental effects in the galaxy population at least to $z \sim 0.7$}

\author[K. Kova\v{c} and zCOSMOS collaboration]{K. Kova\v{c}$^{1}$\thanks{E-mail: kovac@phys.ethz.ch}
 S. J. Lilly,$^{1}$
 C. Knobel,$^{1}$
 T. J. Bschorr,$^{1}$
 Y. Peng,$^{1}$ 
C.~M.~Carollo,$^{1}$
\newauthor
T.~Contini,$^{2,3}$
J.-P.~Kneib,$^{4}$
O.~Le F\'evre,$^{4}$
V.~Mainieri,$^{5}$
A.~Renzini,$^{6}$
M.~Scodeggio,$^{7}$
\newauthor
G.~Zamorani,$^{8}$
S.~Bardelli,$^{8}$
M.~Bolzonella$^{8}$,
A.~Bongiorno$^{9}$,
K.~Caputi$^{19}$,
O.~Cucciati$^{10}$,
\newauthor
S.~de la Torre$^{11}$,
L.~de Ravel$^{11}$,
P.~Franzetti$^{7}$,
B.~Garilli$^{7}$,
A.~Iovino$^{10}$,
P.~Kampczyk$^{1}$,
\newauthor
F.~Lamareille$^{2,3}$,
J.-F.~Le Borgne$^{2,3}$,
V.~Le Brun$^{4}$,
C.~Maier$^{18}$,
M.~Mignoli$^{8}$,
P.~Oesch$^{25}$,
\newauthor
R.~Pello$^{2,3}$,
E.~Perez Montero$^{2,3,12}$,
V.~Presotto$^{10,21}$,
J.~Silverman$^{22}$,
M.~Tanaka$^{13,22}$,
\newauthor
L.~Tasca$^{4}$,
L.~Tresse$^{4}$,
D.~Vergani$^{8,20}$,
E.~Zucca$^{8}$,
H.~Aussel$^{24}$,
A.~M.~Koekemoer$^{15}$,
\newauthor
E.~Le~Floc'h$^{24}$,
M.~Moresco$^{14}$,
and L.~Pozzetti$^{8}$\vspace{0.4cm}\\
$^{1}$Institute for Astronomy, ETH Zurich, Zurich 8093, Switzerland\\
$^2$Institut de Recherche en Astrophysique et Plan\'etologie, CNRS, 14, avenue Edouard Belin, F-31400 Toulouse, France\\
$^3$IRAP, Universit\'e de Toulouse, UPS-OMP, Toulouse, France\\
$^4$Laboratoire d'Astrophysique de Marseille, CNRS/Aix-Marseille Universit\'e, 38 rue Fr\'ed\'eric Joliot-Curie, F-13388 Marseille cedex 13, France\\
$^5$European Southern Observatory, Garching, Germany\\
$^6$INAF-Osservatorio Astronomico di Padova, Vicolo dell'Osservatorio 5, 35122, Padova, Italy\\
$^7$INAF-IASF Milano, Milano, Italy\\
$^8$INAF Osservatorio Astronomico di Bologna, via Ranzani 1, I-40127, Bologna, Italy\\
$^{9}$Max Planck Institut f\"ur Extraterrestrische Physik, Garching, Germany\\
$^{10}$INAF Osservatorio Astronomico di Brera, Milan, Italy\\
$^{11}$Institute for Astronomy, University of Edinburgh, Royal Observatory, Edinburgh, EH93HJ, UK\\
$^{12}$Instituto de Astrofisica de Andalucia, CSIC, Apartado de correos 3004, 18080 Granada, Spain\\
$^{13}$Institute for the Physics and Mathematics of the Universe (IPMU), University of Tokyo, Kashiwanoha 5-1-5, Kashiwa, Chiba 277-8568,\\ Japan\\
$^{14}$Dipartimento di Astronomia, Universit\`a degli Studi di Bologna, Bologna, Italy\\
$^{15}$Space Telescope Science Institute, Baltimore, MD 21218, USA\\
$^{16}$Institut d'Astrophysique de Paris, UMR7095 CNRS, Universit\'e Pierre \& Marie Curie, 75014 Paris, France\\
$^{17}$Insitut d'Astrophysique Spatiale, B\^atiment 121, Universit\'e Paris-Sud XI and CNRS, 91405 Orsay Cedex, France\\
$^{18}$University of Vienna, Department of Astronomy, Tuerkenschanzstrasse 17, 1180 Vienna, Austria\\
$^{19}$Kapteyn Astronomical Institute, University of Groningen, P.O.~Box 800, 9700 AV Groningen, The Netherlands\\
$^{20}$INAF-IASF Bologna, Via P.~Gobetti 101, I-40129 Bologna, Italy\\
$^{21}$Dipartimento di Fisica dell'Universit\`a degli Studi di Trieste - Sezione di Astronomia, via Tiepolo 11, 34143, Trieste, Italy\\
$^{22}$Kavli Institute for the Physics and Mathematics of the Universe, Todai Institutes for Advanced Study, the University of Tokyo, \\Kashiwa, Japan 277-8583 (Kavli IPMU, WPI)\\
$^{23}$Centro de Estudios de F\'{\i}sica del Cosmos de Arag\'on, Plaza San Juan 1, planta 2, 44001 Teruel, Spain\\
$^{24}$Laboratoire AIM Paris-Saclay, UMR 7158, (CEA/IRFU - CNRS/INSU - Universit\'e Paris Diderot), CE Saclay, Bat
709 F-91191\\ Gif-sur-Yvette, France\\
$^{25}$Hubble Fellow; UCO/Lick Observatory, University of California, Santa Cruz, CA 95064, USA}

\begin{document}

\date{Accepted XXXX Received 2013; in original form 2013}

\pagerange{\pageref{firstpage}--\pageref{lastpage}} \pubyear{2013}

\maketitle

\label{firstpage}

\clearpage

\begin{abstract}
We explore the role of environment in the evolution of galaxies over $0.1<z<0.7$ using the final zCOSMOS-bright data set. Using the red fraction of galaxies as a proxy for the quenched population, we find that the fraction of red galaxies increases with the environmental overdensity $\delta$ and with the stellar mass $M_*$, consistent with previous works. As at lower redshift, the red fraction appears to be separable in mass and environment, suggesting the action of two processes: mass $\epsilon_m(M_*)$ and environmental $\epsilon_{\rho}(\delta)$ quenching. The parameters describing these appear to be essentially the same at $z \sim 0.7$ as locally. We explore the relation between red fraction, mass and environment also for the central and satellite galaxies separately, paying close attention to the effects of impurities in the central-satellite classification and using carefully constructed samples well-matched in stellar mass. There is little evidence for a dependence of the red fraction of centrals on overdensity. Satellites are consistently redder at all overdensities, and the satellite quenching efficiency, $\epsilon_{sat}(\delta, M_*)$, increases with overdensity at $0.1<z<0.4$. This is less marked at higher redshift, but both are nevertheless consistent with the equivalent local measurements. At a given stellar mass, the fraction of galaxies that are satellites, $f_{sat}(\delta, M_*)$, also increases with the overdensity. The obtained $\epsilon_{\rho}(\delta) / f_{sat}(\delta, M_*)$ agrees well with $\epsilon_{sat}(\delta, M_*)$, demonstrating that the environmental quenching in the overall population is consistent with being entirely produced through the satellite quenching process at least up to $z=0.7$. However, despite the unprecedented size of our high redshift samples, the associated statistical uncertainties are still significant and our statements should be understood as approximations to physical reality, rather than physically exact formulae.

\end{abstract}


\section{Introduction}

Correlations between various galaxy properties and their environment have been known for many years \citep[e.g.][]{Hubble.1939, Oemler.1974, Davis&Geller.1976, Dressler.1980} and have now been measured up to redshifts of about $z=1-1.5$ \citep[e.g.][]{Balogh.etal.2004, Kauffmann.etal.2004, Blanton.etal.2005, Hogg.etal.2004, Cucciati.etal.2006, Cucciati.etal.2010, Cooper.etal.2007, Cooper.etal.2010, Kovac.etal.2010b, Chuter.etal.2011, Quadri.etal.2012}. Of particular interest for this paper are the processes which evidently lead to the effective cessation, or `quenching', of star-formation in some galaxies. This process, or processes, moves galaxies out of the star-forming Main Sequence population and produces passive red galaxies witch have little or no continuing star formation. We will refer to the fraction of galaxies that have been quenched in this way, as the red fraction.

In a purely empirical approach to quantify the roles of stellar mass and environment in the evolution of galaxies, \citet{Peng.etal.2010} showed that at $z \sim 0$ the differential effects of stellar mass and environment in the quenching of galaxies are independent of each other, in the sense that the red fraction, i.e. the fraction of galaxies that have been quenched at a given stellar mass and at a given environmental overdensity, may be written as a separable function of these two parameters. \citet{Peng.etal.2010} demonstrated that this separability is maintained over 3 decades in the $\sim 1$ Mpc scale environmental overdensity and over 2 decades in stellar mass at $z \sim 0$ in the Sloan Digital Sky Survey \citep[SDSS;][]{York.etal.2000} sample. \citet{Peng.etal.2010} argued that this separability implied the action of two distinct processes, one linked to stellar mass but not environment, called `mass quenching' and the other linked to environment but not mass, `environment quenching'.

Various physical processes have been suggested which would operate only in dense environments (see e.g. \citealt{Boselli&Gavazzi.2006} for an extensive discussion). Among these are strangulation, in which a part or the whole hot gas halo is stripped and the fuel for the future star formation of now satellite galaxy is removed \citep*{Larson.etal.1980, Balogh.etal.2000, Balogh&Morris.2000}. In the case of a galaxy travelling through a sufficiently dense intra-cluster medium, external pressure may remove even the cold gas in the stellar disc, a process commonly referred to as ram pressure stripping \citep{Gunn&Gott.1972, Abadi.etal.1999}. Both strangulation and ram pressure stripping will suppress future star formation of a satellite galaxy, eventually transforming galaxies from blue to red and in the case of a spiral galaxy causing the appearance of spiral arms to fade. Strangulation is expected to affect star formation on the timescales of a few to more than 10 Gyr \citep{McCarthy.etal.2008}, while the ram pressure stripping should quench star formation on a shorter timescale, i.e. tens of Myr (\citealt{Abadi.etal.1999}; \citealt*{Quilis.etal.2000}), where the latter timescale does not take into account the eventual presence of a hot gaseous halo. Moreover, \citet{Rasmussen.etal.2006} have shown that the removal of cold galactic gas can occur also in the environments with a lower intra-medium density, such as groups, but on a greatly increased timescale, by up to two orders of magnitude. Satellites can also experience tidal stripping \citep[e.g.][]{Read.etal.2006}, and their morphologies can be modified by the cumulative effect of tidal forces due to multiple close encounters between galaxies, so called harassment \citep*{Farouki&Shapiro.1981, Moore.etal.1998}. Mergers of galaxies are also expected to be more frequent in dense (but not the most dense) environments, and according to the semi-analytical models (SAMs), the merger rate is 4-20 times higher in regions with a high number overdensity of galaxies than in regions with a $\sim 100$ lower number overdensity \citep{Jian.etal.2012}.

It is clear from the list above that, from a theoretical point of view, the majority if not all of the environmental differences in the overall galaxy population may be caused by the transformation of galaxies after they infall into a larger halo, i.e. after they become a satellite. For years, such theoretical predictions were observationally hard to test as the galaxy-environment studies were based on the local density of galaxies used as an environment indicator (starting from the seminal work of \citealt{Dressler.1980}) or they were based on comparing galaxies in and out of the dense structures. Only with the advent of the large spectroscopic surveys (such as SDSS at $z \sim 0$ \citep{York.etal.2000} and zCOSMOS at $z < 1$ \citep{Lilly.etal.2007, Lilly.etal.2009} it has become possible to reliably separate galaxies into centrals and satellites over a broad range of galaxy and halo masses. In a now growing number of papers, it has been demonstrated that the observations at $z \sim 0$ support the scenario in which environment specific processes indeed act on galaxies upon their infall into a larger halo: at the same stellar mass, satellite galaxies are on average redder and more concentrated than the central galaxies \citep{vandenBosch.etal.2008} while late-type satellites have smaller radii, larger concentration, lower surface brightness and redder colours than the late-type centrals \citep{Weinmann.etal.2009}. About 40-50$\%$ of blue galaxies at any stellar mass transform to red galaxies after being accreted into a larger halo {\citep{vandenBosch.etal.2008, Peng.etal.2012}. This fraction also increases with increasing overdensity and/or decreasing group-centric radius \citep{Peng.etal.2012, Wetzel.etal.2012}.

One could also imagine that environmental effects could also affect the properties of central galaxies, especially if the cooling of gas onto the central galaxy is linked to the mass of the surrounding halo \citep*[e.g.][]{White.etal.1997, McDonald.etal.2011}.

\citet{Peng.etal.2012} showed that, at $z\sim 0$, satellite quenching is sufficient to explain the bulk of the observed environmental dependence of the red fraction in the overall population of galaxies, at least when the local overdensity of galaxies is used as a proxy for environment. In other words, to first order, while all galaxies experience mass quenching, only satellites experience the effects of environmental quenching. Such statements cannot of course be exact, but rather should be taken as good approximations to an underlying, no doubt complex, reality. Not least, several authors \citep[e.g.][]{Woo.etal.2012} have pointed out that, at fixed stellar mass, the red fraction of centrals increases with the halo mass. This however can be explained as a residual effect of the continued increase of dark mass, but not stellar mass, after a galaxy has been quenched (see \citealt{Lilly.etal.2013} for details). There is also evidence \citep{Peng.etal.2010} that the stellar mass function of passive galaxies is modestly altered by subsequent merging in high density environments, producing small changes in the overall mass function of galaxies \citep{Bolzonella.etal.2009} and perturbing the precise separability in red fraction that might otherwise be expected.

At higher redshifts, the picture is less clearly defined, simply because of the absence until recently of sufficiently large samples of galaxies with quantitative environment measures. \citet{Peng.etal.2010} suggested from an initial analysis of the first half of the zCOSMOS-bright survey that separability was maintained to $ z \sim 1$ and that the differential effects of stellar mass and environment, i.e. the mass and environmental quenching efficiencies were also more or less constant. Analysis of the full zCOSMOS-bright group catalogue \citep{Knobel.etal.2012a} indicates that, to $z \sim 0.8$ satellites continue to be redder than centrals of the same mass, with the satellite quenching being, as at low redshift, independent of stellar mass. The mean value of the satellite quenching efficiency $\epsilon_{sat}$ (averaged over all overdensities) has the same value out to $z \sim 0.8$ as locally, i.e. $\epsilon_{sat} \sim 0.5$ (\citealt{Knobel.etal.2012b}, see also \citealt{Quadri.etal.2012, vanderBurg.etal.2013}).

The aim of the current paper is to construct the red fraction-environment relation up to $z=0.7$ using the final so-called ``20k'' sample of the zCOSMOS-bright galaxies and to understand its origin in terms of the central-satellite dichotomy. In particular, we wish to critically test, with a larger data set and with a more sophisticated treatment of potential systematic effects, whether the relations expected from the formalism of \citet{Peng.etal.2010, Peng.etal.2012} which are derived mostly in the local SDSS sample continue to hold at these higher redshifts. Previous studies of environmental effects in the zCOSMOS sample addressed the dependence of colour \citep{Cucciati.etal.2010, Iovino.etal.2010, Knobel.etal.2012b}, morphology \citep{Tasca.etal.2009, Kovac.etal.2010b}, stellar mass function \citep{Bolzonella.etal.2009}, X-ray detected AGNs \citep{Silverman.etal.2009}, infrared-luminous galaxies \citep{Caputi.etal.2009}, radio-emitting galaxies \citep{Bardelli.etal.2010}, post-starburst galaxies \citep{Vergani.etal.2010} and close pairs \citep{Kampczyk.etal.2013} on environment \citep{Kovac.etal.2010a, Knobel.etal.2009, Knobel.etal.2012a}. Using only photometric redshifts, \citet{Scoville.etal.2007b, Scoville.etal.2013} extended the environment reconstruction in the COSMOS field up to $z=3$.

The layout of the paper is as follows. We introduce the data set and the derived products needed for the analysis in Section~\ref{sec_data}. In Section~\ref{sec_envquench}, we measure the relation between the red fraction of all galaxies and their environments, and in Section~\ref{sec_censat}  we investigate the equivalent relations obtained when separating population of galaxies into centrals and satellites. The comparison between different environment estimators and with previous results is presented in Section~\ref{sec_whichenv}. Final conclusions are presented in Section~\ref{sec_concl}. A concordance cosmology with $\Omega_m = 0.25$, $\Omega_{\Lambda} = 0.75$, and $H_0 = 70$ \kms Mpc$^{-1}$ is assumed throughout. Magnitudes are given in the AB system. We use the term "dex" to express the antilogarithm, where 0.1 dex equals to 1.259.

\section[]{Data}
\label{sec_data}

\subsection[]{zCOSMOS survey}
\label{sec_zCOSMOS}

zCOSMOS \citep{Lilly.etal.2007, Lilly.etal.2009} is a redshift survey conducted in the 2 deg$^2$ area of the large legacy HST programme COSMOS \citep{Scoville.etal.2007a}. The observations were carried out using the VIMOS spectrograph \citep{LeFevre.etal.2003} on the 8m  VLT/UT3 telescope spread over 600 hrs of service dark time. As a part of the zCOSMOS campaign, the spectra of about 20,000 galaxies down to $I_{AB}<22.5$ were obtained over the whole COSMOS field producing a sample in the redshift range $z<1.2$ (zCOSMOS-bright). In addition, about 10,000 targets in the inner 1 deg$^2$ were selected by a combination of the colour-colour criteria aimed to select galaxies in $1.8<z<3$ (zCOSMOS-deep).

In this paper we use galaxies with reliable redshifts from the final zCOSMOS-bright catalogue, the so called 20k sample. This catalogue is obtained by removing galaxies with confidence classes 0,1.1-1.4,2.1,2.3, 9.1, and also the secondary and broad-line objects with corresponding confidence classes (see \citealt{Lilly.etal.2007, Lilly.etal.2009} for a description), leaving us with 17042 objects with reliable redshifts. On average, a sampling rate of $\sim 50\%$ is achieved (e.g. Figure 1 in \citealt{Knobel.etal.2012a}) with the redshift uncertainty about 100 \kms.

\subsection[]{Photometric redshifts and properties of galaxies}
\label{sec_galprop}

\subsubsection{Photometric redshifts}

The targeted COSMOS field has a multitude of follow-up photometric observations by a range of ground based and spaced based facilities covering the UV-NIR range \citep[e.g.][]{Capak.etal.2007}, and the Spitzer IRAC bands \citep{Sanders.etal.2007}. The large amount of multi-wavelength data and the precision of the HST ACS data are utilised to derive state-of-the-art photometric and structural properties of all zCOSMOS galaxies.

Although the primary analysis is based on the spectroscopic sample, we use the photometric redshifts (both the photometric redshift probability distribution function, i.e. $PDF(z)$, and the maximum likelihood photometric redshift) of zCOSMOS galaxies at two points (see Sections~\ref{sec_masscompl} and~\ref{sec_density}). These were generated using ZEBRA+ \citep{Oesch.etal.2010, Knobel.etal.2012a}, an extension of the ZEBRA code \citep{Feldmann.etal.2006}. The code is based on the spectral energy distribution (SED) fitting to the photometric data allowing both Maximum Likelihood and Bayesian approach. The photometric redshifts which we use are obtained from the Bayesian ZEBRA+ run based on the \citet{Bruzual&Charlot.2003} models with the emission lines, with the template correction module of ZEBRA based on a randomly selected subset of zCOSMOS spectroscopic redshifts. The fitting is carried out on the 26 photometric bands ranging from the CFHT ${\it u^*}$ to the Spitzer IRAC 4.5 $\mu$m, where 12 of the used bands are broad, 12 bands are intermediate and 2 bands are narrow. The uncertainty of the photometric redshifts is $0.01(1+z)$ with a catastrophic failure rate of 2-3\%.

\subsubsection{Stellar masses and luminosities}

\begin{figure}
\includegraphics[width=0.43\textwidth]{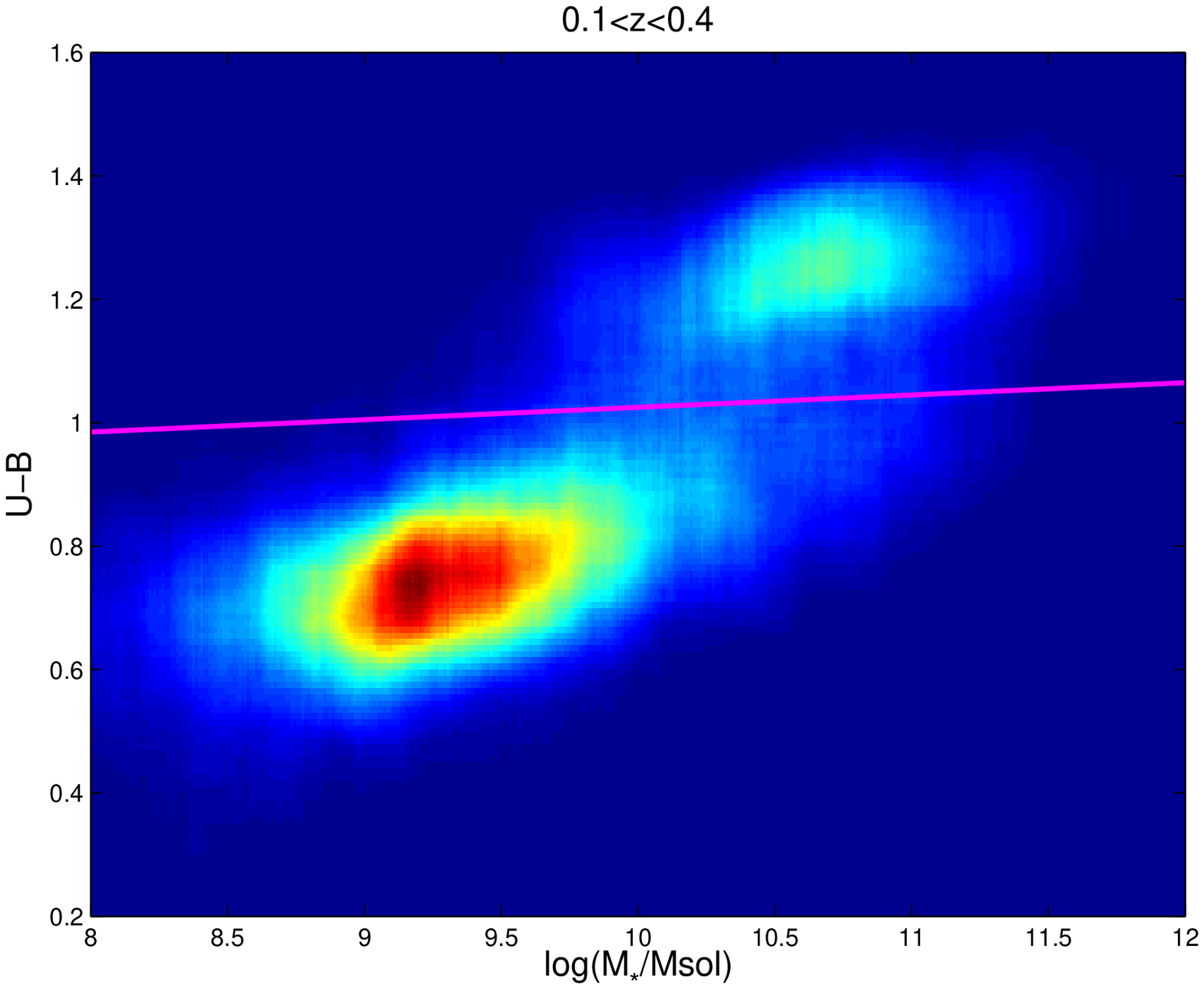}
\includegraphics[width=0.43\textwidth]{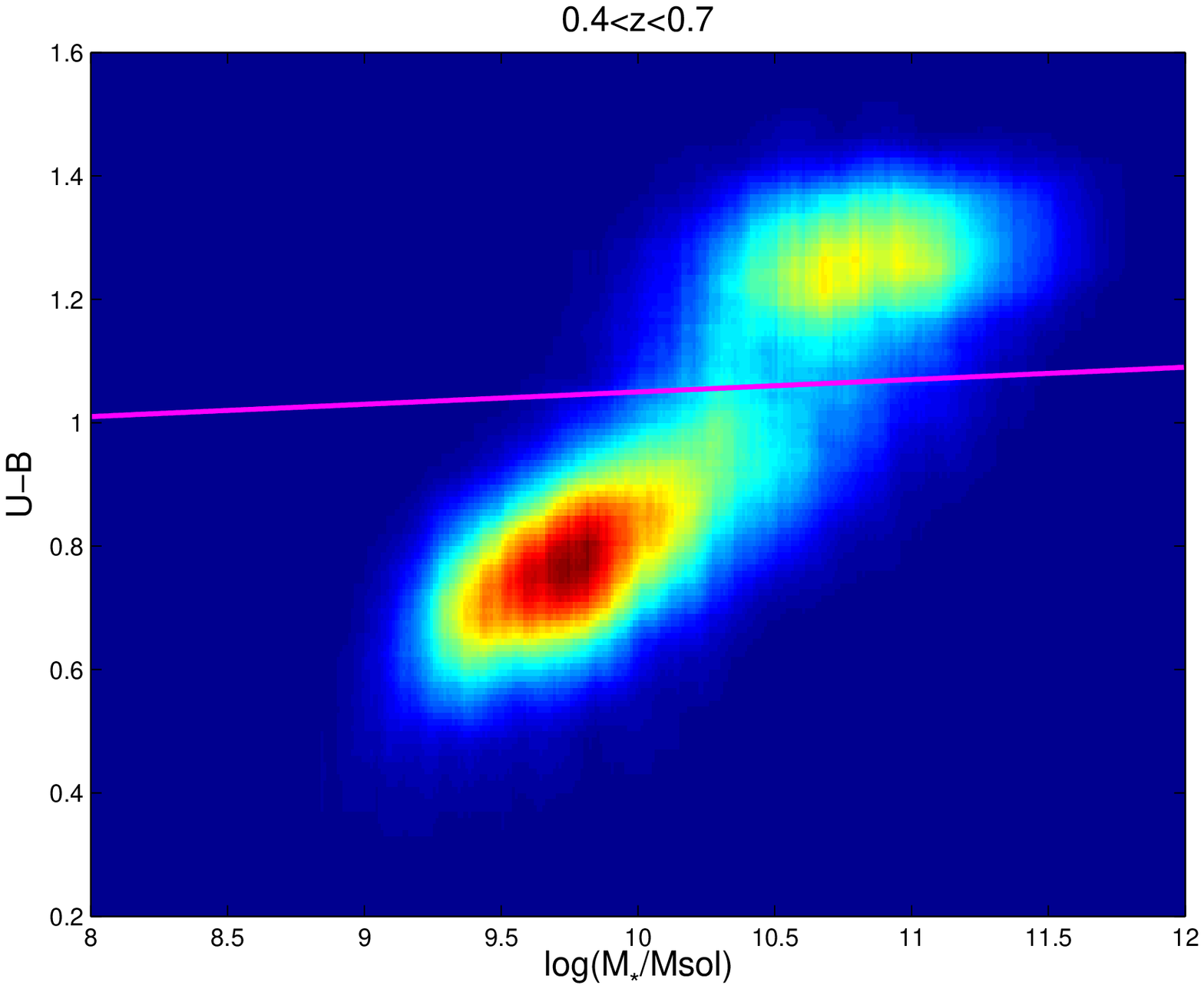}
\caption{\label{fig_colmasscut}Rest-frame $U-B$ colour versus stellar mass for zCOSMOS galaxies in $0.1<z<0.4$ (top) and $0.4<z<0.7$ (bottom). Panels are colour coded according to the numbers of galaxies calculated in a sliding box $\Delta (U-B) = \Delta \log(M_*/$\Msol$) = 0.2$ with a step of 0.01 in both parameters. The continuous line marks the adopted division between the blue and red galaxies. In this work  we will refer to all galaxies above the division line as red (i.e. quenched).}
\end{figure}

The same code is also utilised to derive stellar masses $M_*$ and luminosities of all zCOSMOS galaxies but using only the 12 broad band filters and the \citet{Bruzual&Charlot.2003} templates without emission lines, assuming the \citet{Chabrier.2003} initial stellar mass function (IMF) and \citet{Calzetti.etal.2000} dust model. During the fitting process the redshift of a galaxy is fixed to the spectroscopic one for galaxies with reliable spectroscopic redshifts, otherwise the maximum likelihood photometric redshift is used. The stellar masses which we use are obtained by integrating the star formation rate (SFR) and they include the mass of gas processed by stars and returned to the interstellar medium (this is different than in the previous results by our group based on the 10k zCOSMOS sample). This is particularly useful when comparing quiescent galaxies over different redshifts, as their masses will remain the same. On average, the total integrated stellar masses are higher for about 0.2 dex than the stellar masses with the mass return subtracted.

\subsubsection{Colour and SFR bimodality}

The rest-frame $U-B$ colour versus stellar mass diagrams of the zCOSMOS 20k galaxies in $0.1<z<0.4$ and $0.4<z<0.7$ are shown in Figure~\ref{fig_colmasscut} in the top and bottom panels, respectively. The continuous line corresponds to our adopted division of galaxies into blue and red (consistent with no redshift evolution in the colour division cut, mirroring \citealt{Knobel.etal.2012b}). The division is somewhat arbitrary, as there is no clear separation between the blue and red galaxies in our data, i.e. when plotting the colour histogram in narrow bins of mass the minimum in the histogram does not show a clear trend with the mass (probably also due to the small numbers of galaxies). The division which we use is such that the blue cloud galaxies are below this division line and a slope of the division is similar to the slope of red sequence galaxies.

We will use red galaxies as a proxy for the quenched population as it is measured in a simple well-defined way for all zCOSMOS galaxies. A weakness of such an approach, commonly pointed out, is the presence of dust in star-forming galaxies, which will redden galaxies and put some of the dusty star formers into the population of the colour-defined quenched galaxies. To understand the influence of different selection of the quenched population, we derive the fraction of red and non-star-forming galaxies and compare the two as a function of stellar mass.

For this, we checked the different SFR estimators and selected to use SFR derived from IR+UV, for galaxies detected in 24 $\mu$m, and, otherwise, an SFR derived from the SED fit. There is a good agreement between these two estimators, while the [OII] based SFR show a rather large scatter with the other two SFR measurements. We use the 24 $\mu$m data down to $S_{24 \mu m} \ge 80$ $\mu$Jy obtained from the Spitzer observations \citep{LeFloch.etal.2009}. The total IR luminosity is obtained following \citet{Wuyts.etal.2011} and, combining it with the UV from the SED fit, the total SFR is calculated following \citet{Kennicutt.1998} assuming \citet{Chabrier.2003} IMF.

\begin{figure}
\includegraphics[width=0.43\textwidth]{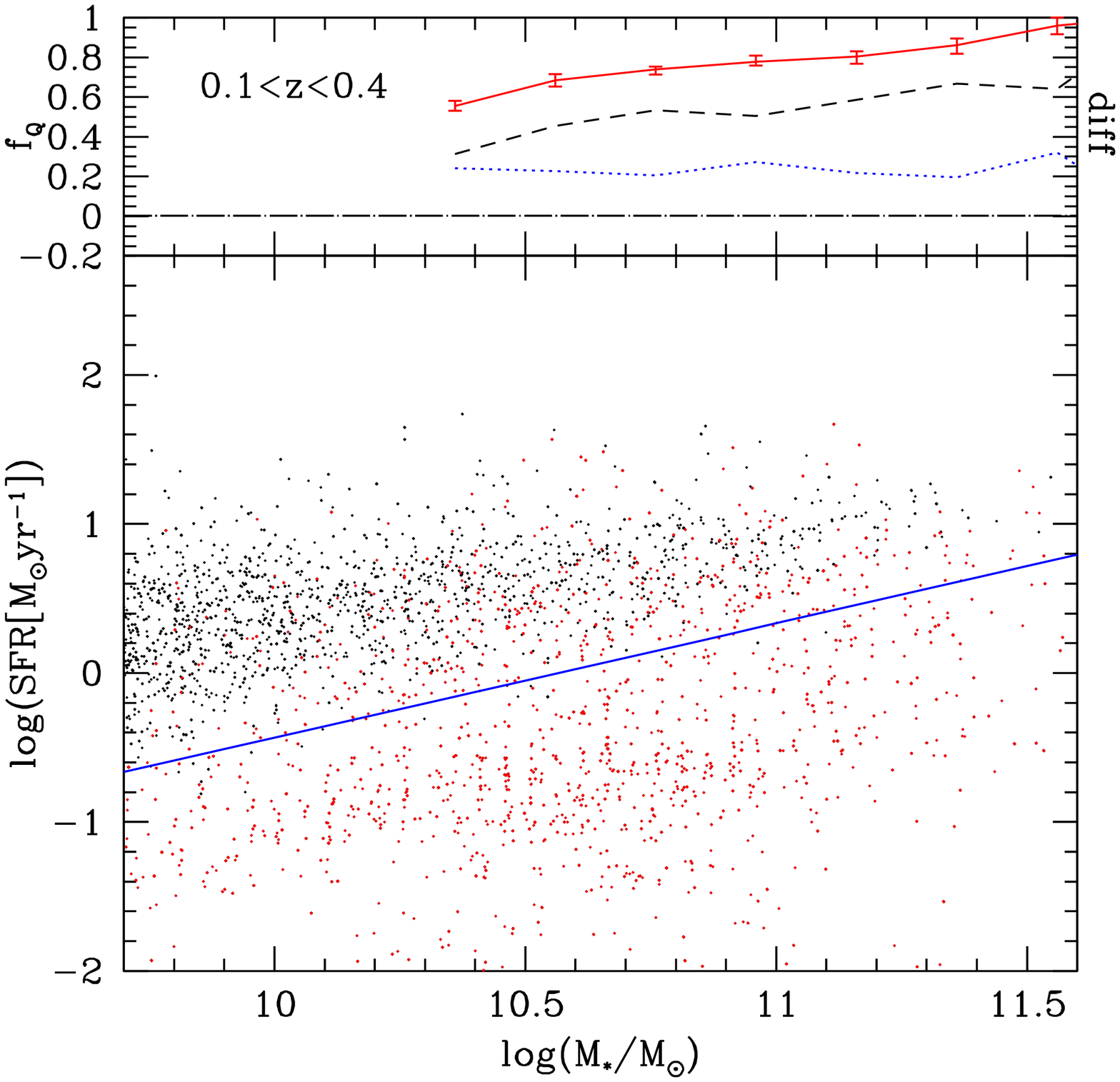}
\includegraphics[width=0.43\textwidth]{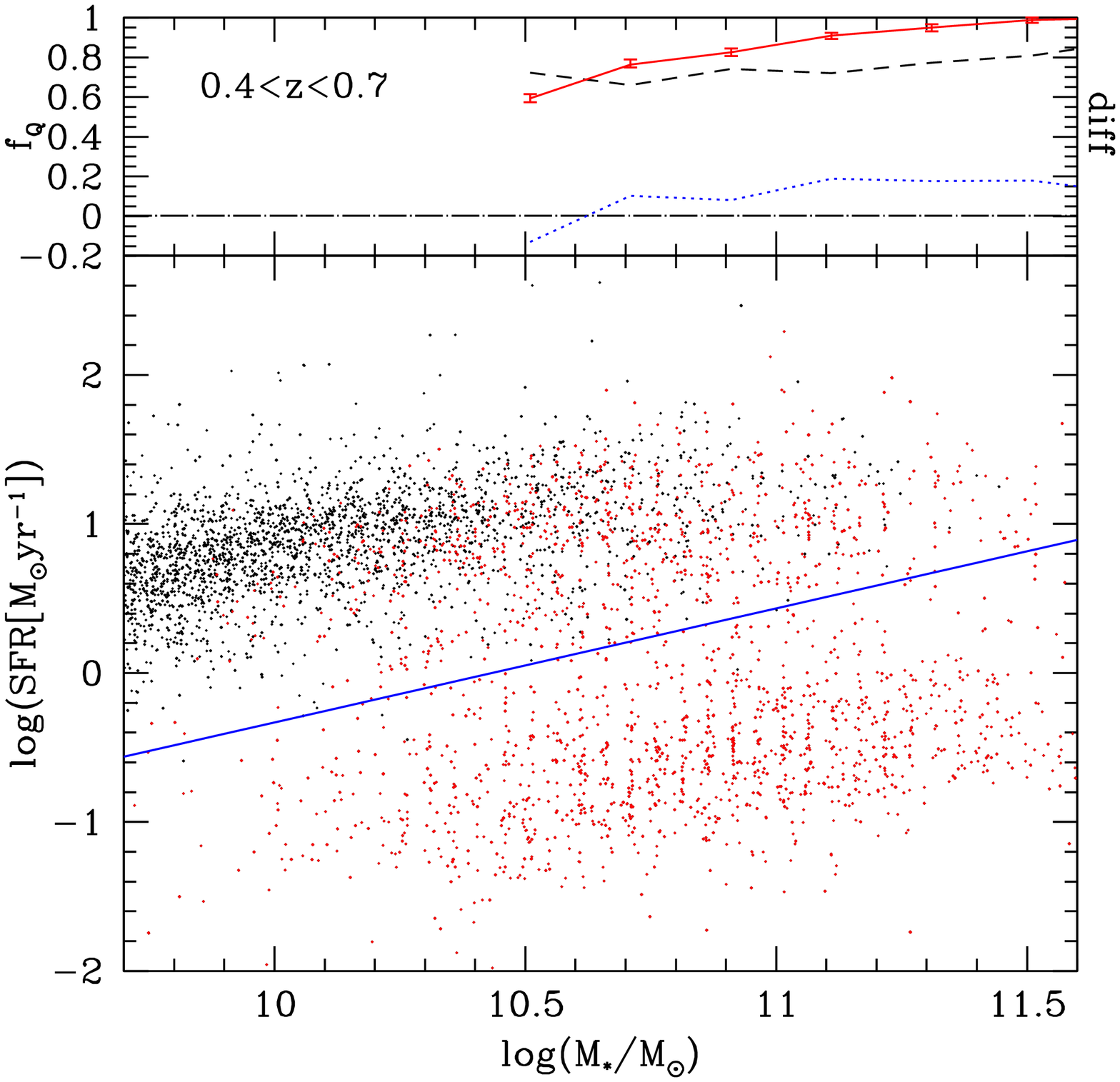}
\caption{\label{fig_sfrvsmass}Difference in selecting the quenched population of galaxies by colour and by SFR in $0.1<z<0.4$ (top figure) and $0.4<z<0.7$ (bottom figure). The SFR-stellar mass relations are shown in the bottom panels, marking in red galaxies termed quenched according to the colour bimodality. The continuous lines correspond to the adopted division between the star-forming and non star-forming galaxies. The quenched fractions of galaxies $f_Q$ are shown in the top panels, derived when separating galaxies using colour (red continuous lines) and SFR (black dashed lines). The difference between those two fractions is shown as the blue dotted lines. The long-dashed-dotted lines mark the zero values. The error bars are estimated by bootstrapping (16-84$\%$ interval) and for clarity they are shown only for the red fractions.}
\end{figure}

The SFR-stellar mass plots are shown in the lower two panels in each part of Figure~\ref{fig_sfrvsmass}, where the continuous line is used to separate star-forming and non star-forming (quenched) galaxies, following the division by \citet{Knobel.etal.2012b}. We colour-code galaxies classified to be red according to the previously discussed colour bimodality of zCOSMOS galaxies. As expected, some of the red galaxies lie on the star-forming sequence. This is quantified in the top panels, where we show the fraction of red (continuous lines) and non star-forming (dashed lines) galaxies and the difference in these fractions (dotted lines) as a function of stellar mass, showing only the points where the red galaxies are complete above 40$\%$ with respect to the fully sampled deeper, photometric data set (mass completeness will be discussed in Section~\ref{sec_masscompl}; here we do not correct the measured fractions for the incompleteness in stellar mass). While the red fractions are systematically higher than the fractions of non star-forming galaxies (except for one point), the difference is not a strong function of stellar mass. The difference is constant at about 20$\%$ in $0.1<z<0.4$, and it changes from 15-19$\%$ in $0.4<z<0.7$, ignoring the lowest mass bin at higher redshift where there is a higher fraction of non star-forming than red galaxies, but mass incompleteness is also highest. We suspect that the apparent weak mass dependence in the difference between the two fractions in $0.4<z<0.7$ is a result of the change in SFR tracer among lower mass non star-forming galaxies, as we are complete only to about $\log(SFR[$\Msol$yr^{-1}])=0.4$ for the IR+UV measurement in this redshift bin. The IR+UV based SFR tends to empty the so-called green valley in colour-mass diagram and make the fraction of non star-forming galaxies lower.

We conclude that even though there are systematic differences in fractions of quenched galaxies when defined by colour and by SFR, they are almost independent of stellar mass, and therefore, qualitatively our conclusions should be also applicable when separating galaxies by their SFR. Finally, there are many other ways to define the quenched population of galaxies (discussed extensively in the case of zCOSMOS in \citealt{Moresco.etal.2013}), and one needs to take care of these differences when comparing results reported in the literature.

\subsection{Mass completeness}
\label{sec_masscompl}

To understand the mass completeness in the zCOSMOS sample we make use of the available, much deeper, photometric data. For this purpose we use an independently estimated catalogue with photometric redshifts, masses and luminosities of the COSMOS galaxies down to $i=24$, also generated using the ZEBRA+ code. The completeness is calculated as the fraction of zCOSMOS $i<22.5$ galaxies with reliable spectroscopic redshifts in the sample of $i=24$ galaxies, as a function of stellar mass in the zCOSMOS sampled area. The $i=24$ sample can be assumed to be complete at the masses of interest. We calculate it separately for the red and blue galaxies in the two bins of consideration. By comparing the reliable spectroscopic sample of the zCOSMOS galaxies to the overall $i=24$ sample we take into account the effects of any residual incompleteness stemming from the dependence on the galaxy colour and other properties of the ability to obtain its spectroscopic redshifts, even those are in principle small (Figures 2 and 3 in \citealt{Lilly.etal.2009} for the 10k zCOSMOS).

Throughout this paper we use the inverse values of these fractions  $w_c(M_*, z, U-B)$ to weight galaxies in order to correct for the colour-dependent mass incompleteness, using different weights at each mass depending on the colour of a galaxy and its redshift. Furthermore, we only consider masses above mass-completeness limits, taken to be the mass where the zCOSMOS red galaxies reach 30$\%$ completeness relative to the deeper $i=24$ sample. In $0.1<z<0.4$ this completeness limit is at $\log(M_*/$\Msol$) = 9.82$ and in $0.4<z<0.7$ it is at $\log(M_*/$\Msol$) = 10.29$.

\subsection[]{Environmental overdensity}
\label{sec_density}

We quantify environment by measuring the overdensity of tracer galaxies at the position of each galaxy in $0.1<z<0.7$, as well as a central/satellite classification (Section~\ref{sec_groups}). The overdensity $\delta$ is defined as

\be \delta = \frac{\rho - \bar{\rho}}{\bar{\rho}} \ee

\noindent
where $\rho$ is the density of tracer galaxies in a given aperture, and $\bar{\rho}$ is the mean density at the redshift of a galaxy whose environment is being measured. \citet{Kovac.etal.2010a} discuss in great detail the quantification of galaxy environments, focusing particularly on the optimisation of the density field reconstruction for the zCOSMOS survey. In that paper a new algorithm ZADE was developed to take into account galaxies without spectroscopic redshifts. In essence, ZADE is based on modifying the PDF(z) of galaxies not yet observed spectroscopically by the number counts of galaxies with reliable redshifts within some spherical $R_{ZADE}$ apertures along the line of sight. In this way, the precise R.A.-dec positions of all galaxies are preserved and in a statistical sense the mean intergalactic separation is equal to that of the total sample of tracer galaxies, and not only of those targeted spectroscopically. Importantly, any of the complicated spatial selection function is taken automatically into account. With the ZADE algorithm any of the standard density measures can be applied, resulting in the density field reconstructed on a smaller scale and with less noise then in the case of otherwise used statistical weighting for the objects without spectroscopy. 

Following \citet{Kovac.etal.2010a}, we use the ZADE algorithm to produce modified PDF(z) of $i<22.5$ galaxies in the zCOSMOS field which are not part of the 20k zCOSMOS sample with reliable redshifts (described in Section~\ref{sec_zCOSMOS}). Though with the 20k sample we have about twice as many galaxies with reliable redshifts, than in the 10k sample, the spectroscopic coverage increased towards the edges of the survey area, as well as the survey's area itself and we use the same $R_{ZADE}=5$ \hh Mpc as for the density reconstruction with the 10k sample (checked on the mock catalogues). To measure densities we use circular apertures defined by the distance to the 5th nearest neighbour projected within $\pm 1000$ \kms\ from the sample of tracer galaxies. We use the  $M_B<-19.3-z$ set of tracer galaxies which is luminosity complete up to $z=0.7$. These are the aperture and the tracer set used for many of the analyses with the 10k sample and $z \sim 0$ SDSS analysis by \citet{Peng.etal.2010, Peng.etal.2012} and allow direct comparison with the previous results. The mean density is obtained by adding the $\Delta V(z)/V_{max}$ contribution of all tracer galaxies (both with spectroscopic and photometric redshifts) at a given redshift, following the method outlined in \citet{Kovac.etal.2010a}. According to the tests that were carried out on the mock catalogues from \citet{Kitzbichler&White.2007}, the error interval ($\pm 1 \sigma$) in such reconstructed overdensities at the positions of galaxies is in the range $-0.2<\log(1+\delta)<0.1$, without systematic offset in $-1<\log(1+\delta)<0.75$, while in more dense regions the reconstructed observed overdensity is systematically lower than the true mock value, by about $\log(1+\delta)=0.1$ at $\log(1+\delta) \approx 2$.

\subsection[]{Group catalogue and classification of galaxies into centrals and satellites}
\label{sec_groups}

The group catalogue employed in this work is reconstructed from the 20k zCOSMOS sample as described in \citet{Knobel.etal.2012a}. The final catalogue includes 1498 groups in the redshift range $0.1 \lesssim z \lesssim 1.0$, where the fraction of galaxies in the groups decreases from 35$\%$ to 10$\%$ over the same redshift interval. The reconstruction is based on the flux limited sample of galaxies with reliable redshifts, utilising the friends-of-friends algorithm in the multi-run mode introduced in \citet{Knobel.etal.2009}, where different group finding parameters are used depending on the group richness. The optimisation of the group finding parameters is carried out on the \citet{Kitzbichler&White.2007} mock catalogues, but the final assessment of quality of the reconstructed groups is recalculated on the updated COSMOS cones described in \citet{Henriques.etal.2012}. According to these, the zCOSMOS 20k groups have both completeness and purity of around 83$\%$. Typical halo masses of the groups are estimated using the correlation between the number of group members, corrected for the spatial sampling and the redshift success rate, and the halo mass calibrated on the mock catalogue.

\citet{Knobel.etal.2012a} developed a scheme to separate galaxies into centrals and satellites, where by definition the most massive galaxy in a group is the central, and all other group galaxies are satellites. To improve the quality of the central/satellite classification in the overall 50$\%$ zCOSMOS spectroscopically complete sample, also the galaxies with only photometric redshifts (down to the flux limit of the zCOSMOS sample) are included. The classification of a galaxy to be a central or a satellite is given in a probabilistic manner through the combination of values of the three parameters: $p$, $p_M$, and $p_{MA}$. These parameters correspond to a probability of a galaxy to be a member of a group, a probability that a given galaxy is the most massive in a group, and a probability that a galaxy is the most massive normalised by the area of the projected Voronoi cell of that galaxy, respectively (see \citealt{Knobel.etal.2012a} for more details).

Following \citet{Knobel.etal.2012a} we separate zCOSMOS galaxies into dichotomous samples of centrals and satellites identifying the satellites as galaxies satisfying the following criteria simultaneously: $p>0.1, p_M<0.5$, and $p_{MA}<0.5$. The central galaxies are considered to be galaxies not satisfying at least one of these criteria. Such defined samples of zCOSMOS centrals and satellites in $0.1<z<0.8$ are complete at the level of 93 and 54$\%$, respectively, and their purities are at the level of 84 and  74$\%$, respectively \citep{Knobel.etal.2012a}. This includes the uncertainty that in the mocks the most massive galaxy in a group is not always the central galaxy.\footnote{In the mocks, central galaxies (i.e. `type 0') are not identified by their stellar mass; they are the central galaxies of main subhaloes \citep{Guo.etal.2011}.} Finally, the number of galaxies in different samples used in the subsequent analysis is given in Table~\ref{tab_galsamples}.

\begin{table}
\caption{\label{tab_galsamples}Galaxy samples.}
\begin{tabular}{c|c|c|c|c|c}
\hline
&$z1<z<z2$ &  All &  Centrals &  Satellites \\
\hline
&$0.1 < z < 0.4$ & 2340 & 1730 & 610 \\
&$0.4 < z < 0.7$ & 2448 & 2062 & 386 \\
\hline
\end{tabular}

\medskip
Number of galaxies in different samples used in the analysis. The redshift bins $z_1 < z < z_2$ are given in the first column, and the numbers of all, central and satellite galaxies above the adopted stellar mass completeness limit in a given redshift bin are given in the second, third, and forth column, respectively. 
\end{table}

\section[]{Colour-density relation for all galaxies}
\label{sec_envquench}

\subsection{Red fractions of zCOSMOS galaxies}
\label{sec_redfracall}

\begin{figure*}
\includegraphics[width=0.4\textwidth]{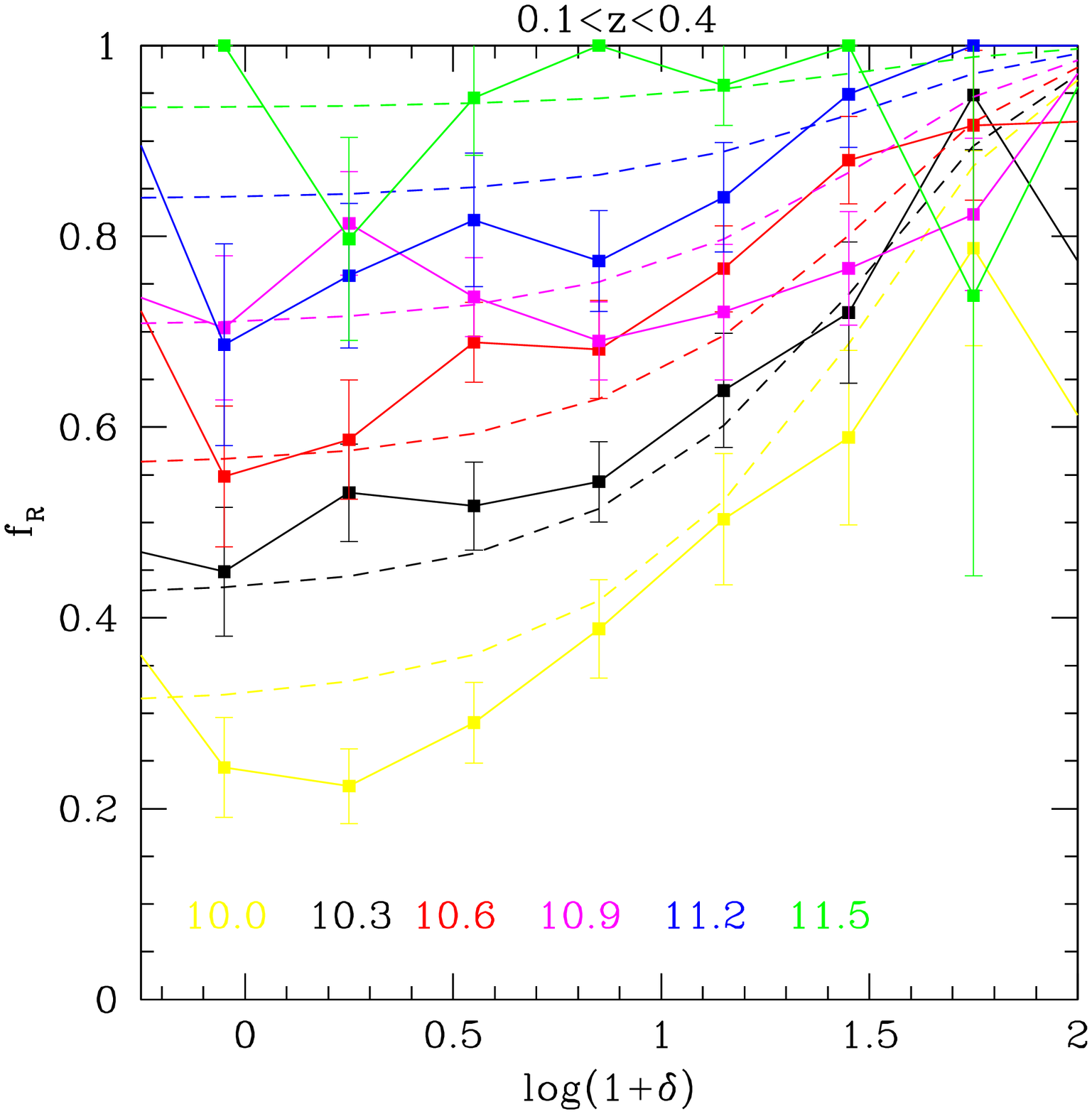}
\includegraphics[width=0.4\textwidth]{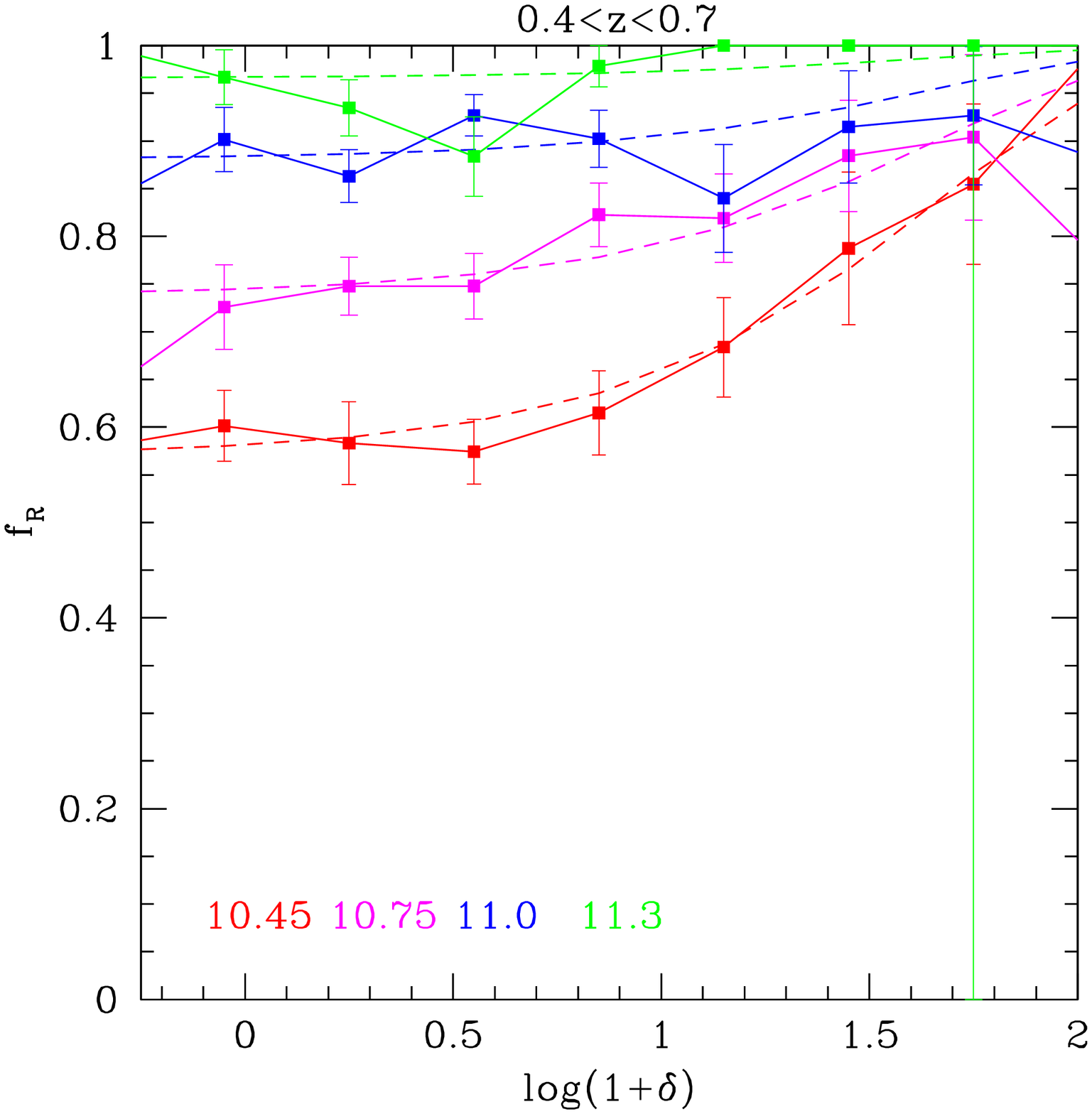}
\caption{\label{fig_redfrvsod}Red fraction as a function of overdensity in different mass bins. Results in the left and right panels are for $0.1<z<0.4$ and $0.4<z<0.7$, respectively. The measurement points connected by the continuous lines and the dashed curves are for the 0.3 dex bins in stellar mass, with the stellar mass increasing from the bottom to the top. The lowest mass bin starts at the adopted mass-completeness value in that redshift bin, and the rounded values of the centres of the used mass bins are given in the increasing order at the bottom of each panel. The symbols and the continuous lines correspond to the zCOSMOS measurements (Section~\ref{sec_redfracall}), and the dashed lines correspond to the best-fit model given by equation (\ref{eq_redfrmodel}) (Section~\ref{sec_redallmodel}). The error bars correspond to half of the 16-84$\%$ interval of the ordered red fractions from 100 bootstrapped samples in each redshift bin.}
\end{figure*}

\begin{figure*}
\includegraphics[width=0.4\textwidth]{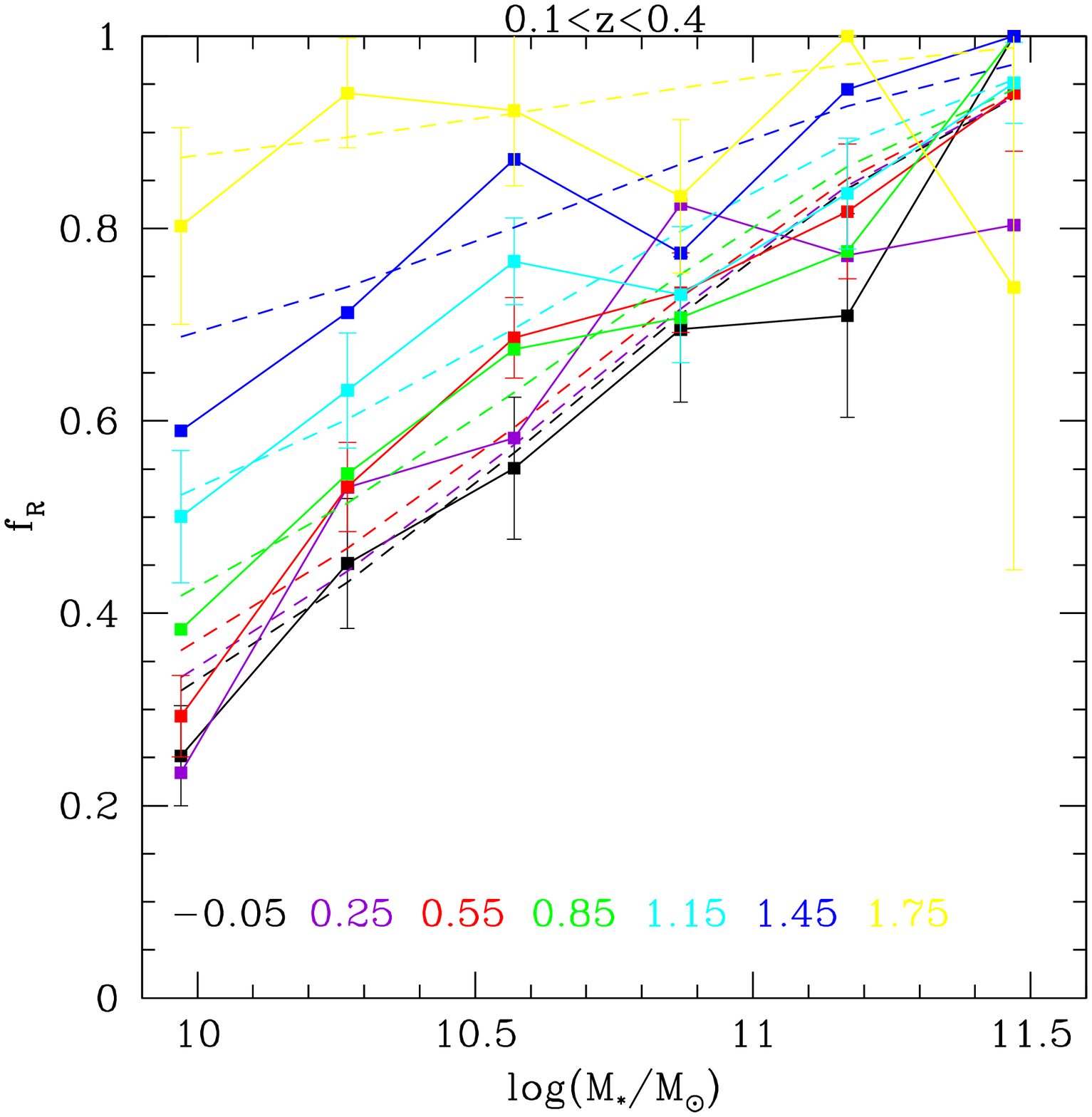}
\includegraphics[width=0.4\textwidth]{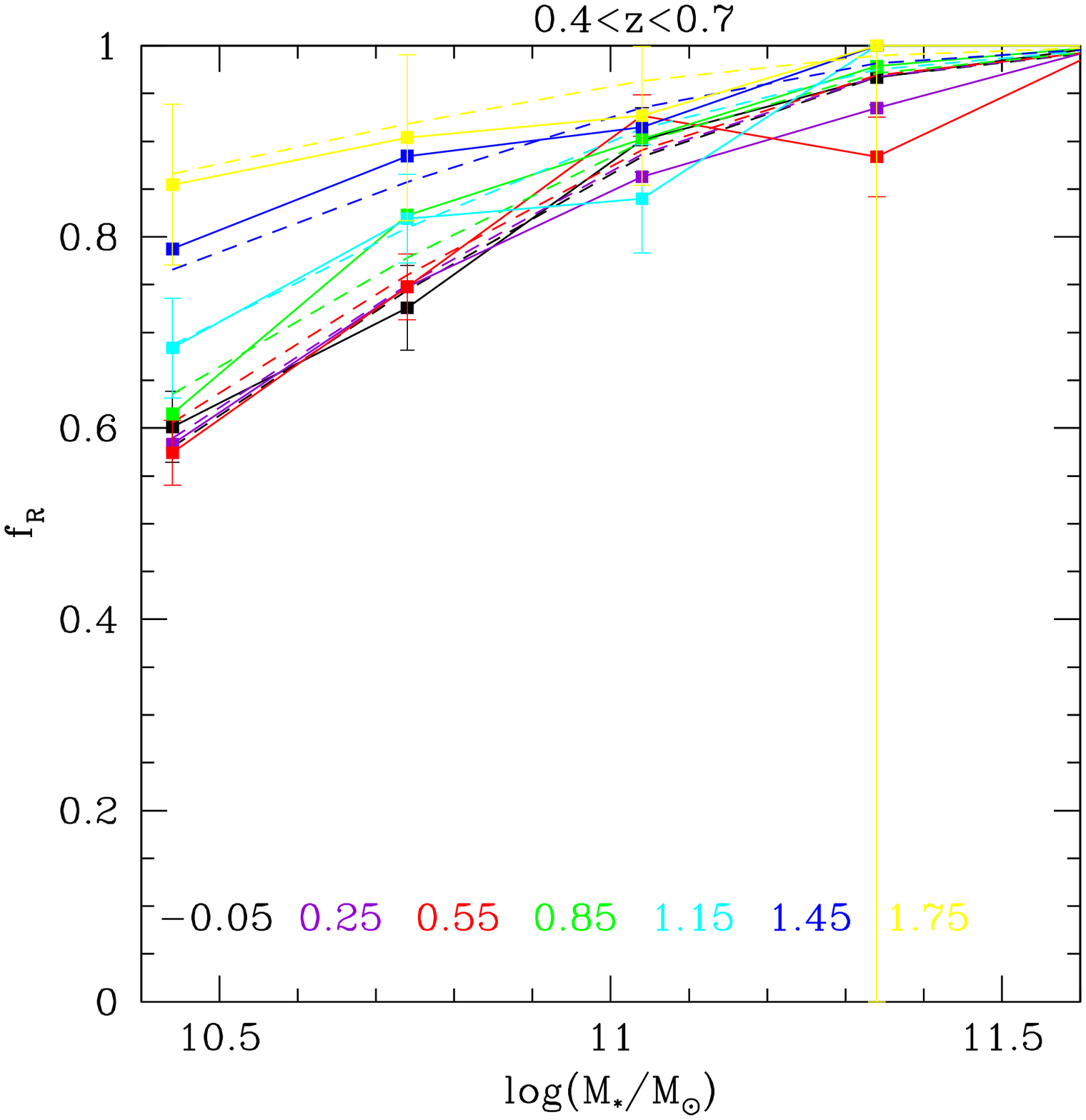}
\caption{\label{fig_redfrvsmass}Red fraction as a function of stellar mass in different bins of overdensity. Results in the left and right panels are for $0.1<z<0.4$ and $0.4<z<0.7$, respectively. The measurement points connected by the continuous lines and the dashed curves correspond to the 0.3 dex bins in $\log(1+\delta)$, where the environmental overdensity increases from the bottom to the top. The values of the centres of the used environment bin, in an increasing order, are shown at the bottom of each panel. The horizontal axis starts at the adopted mass-completeness value in a given redshift bin. The symbols and the lines correspond to the zCOSMOS measurements (Section~\ref{sec_redfracall}), and the dashed lines correspond to the best-fit model given by equation (\ref{eq_redfrmodel}) (Section~\ref{sec_redallmodel}). The error bars correspond to half of the 16-84$\%$ interval of the ordered red fractions from 100 bootstrapped samples in each redshift bin, shown for clarity only for every second $\log(1+\delta)$ bin.}
\end{figure*}

Numerous studies, both in the local Universe and at higher redshifts, have shown that the properties of galaxies depend on both their stellar mass and on their environment \citep[e.g.][]{Kauffmann.etal.2003b, Baldry.etal.2006, Peng.etal.2010}. In order to isolate the effects of the latter, it is necessary to carefully control the selection of the sample in terms of the former, especially given the evidence that the mass function of galaxies may be different in different environments \citep[e.g.][]{Bolzonella.etal.2009, Kovac.etal.2010b}.

In this section we explore the relation between the red fraction of zCOSMOS galaxies, using the red galaxies as a proxy for the quenched population, and their environment and stellar mass. In Section~\ref{sec_censat}, we will split the sample into centrals and satellites. We calculate the fraction of red galaxies in bins of $\Delta \log(M_*/$\Msol$) = \Delta \log(1+\delta)=0.3$ above the mass completeness limits (Section~\ref{sec_masscompl}) in two redshift intervals $0.1<z<0.4$ and $0.4<z<0.7$. We weight galaxies by their individual colour-dependent mass-completeness $w_c(M_*, z, U-B)$ computed in Section~\ref{sec_masscompl}. The statistical uncertainties are taken to be  half of the $16 - 84\%$ intervals obtained by bootstrapping the samples 100 times.

The resulting colour-density relations are shown in Figure~\ref{fig_redfrvsod} for independent stellar mass bins. Although there is some scatter, there is a clear trend that, at a given stellar mass, the red fraction of galaxies increases with the environmental overdensity. At a given overdensity, there is also a trend for the red fraction to increase with the stellar mass (Figure~\ref{fig_redfrvsmass}). These trends are consistent, but with improved statistics, with the previously published results from the smaller 10k zCOSMOS sample (\citealt{Cucciati.etal.2010}, see also \citealt{Tasca.etal.2009} for a similar analysis based on morphologies) and are also in qualitative agreement with similar studies from other surveys up to redshifts of unity or slightly higher \citep[e.g.][]{Baldry.etal.2006, Peng.etal.2010, Cooper.etal.2010}.

\subsection{Mass and environmental quenching}
\label{sec_redallmodel}

It was proposed by \citet{Baldry.etal.2006} and further explored by \citet{Peng.etal.2010, Peng.etal.2012} that the dependence of the red fraction on the mass $M_*$ and environment $\delta$ at $z \sim 0$ can be quantitatively described by the separable functional form

\be f_{red} (M_*, \delta) =  1 - \exp[-(\delta/ p_1)^{p2} - (M_*/p_3)^{p4}] 
\label{eq_redfrmodel}
\ee

\noindent 
where the $p1-p4$ parameters are linked to the quenching mechanisms building the red population of galaxies \citep{Peng.etal.2010}. Using the limited zCOSMOS 10k data set in $0.3<z<0.6$, \citet{Peng.etal.2010} also showed that there is a good indication that such fitting formulae can describe the measured red fraction up to $z \sim 1$. We now reexamine this with the improved 20k sample.

Our best fit $p1-p4$ parameters are given in Table~\ref{tab_fitpar}, using for the fit $f_{red}(M_*,\delta)$ measured in the $-0.25 < \log(1+\delta) < 2$ and $M_{compl}(30\%) < \log(M_*/$\Msol$) < 11.5$ intervals, where the environment-mass plane is well sampled by the zCOSMOS data. In the fitting procedure, we use the overdensity and stellar mass in $\log(1+\delta)$ and $\log(M_*/$\Msol$)$ units. The red fractions resulting from our best fit model are over-plotted in Figures~\ref{fig_redfrvsod} and~\ref{fig_redfrvsmass} and agree with the majority of points within 2$\sigma$. The same result is also shown in Figure~\ref{fig_thfit}, where the red fraction across the $(M_*,\delta)$ plane is plotted. The color background shows the fitted model, while the contours show the data. The differences between the modelled and observed red fractions are in the range of 5-10$\%$ (in the extreme points 15$\%$).  Most importantly, the residuals show no systematic trends across the $(M_*,\delta)$ plane when normalized by the uncertainties (as shown in Figure~\ref{fig_fiterr} in Apendix~\ref{appendix_fiterr}). We conclude that overall a separable $f_{red}$ function provides a good representation of the zCOSMOS data in both redshift intervals (the errors on the fit will be shown below).

\begin{table}
\caption{\label{tab_fitpar}Best fit parameters to the red fraction model.}
\begin{tabular}{c|c|c|c|c|c}
\hline
&$z1<z<z2$ &  p1 &  p2 &  p3 & p4 \\
\hline
&$0.1 < z < 0.4$ &1.54 &1.10 &10.72 &0.58  \\
&$0.4 < z < 0.7$ &1.68 &0.94 &10.55 &0.67 \\
\hline
\end{tabular}

\medskip
The lower $z_1$ and upper $z_2$ limits of the redshift bin to which the results refer to are given in the first column, and the best-fit $p1$, $p2$, $p3$, and $p4$ parameters to the equation~\ref{eq_redfrmodel} are given in the second to the fifth columns. The fit is obtained with overdensity and mass expressed in $\log(1+\delta)$ and $\log(M_*/$\Msol$)$ units.
\end{table}

\begin{figure*}
\includegraphics[width=0.4\textwidth]{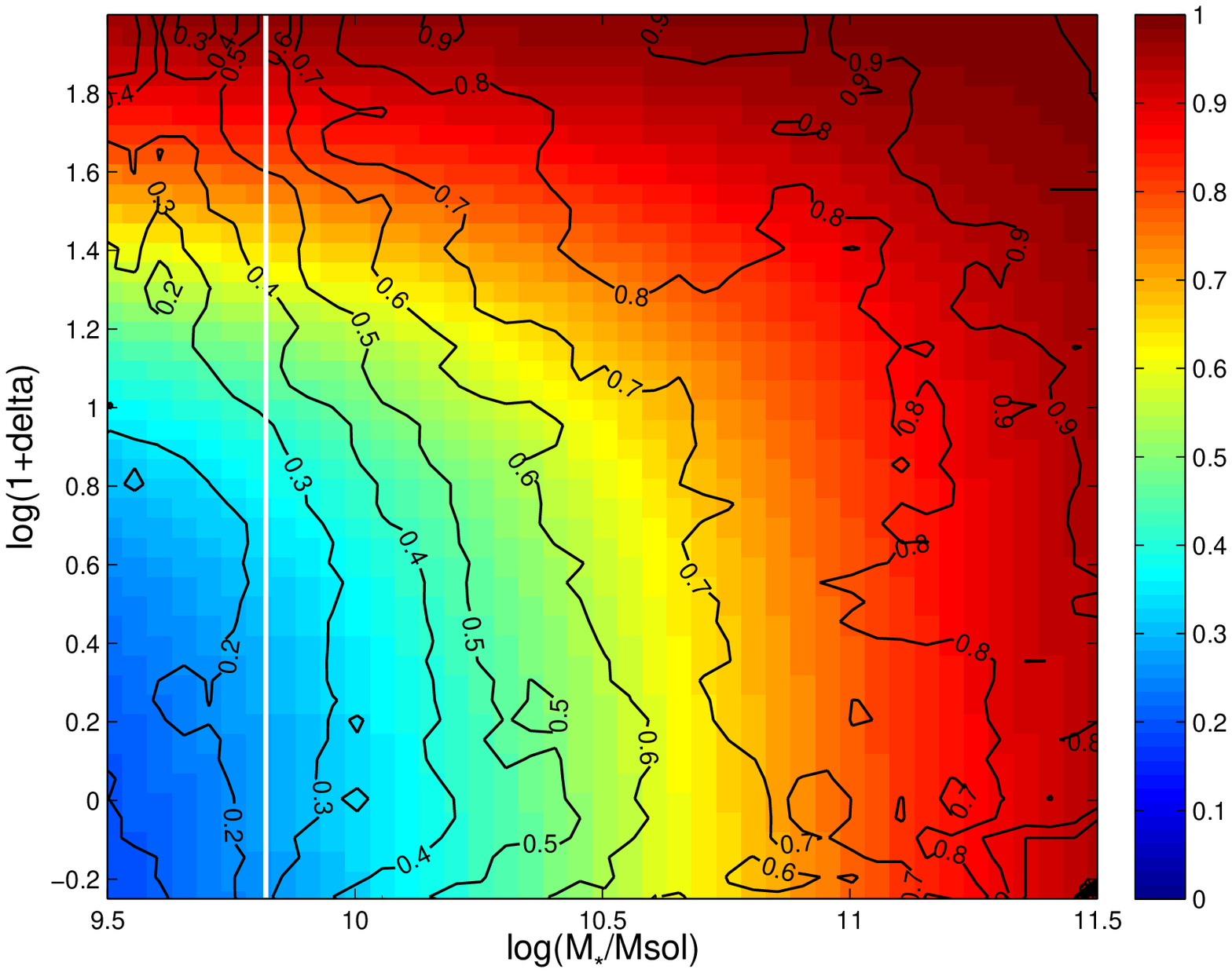}
\includegraphics[width=0.4\textwidth]{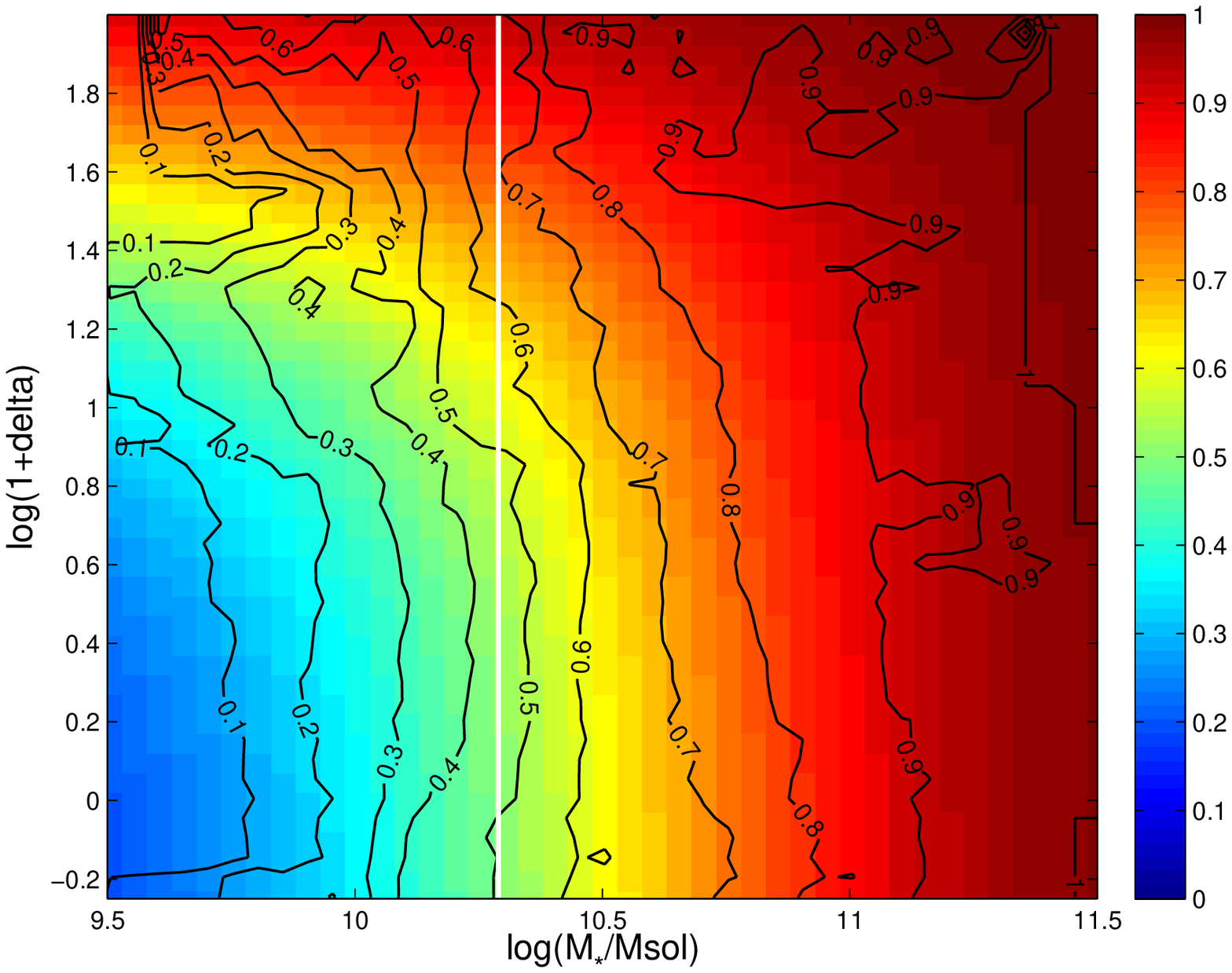}
\caption{\label{fig_thfit}Red fractions from the model and from the data in the $\log(1+\delta)-\log(M_*/$\Msol$)$ plane. The panels are for $0.1<z<0.4$ and $0.4<z<0.7$ on the left and the right, respectively. The colour shading corresponds to the red fraction from the best-fit model given by equation (\ref{eq_redfrmodel}). The black contours correspond to the red fractions measured from the zCOSMOS data boxcar smoothed with the interval $\Delta \log(M_*/$\Msol$) = \Delta \log(1+\delta)=0.45$ with the step 0.05 in both parameters. The white vertical lines mark the adopted mass-completeness limits. Our analysis is based only on the data in the mass-complete samples.}
\end{figure*}

As argued in \citet{Peng.etal.2010}, this suggests that the red fraction is dominated by two distinct processes, one a function of mass and not environment, and the other a function of environment and not mass. This is because the blue fraction $f_{blue}$, which is the chance that a galaxy has survived as a star-forming galaxy can be written as the product of two functions, one of mass alone and the other of environment alone

\be 
f_{blue} = 1 - f_{red} (M_*, \delta) = (1-\epsilon_{\rho} (\delta, \delta_0)) \times (1- \epsilon_m (M_*, M_{*0}))  
\label{eq_redfreps}
\ee

\noindent
where $\epsilon_{\rho} (\delta, \delta_0) = 1 - \exp(-(\delta/ p_1)^{p2})$ and $\epsilon_m (M_*, M_{*0}) = 1 - \exp(-(M_*/p_3)^{p4})$ \citep{Peng.etal.2010}. $\epsilon_{\rho}$ and $\epsilon_m$ are the so called relative environmental quenching and mass quenching efficiencies, respectively.

These may be calculated from the data as

\be \epsilon_{\rho} (\delta, \delta_0, M_*) = \frac{f_{red}(\delta, M_*) - f_{red}(\delta_0, M_*)}{f_{blue}(\delta_0, M_*)} \ee

\noindent
and 

\be \epsilon_{m} (M_*, M_{*0}, \delta) = \frac{f_{red}(M_*, \delta) - f_{red}(M_{*0}, \delta)}{f_{blue}(M_{*0}, \delta)}, \ee

\noindent
where $\delta_0$ is the overdensity at which $\epsilon_{\rho}$ is negligible and $M_{*0}$ is the mass at which $\epsilon_{m}$ is negligible.  

$\epsilon_{\rho}$ and $\epsilon_m$ can be interpreted as the net fraction of galaxies that have been quenched by the environment- or mass-related processes, respectively, and their functional form provides an important constraint on the physical nature of these processes. In \citet{Peng.etal.2010, Peng.etal.2012} it was shown that the continuity equation for galaxies plus the observed constancy of the shape of the mass function of star-forming galaxies requires that $\epsilon_m$ should have a particular form related to the faint-end slope and M* of the Schechter-fit to the star-forming mass function and the logarithmic slope of the $SFR-M_*$ relation of Main Sequence galaxies. This is different in detail from the exponential form of equation~\ref{eq_redfrmodel}. The focus of this paper is on $\epsilon_{\rho}$, and we prefer to use the functional form of $\epsilon_m$ given above with fitted $p3$ and $p4$ parameters to self-consistently define $\epsilon_m$.

One of the concerns when fitting $f_{red}$ over the two dimensional overdensity-mass surface is that we are working with a relatively small number of galaxies covering a rather narrow range of stellar mass. To estimate the uncertainties on the fit (i.e. on the mass and environmental quenching efficiencies) we generate mock samples of 100,000 galaxies in each redshift bin whose distribution in mass-overdensity matches the zCOSMOS sample. Each simulated galaxy is then assigned to be red or blue according to the zCOSMOS best fit $p1-p4$ model. We then repeat the fitting procedure for 100 randomly chosen sub-samples that each contain the same number of galaxies as the actual zCOSMOS sample in that redshift bin.  We find that there are no systematic effects due to the reduced sampling in the sense that the median $\epsilon_\rho$ and $\epsilon_m$ functions from the 100 re-samples closely matches the input functions. These 100 re-samples are then used to define the 16-84$\%$ interval of the distribution of returned $\epsilon_\rho$ and $\epsilon_m$, which are then taken as the  $\pm 1\sigma$ uncertainty interval on the fits returned from the actual zCOSMOS data.

We show in Figures~\ref{fig_epsrovar} and~\ref{fig_epsmvar} the best-fit $\epsilon_\rho$ and $\epsilon_m$ quenching efficiency curves in the $0.1<z<0.4$ and $0.4<z<0.7$ ranges, with the adopted error intervals estimated as above. For reference, we also show the corresponding $z \sim 0$ functions measured from the SDSS by \citet{Peng.etal.2010}, slightly shifted in mass to account for different definition of mass used. Comparing the curves at different redshifts, we do not find a significant evolution in either the  $\epsilon_{\rho}$ and $\epsilon_{m}$ functions with redshift to $z=0.7$.\footnote{\citet{Peng.etal.2010} use properties of the zCOSMOS 10k galaxies estimated differently than in this work, and they adopt the redshift-evolving cut in the red/blue division of galaxies. The $\epsilon_m$ measured from the 10k zCOSMOS data set in $0.1<z<1.0$ does not evolve with redshift (Peng et al. 2010), consistent with our results.} The constancy of $\epsilon_m$ is directly related, via the continuity equation, to the observed constancy of the M* of the star-forming galaxy population \citep[see ][]{Peng.etal.2010}. The constancy of $\epsilon_{\rho}$, which was also noted with our earlier data \citep{Peng.etal.2010} is in a sense more interesting: it tells us that the physical processes responsible for the environmental differentiation of galaxies (or, at least, the environmentally caused quenching) must act in the same way as a function of environment over the range of epochs studied. Our measure of environment is a nominal overdensity. We explore in Section~\ref{sec_whichenv} the links between this and other environment measurements.

Separability will be broken if there is any quenching channel that depends on a coupling between mass and environment, or if there are ways for galaxies to e.g. change their mass differentially (e.g. through merging) in different environments.  With the uncertainties in the data, we can not rule out the presence of any such effects: our point would simply be that they must at most represent  small perturbation to the simple picture implied by separability.

Finally, it should be noted that a constancy of $\epsilon_{\rho}$ does not imply that environmental quenching is not continuing over time. Galaxies will be migrating to denser environments and, in addition, environmental quenching must continuously operate to quench some of the lower mass galaxies whose mass increases by star formation, as will be discussed in a future paper (Peng et al. in preparation).

\begin{figure}
\includegraphics[width=0.45\textwidth]{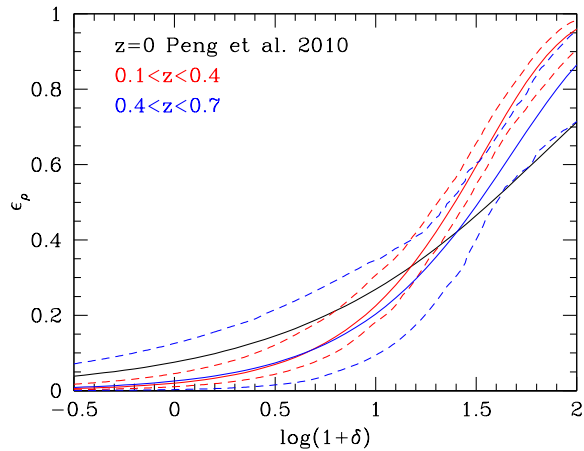}
\caption{\label{fig_epsrovar}Redshift evolution of the $\epsilon_{\rho}$ function. The environmental quenching efficiencies obtained from the fit to the measured red fractions in the zCOSMOS data in $0.1<z<0.4$ and $0.4<z<0.7$ are shown as continuous lines in red and blue, respectively. The dashed lines in the same colour mark the 16-84$\%$ error interval. For comparison, the $\epsilon_{\rho}$ function from the SDSS $z \sim 0$ data measured by \citet{Peng.etal.2010} is shown in black.}
\end{figure}

\begin{figure}
\includegraphics[width=0.45\textwidth]{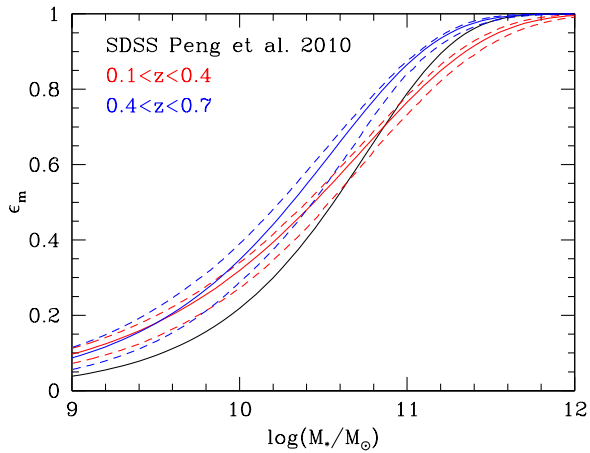}
\caption{\label{fig_epsmvar}Redshift evolution of the $\epsilon_{m}$ function. The mass quenching efficiencies obtained from the fit to the measured red fractions in the zCOSMOS data in $0.1<z<0.4$ and $0.4<z<0.7$ are shown as the continuous lines in red and blue, respectively. The dashed lines in the same colour mark the 16-84$\%$ error interval. For comparison, the $\epsilon_{m}$ function from the SDSS $z \sim 0$ data measured by \citet{Peng.etal.2010} is shown in black. To correct for the different definition of the stellar mass used, the $\log(p3)$ parameter from \citet{Peng.etal.2010} is increased by 0.2.}
\end{figure}

\section[]{Central-satellite dichotomy (Satellite quenching)}
\label{sec_censat}

\subsection{Fraction of satellites in different overdensities}
\label{sec_satfrac}

In this section we investigate the overall relations between the zCOSMOS centrals and satellites and their environments defined by the nearest neighbour counts (described in Section~\ref{sec_density}). As a first step we show in Figure~\ref{fig_censatod} the distribution of overdensities for the zCOSMOS centrals and satellites. As before, we consider two redshift bins, $0.1<z<0.4$ and $0.4<z<0.7$, and only galaxies above the mass-completeness limit. The satellite galaxies are systematically shifted to higher overdensity regions in both redshift bins (see also Figure 22 in \citealt{Kovac.etal.2010a}). Even though centrals make on average a larger fraction in the total galaxy population (Figure 1 in \citealt{Knobel.etal.2012b} for the zCOSMOS 20k sample), satellites become the dominant population of galaxies at the highest overdensities (green curve in Figure~\ref{fig_censatod}).

\begin{figure}
\includegraphics[width=0.45\textwidth]{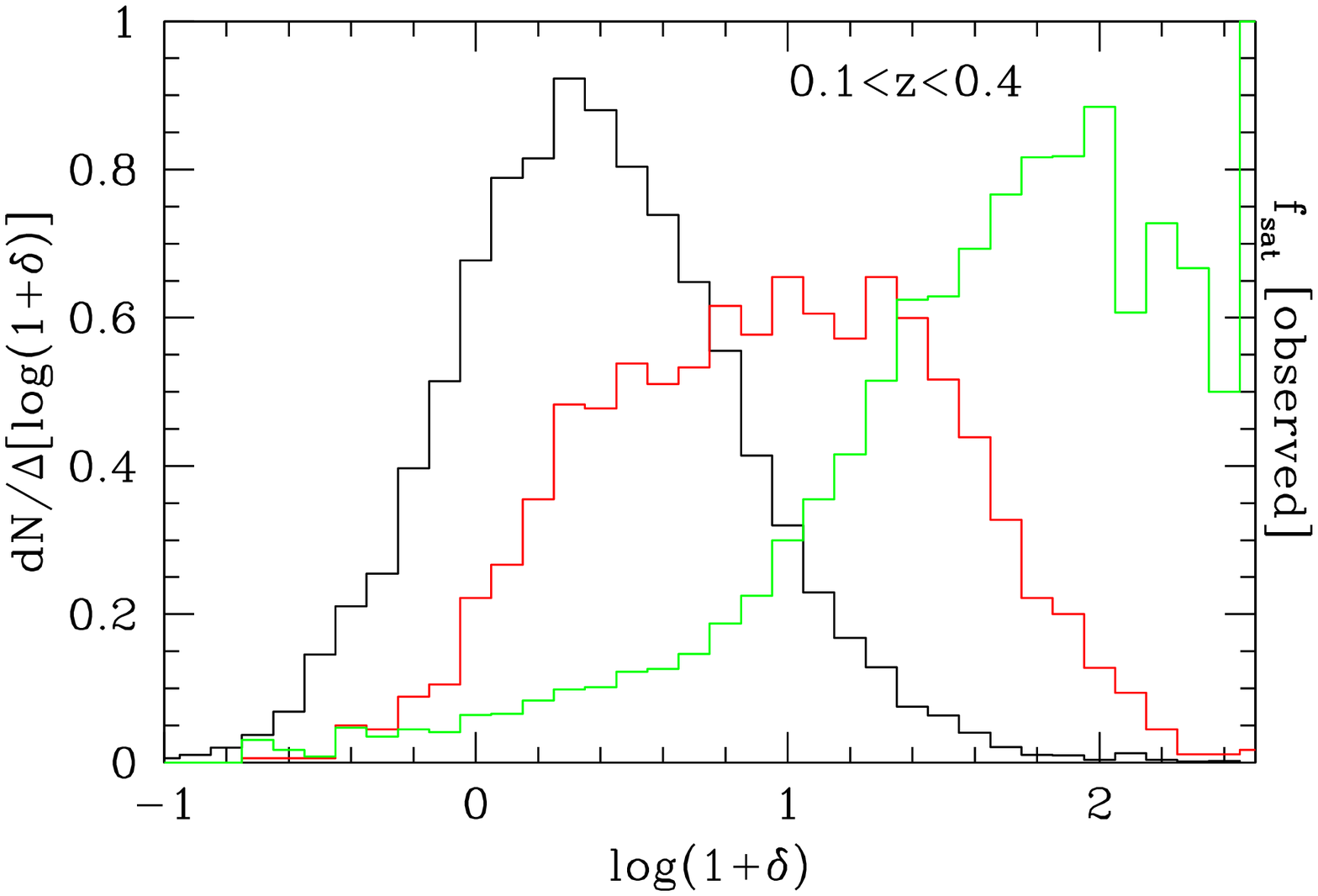}
\includegraphics[width=0.45\textwidth]{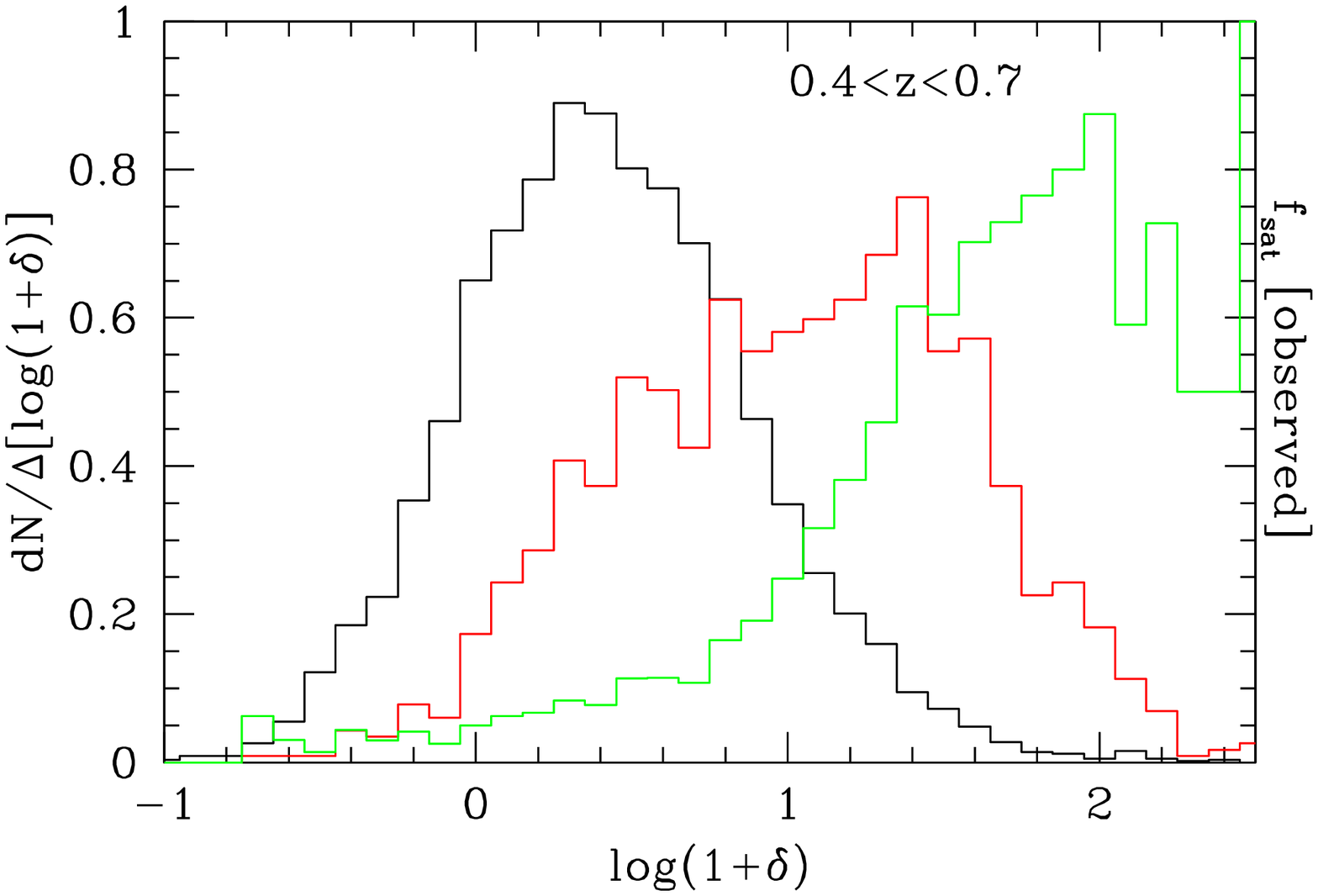}
\caption{\label{fig_censatod}Overdensities of central and satellite galaxies. The distributions for the mass-complete samples in $0.1<z<0.4$ and $0.4<z<0.7$ are shown in the top and bottom panels, respectively. The black and red curves correspond to the overdensity distributions of central and satellite galaxies, respectively, normalised by their respective observed number in a given redshift bin and scaled by the bin size. The green curves correspond to the observed fraction of satellites in the total population of galaxies in a given redshift bin as a function of overdensity.}
\end{figure}

In order to determine the role of centrals and satellites in shaping the overall environmental relations, we need to calculate the satellite fraction as a function of mass and overdensity, correcting the observed numbers for the mass incompleteness and for the miss-classification of centrals and satellites.

\citet{Knobel.etal.2012b} argued that the purity of central-satellite classification is rather high, about 80$\%$, for the mass-complete samples of zCOSMOS galaxies up to $z<0.8$, and only weakly dependent on stellar mass, based on the 24 \citet{Henriques.etal.2012} mocks. Here, we use the same \citet{Henriques.etal.2012} mocks to test whether there should be any dependence of purity on the overdensity. For simplicity, we quantify the environment around each mock galaxy (satisfying $i<22.5$ and $0.1<z<0.7$) by calculating the local overdensity of galaxies using the distance to the fifth nearest $M_B<-19.3-z$ neighbour projected within $\pm 1000$ \kms. By applying the same central/satellite reconstruction procedure on the mock galaxies as used for the real data \citep[developed by ][]{Knobel.etal.2012a}, we find that the purity $P$ of the mass-complete samples of centrals and satellites is strongly dependent on the overdensity because $f_{sat}$ increases with the overdensity. This is shown in Figure~\ref{fig_purityvsdelta} for both redshift samples considered in this work. The purity of the central galaxies is more than 95$\%$ in the lowest density environments, and then slowly decreases with increasing environmental overdensity to 78$\%$ at $\log(1+\delta)=0.65$, after which the purity sharply decreases to values of about 30-32$\%$ at $\log(1+\delta) \sim 2$. As shown in Figure~\ref{fig_censatod}, there are very few centrals at these high overdensities. This is rather obvious, as at high overdensities most galaxies are in large groups with many members, only one being the central. The purity of satellites shows the opposite behaviour. It is roughly constant at about 80$\%$ and 70$\%$ up to $\log(1+\delta) \sim 1.25$ in the lower and higher redshift bins, respectively, and then it increases to more than 90$\%$ for $\log(1+\delta) > 2$ in both redshift bins. We obtain similar results when repeating the calculation within bins of stellar mass. The strong dependence of purity on the overdensity of the reconstructed samples of centrals and satellites, and the opposite direction of this dependence, make it imperative to correct for these impurities for any quantities measured from the observed samples of centrals and satellites.

\begin{figure}
\includegraphics[width=0.4\textwidth]{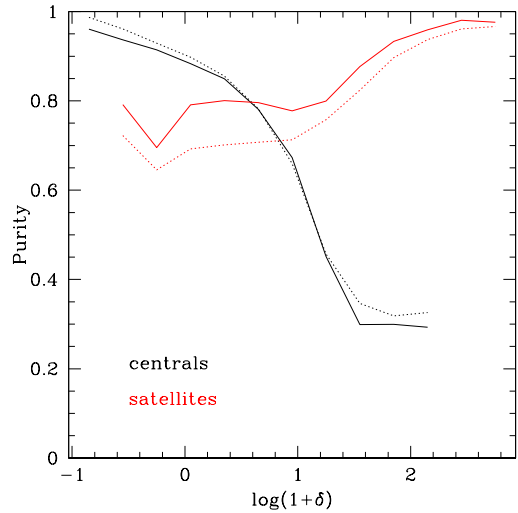}
\caption{\label{fig_purityvsdelta}Purity of the central and satellite galaxies as a function of overdensity in the reconstructed mock samples. The continuous and dotted curves are for the redshift bins $0.1<z<0.4$ and $0.4<z<0.7$, respectively. The curves in black and red are the results for the central and satellite galaxies, respectively. Only the mock galaxies with stellar masses above the adopted zCOSMOS mass completeness limit in a given redshift bin are used.}
\end{figure}

We use these $P(M_*, \delta)$ purity values to calculate the fraction of satellites as a function of overdensity. For the dichotomously defined samples consisting of $\tilde{N}_{cen}$ centrals and $\tilde{N}_{sat}$ satellites (i.e. every galaxy in the full zCOSMOS sample is either a central or a satellite) the  corrected (i.e. true) number of centrals $N_{cen}$ and satellites $N_{sat}$ can be calculated as  

\be
N_{cen} = \tilde{N}_{cen} - (1 - P_c) \tilde{N}_{cen} + (1 - P_s)\tilde{N}_{sat} 
\ee
\be
N_{sat} = \tilde{N}_{sat} - (1 - P_s) \tilde{N}_{sat} + (1 - P_c)\tilde{N}_{cen} 
\ee

\noindent
and then the `real' satellite fraction is simply given as $f_{sat}(M_*, \delta) = N_{sat} / (N_{sat} + N_{cen})$. The purity-corrected fraction of zCOSMOS satellites as a function of overdensity in the $\log(M_*/$\Msol$)=0.5$ bins is shown in Figure~\ref{fig_satfrac} for the lower and higher redshift bins in the panels on the top and bottom, respectively. The error interval corresponds to the $16-84\%$ range in the satellite fractions from the 100 bootstrapped samples. Given that the samples of centrals and satellites are dichotomous, the fraction of centrals at any overdensity is simply $1-f_{sat}(M_*, \delta)$.

\begin{figure}
\includegraphics[width=0.45\textwidth]{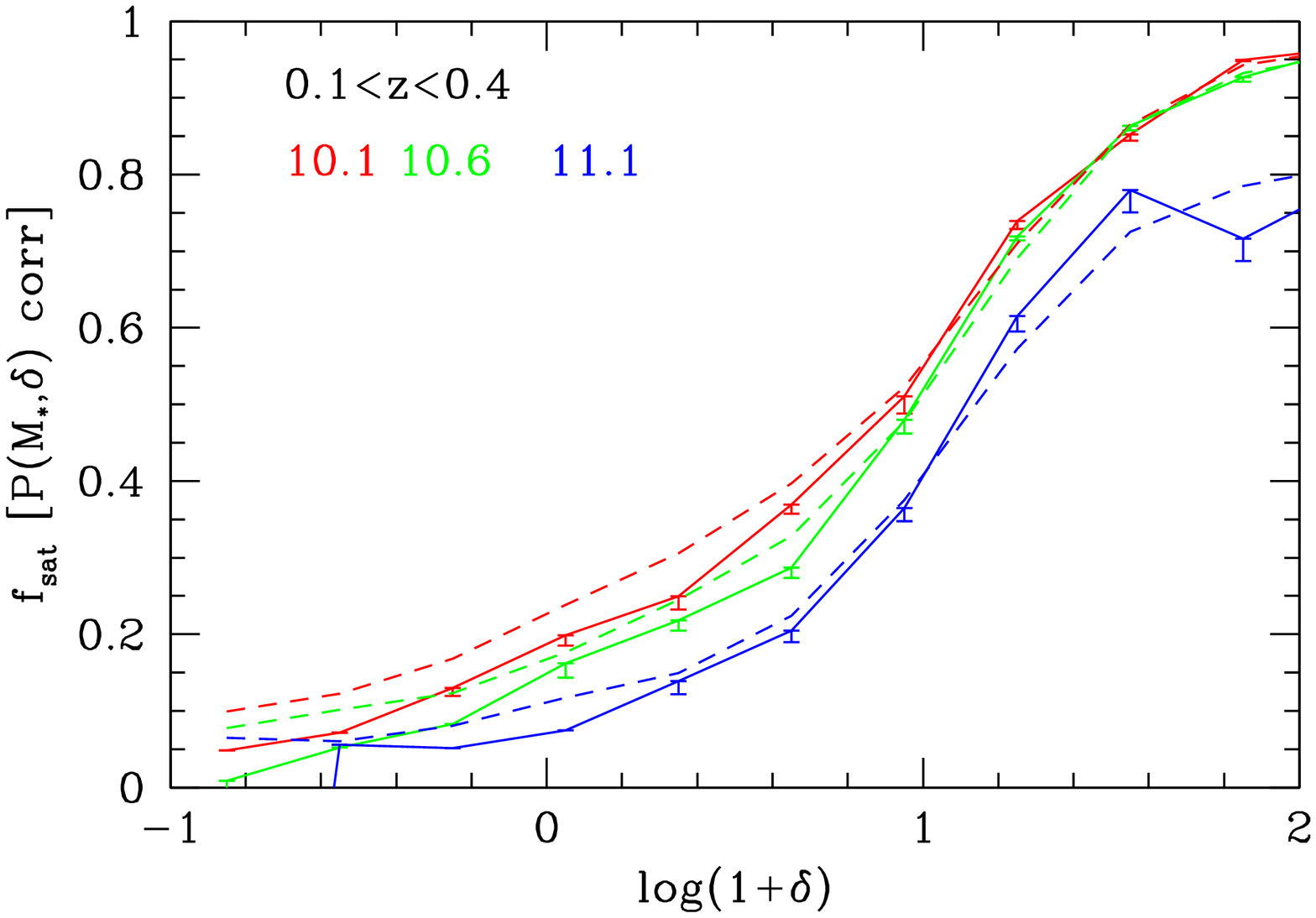}
\includegraphics[width=0.45\textwidth]{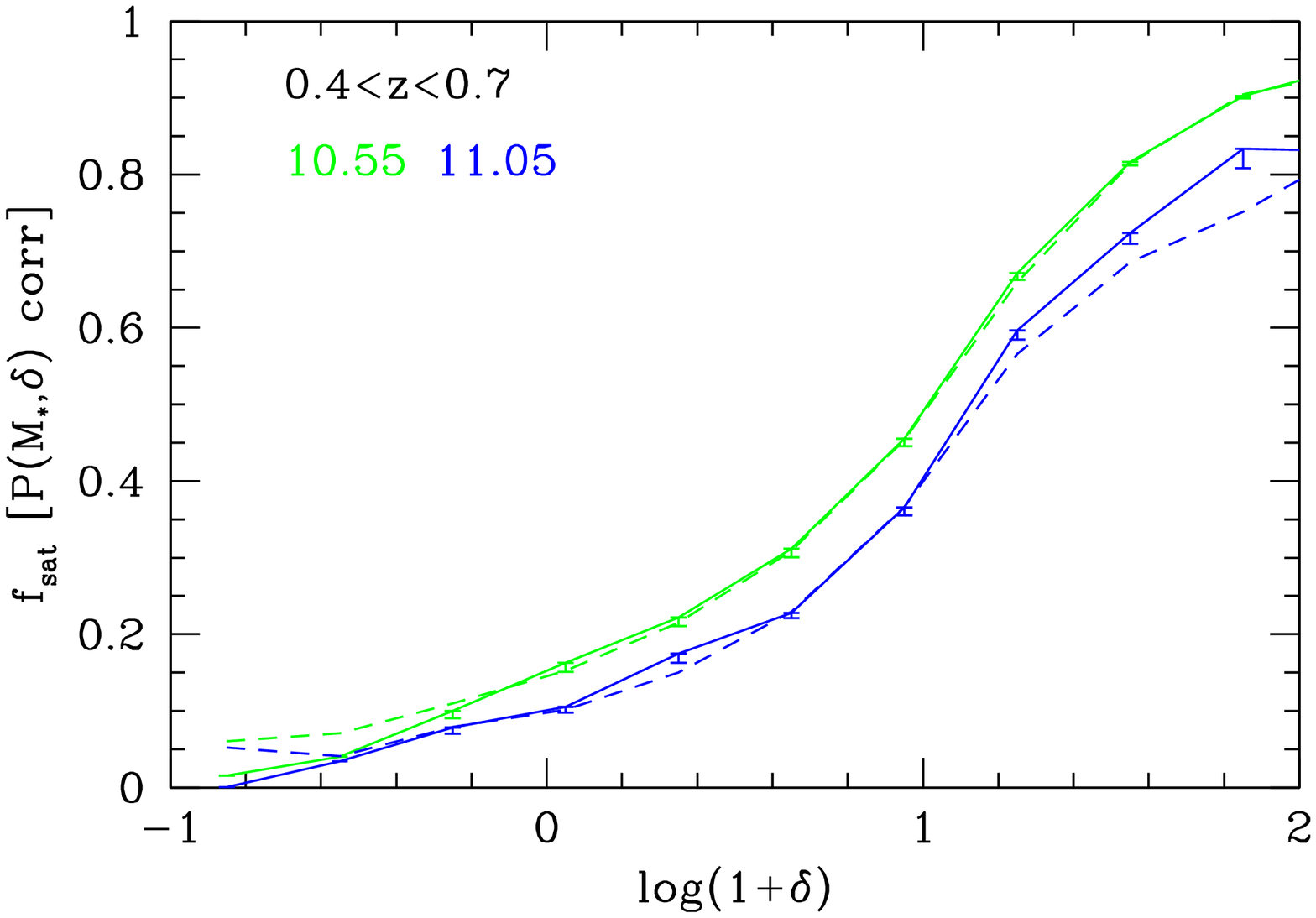}
\caption{\label{fig_satfrac}Fraction of satellites as a function of overdensity. The results in $0.1<z<0.4$ and $0.4<z<0.7$ are shown in the top and bottom panels, respectively. Differently coloured curves correspond to the fractions measured in 0.5 dex bins of stellar mass. The lowest mass bins start at the adopted mass completeness limit. The rounded values of the centre of the mass bins are marked in the second row of the label in the top left corner in the panels. The symbols (connected by the continuous lines) represent the fractions measured from the zCOSMOS data, and the dashed lines represent the equivalent fractions measured from the \citet{Henriques.etal.2012} mocks. The errors on the data points correspond to the $16-84\%$ range in the satellite fractions from the 100 bootstrapped samples. As the samples of centrals and satellites are dichotomous, the fraction of centrals at any overdensity is simply $1-f_{sat}(M_*, \delta)$.}
\end{figure}

Following this correction, the fractional contribution of centrals and satellites to the overall galaxy population remains a strong function of overdensity. At the lowest overdensities, more than 90$\%$ of galaxies are centrals. With increasing overdensity, the fraction of satellites increases, and satellites become the dominant population at $\log(1+\delta) \gtrsim 1$. The fraction of satellites is also a weak function of stellar mass: at a given overdensity,  $f_{sat}(M_*, \delta)$ decreases with stellar mass, but the difference is less than 10$\%$ between the mass bins in the well-sampled range of overdensities. Moreover, our data do not indicate any strong redshift evolution in $f_{sat}(M_*, \delta)$. For the 0.5 dex bin in stellar mass centred at $\log(M_*/$\Msol$)=10.54$, which is observable in both redshift bins, we do not find an evolution in the fraction of satellites at a given overdensity between the two redshifts bins.

We also compare the measured satellite fractions with the equivalent fractions in the mock catalogues. Overall, the fractions of satellites in the zCOSMOS and mocks agree well both in the trends with the overdensity and with stellar mass. This can be seen in Figure~\ref{fig_satfrac}. The $f_{sat}(M_*, \delta)$ curves in \citet{Peng.etal.2010}, which were based on the earlier \citet{Kitzbichler&White.2007} mocks, show a much weaker dependence on stellar mass. As a check we measure the environments of $i<22.5$ galaxies in the \citet{Kitzbichler&White.2007} mocks, calculating the overdensities in exactly the same manner as for the \citet{Henriques.etal.2012} mocks, as part of the difference could stem from differently defined mean densities. We also find that $f_{sat}(M_*, \delta)$ in the \citet{Kitzbichler&White.2007} mocks shows much weaker dependence on stellar mass, finding the decrease in $f_{sat}(M_*, \delta)$ only for the highest mass $\sim 11.1$ bin in $0.1<z<0.4$. As \citet{Peng.etal.2012} used only the mock defined $f_{sat}(M_*, \delta)$, some of the differences in the interpretation of the results on satellite quenching in \citet{Peng.etal.2012} and in this paper will reflect the weak mass dependence of $f_{sat}(M_*, \delta)$ in our data.

\subsection{Red fractions of central and satellite galaxies}
\label{sec_redfraccenandsat}

In this section we investigate the colour-density relations obtained when separating the overall galaxy population into central and satellite galaxies. Due to the opposite dependence of the purity on the overdensity for the centrals and satellites (shown in Figure~\ref{fig_purityvsdelta}), any relation between a given property and the overdensity that is measured separately for the centrals and satellites, and the differences between the two, may be strongly affected by the impurity. We can however correct for this statistically, as the observed red fractions of centrals $\tilde{f}_{r,cen}$ and satellites $\tilde{f}_{r,sat}$ are related to the `true' respective fractions $f_{r,cen}$ and $f_{r,sat}$ by simple relations:

\be 
\tilde{f}_{r,sat} =P_{sat}f_{r,sat} + (1 - P_{sat}) f_{r,cen}
\label{eq_impredfrsat}
\ee
\be
\tilde{f}_{r,cen} =P_{cen}f_{r,cen} + (1 - P_{cen}) f_{r,sat}. 
\label{eq_impredfrcen}
\ee

\noindent
Using the $P_{sat}$ and $P_{cen}$ (as a function of mass and environment) as estimated above in Section~\ref{sec_satfrac}, one can easily obtain the statistically corrected, i.e. `true', red fractions of centrals and satellites from the observed red fractions of apparent centrals and satellites.

Ideally, we would like to carry out the analysis for the centrals and satellites in the same way as for the overall galaxy population, presented in Section~\ref{sec_envquench}. However, given the small number of satellites (totalling about 1000 in both redshift bins, see Table~\ref{tab_galsamples}), the systematic shift in the overdensities occupied by the central and satellite galaxies, and the necessity to derive any quantity in the equivalent bins for both centrals and satellites in order to correct for impurities, we need to modify our approach to obtain reliable results. Instead of working in narrow bins of stellar mass, we derive the red fractions of centrals and satellites as a function of overdensity in mass-matched samples.

We create mass-matched samples of both centrals and satellites, as follows. Starting from the sample of centrals in each of the two redshift ranges, we first divide the galaxies lying above the mass completeness limit into four bins in overdensity, defined by the quartiles in the overdensity distribution of the centrals. As a reference sample, we define a fifth overdensity bin which contains central galaxies around the median overdensity, i.e. with overdensities in the 37.5-62.5$\%$ range of sorted overdensities. The stellar mass-matched samples are then built by picking randomly a galaxy from each of the four overdensity quartiles which has a stellar mass close to the stellar mass of a galaxy from the central reference sample, for each galaxy in the reference sample. To obtain the errors we repeat this process 20 times. More details of the matching process and the (weighted) stellar mass distributions of the matched samples are given in Appendix~\ref{appendix_massdist}.

Using the adopted colour cut, we then measure the red fraction of centrals at the median weighted stellar mass in the matched samples in the four overdensity bins. The median mass is $\log(M_*/$\Msol$)=10.47$ and 10.71 in the lower and higher redshift bins, respectively. We then combine equations (\ref{eq_impredfrsat}) and (\ref{eq_impredfrcen}) to correct for impurity, calculating $\tilde{f}_{r,sat}$ using the satellites residing in the same range of overdensities and within the 0.75 dex mass interval centred on the median mass of the central galaxies. This approach to derive the red fractions of satellites, which are needed for the purity correction for the centrals, is driven by the small number of satellite galaxies in the regions of lowest overdensity.

We take the median in the distribution of purity corrected red fractions of centrals from the 20 mass matched samples to be our final result, and the 16-84$\%$ interval in the same distribution as the $\pm1 \sigma$ error interval. The obtained red fractions and the errors on the purity corrected measurements are shown in Figure~\ref{fig_redfraccen}.

\begin{figure}
\includegraphics[width=0.45\textwidth]{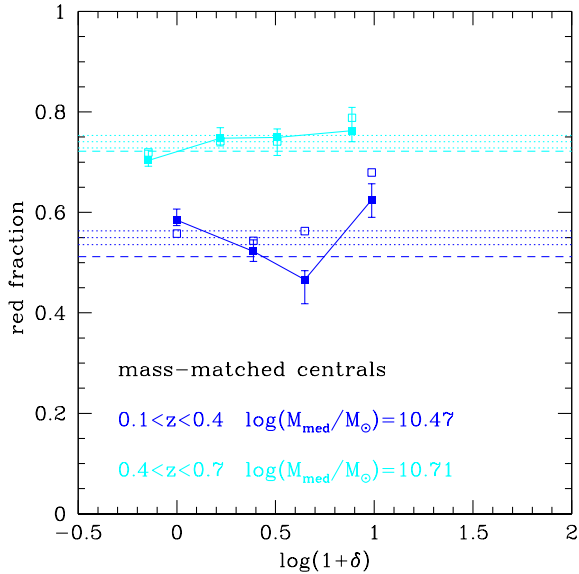}
\caption{\label{fig_redfraccen}Red fractions of central galaxies as a function of overdensity. Symbols correspond to the red fraction measured at the median stellar mass $\log(M_{med}/$\Msol$)$ of the mass-matched centrals in the four quartiles in overdensity. The error-bars encompass the 16-84$\%$ interval in the purity corrected red fractions of centrals from 20 realisations of the mass matching. The results are plotted at the median overdensity in each bin, where the empty and solid symbols correspond to the purity uncorrected and corrected quantities, respectively. The dotted lines mark the mean and the $\pm 1 \sigma$ interval of the four $f_{r,cen}$ values. The dashed lines correspond to the red fraction from the best-fit model (Table~\ref{tab_fitpar}) at the median mass of centrals when assuming that only the mass quenching operates. The blue and cyan colours correspond to the results in $0.1<z<0.4$ and $0.4<z<0.7$, respectively.}
\end{figure}

\begin{figure}
\includegraphics[width=0.45\textwidth]{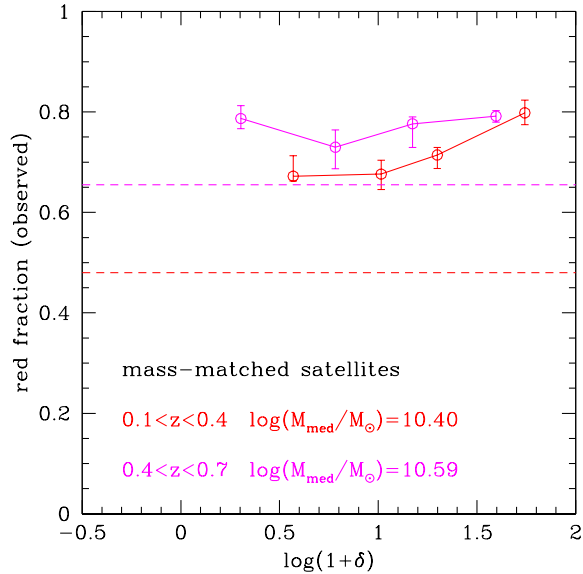}
\caption{\label{fig_redfracsat}Red fractions of satellite galaxies as a function of overdensity. Symbols correspond to the red fractions measured at the median stellar mass $\log(M_{med}/$\Msol$)$ of the mass-matched satellites in the four quartiles in overdensity. The error-bars encompass the 16-84$\%$ interval in the red fractions of satellites from 20 realisations of the mass matching. The results are uncorrected for purity, plotted at the median overdensity in each bin. The dashed lines correspond to the red fraction from the best-fit model (Table~\ref{tab_fitpar}) at the median mass of satellites when assuming that only the mass quenching operates. The red and magenta colours correspond to the results in $0.1<z<0.4$ and $0.4<z<0.7$, respectively.}
\end{figure}

We then repeat the mass-matching process for the sample of satellite galaxies in the equivalent way as for the centrals, defining the reference sample and the overdensity quartiles using the mass complete sample of satellites in the two redshift bins. For the moment, we measure only the purity-uncorrected red fractions of satellites, as there are only a handfull of centrals of similar mass at the highest overdensities that are occupied by satellites. As for the centrals, we take for $\tilde{f}_{r,sat}$ the median in the distribution of red fractions from the 20 mass-matched samples of satellites and the 16-84$\%$ interval in this distribution to be the $\pm1 \sigma$ error. This is shown in Figure~\ref{fig_redfracsat}. For reference, the weighted cumulative distribution functions (wCDFs) of stellar mass of the mass-matched samples of satellite galaxies are shown in Appendix~\ref{appendix_massdist} in Figure~\ref{fig_masssatmatch}.

We analyse now the obtained colour-density relations for centrals and satellites, as shown in Figures~\ref{fig_redfraccen} and~\ref{fig_redfracsat}, respectively. Even though the low redshift measurements of the red fraction of centrals $f_{r,cen}$ are quite noisy, there is little evidence for a strong trend with the overdensity in any redshift bin. To check whether environment quenching is indeed negligible for centrals, as would be expected from \citet{Peng.etal.2012}, we calculate the red fraction of centrals that would be predicted from mass-quenching alone, i.e as given by equation (\ref{eq_redfreps}) with the best-fit parameters for $\epsilon_m$ from Table~\ref{tab_fitpar} and $\epsilon_{\rho} = 0$. The expected red fractions, calculated at the median mass of central galaxies in the mass-matched samples, are 0.51 and 0.72 in the lower and higher redshift bins, respectively. These are plotted as the horizontal dashed lines in Figure~\ref{fig_redfraccen}. The average $f_{r,cen}$ values and their error intervals are calculated to be $0.55\pm0.01$ and $0.74\pm0.01$ in the equivalent redshift bins, and these are plotted as the dotted lines in Figure~\ref{fig_redfraccen}. The $\epsilon_m$-predicted red fractions differ by only a few percent from the corrected red fraction of central galaxies derived from the data, leading us to conclude that the build up of the red central galaxies is almost entirely due to the mass quenching process. Any residual environmental quenching is increasing the red fraction of centrals by at most a few percent (i.e. 2-4$\%$).

In contrast, the red fraction of satellites (Figure~\ref{fig_redfracsat}) shows rather clear dependence on overdensity in $0.1<z<0.4$, with $\tilde{f}_{r,sat}$ increasing (by more than 10$\%$) with the increasing overdensity, while in $0.4<z<0.7$ the red fraction of satellites does not show a clear trend with overdensity. However, in both redshift ranges, the satellites are consistently redder than the centrals. The red fractions of galaxies with $\log(M_*/$\Msol$)=10.4$ and $\log(M_*/$\Msol$)=10.59$, which are the median stellar masses of galaxies in the mass-matched samples of satellites in the lower and higher redshift bins respectively, that would be expected to be produced by the mass-quenching process are 0.48 and 0.66. These are plotted as the dashed lines in Figure~\ref{fig_redfracsat} and are consistently lower than the observed red fractions of satellites. Environmental quenching has increased the red fraction of satellites by, on average, 23$\%$ and 13$\%$, in the lower and higher redshift bins, respectively. Even though the conclusions which we presented so far for satellites are based on the impure quantities, due to the small number of equally massive centrals at the highest overdensity bins, this is of lesser concern than in the case of centrals, as the purity of satellites shows less change with overdensity, and is at 70-80$\%$ even well below $\log(1+\delta)=1$.

\subsection{Satellite quenching}

In the previous section we have explored the dependence of the red fractions of centrals and satellites on the overdensity for galaxies with particular mass distributions. The results for centrals and satellites were obtained for different ranges of overdensity, as the reference samples of centrals cover a lower range in overdensity than that of the satellites. As a result, we could not correct the measured red fractions of satellites for impurities. It should be noted that the mass distribution of centrals is also systematically shifted to higher stellar masses compared with that of the satellites, and so the red fractions can not be immediately compared since the different masses will have different red fractions from mass quenching.

We now analyse the environmental dependence of the {\it excess} in the red fraction of satellites with respect to the centrals. We create new samples of mass-matched centrals by matching their stellar masses (in the same four quartiles of overdensity, as defined by the centrals) to the mass distribution of the satellite galaxies in a given redshift bin.  We follow the same matching procedure and the same estimation of red fractions and associated uncertainties as were described in Section~\ref{sec_redfraccenandsat}. The resulting $f_{r,cen}$ in the four overdensity quartiles are shown as solid squares in Figure~\ref{fig_redfraccensat} in the top and bottom panels for the intervals $0.1<z<0.4$ and $0.4<z<0.7$, respectively. It should be noted that the obtained values are somewhat lower than when masses are matched to the centrals (i.e. as in Figure~\ref{fig_redfraccen}), because of the systematically lower stellar masses of satellites. It is however reassuring that, as already shown previously, the red fractions of centrals do not show any strong trend with the overdensity, i.e. that we are not introducing some artifical colour-density trend by adopting a satellite mass distribution also for the central galaxies. As before, the average red fractions of centrals (estimated to be 0.53 and 0.64 in the lower and upper redshift bins, respectively, and marked as the dashed lines in Figure~\ref{fig_redfraccensat}) are very close to the red fractions expected from purely mass quenching at the median stellar mass of satellites in the reference sample (estimated to be 0.48 and 0.66 in the lower and higher redshift bin, respectively).

\begin{figure}
\includegraphics[width=0.45\textwidth]{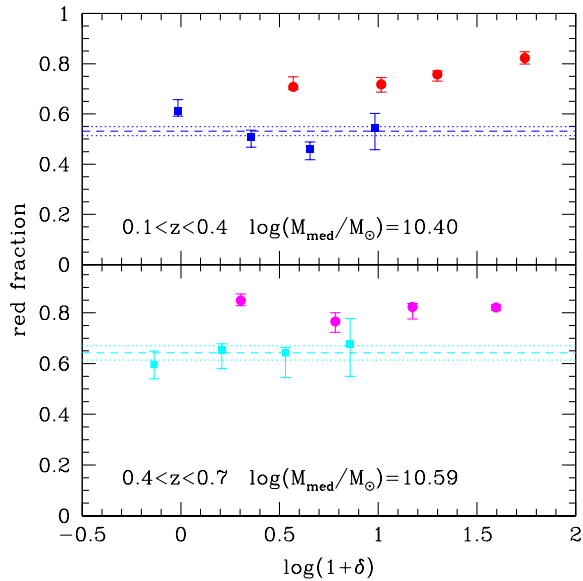}
\caption{\label{fig_redfraccensat}Red fractions of centrals (squares) and satellites (circles) mass-matched to the same distribution as a function of overdensity. Only the purity corrected quantities are shown, measured at the median stellar mass $\log(M_{med}/$\Msol$)$ in $0.1<z<0.4$ and $0.4<z<0.7$ in the top and bottom panels, respectively. The results are plotted at the median overdensity in the bins of overdensity quartiles. The respective errors encompass the 16-84$\%$ interval of the purity corrected red fractions of centrals or satellites in the 20 realisations of the mass matching. The dashed and dotted lines correspond to the average and $\pm 1 \sigma$ values of $f_{r,cen}$.}
\end{figure}

\begin{figure*}
\includegraphics[width=0.45\textwidth]{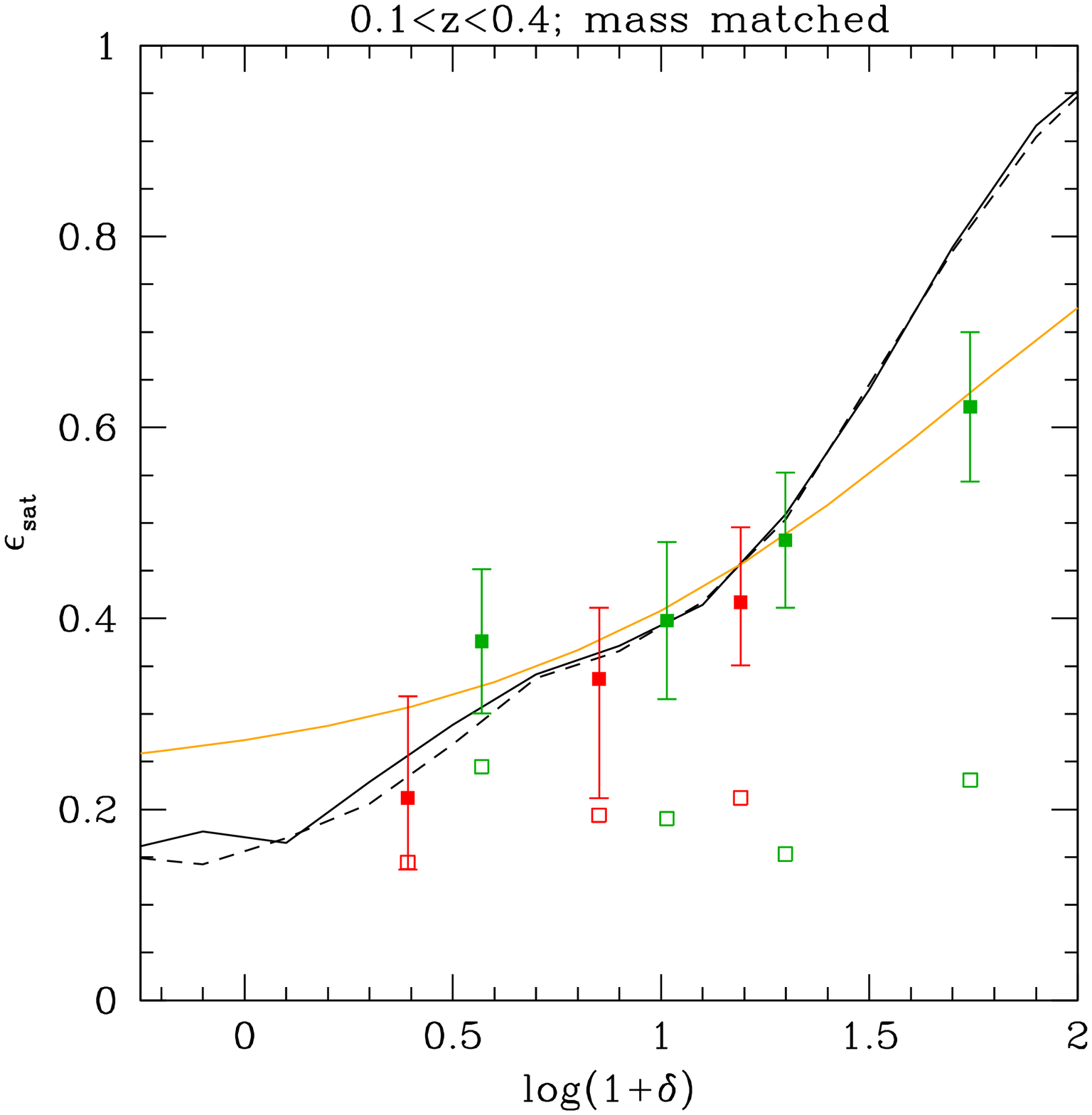}
\includegraphics[width=0.45\textwidth]{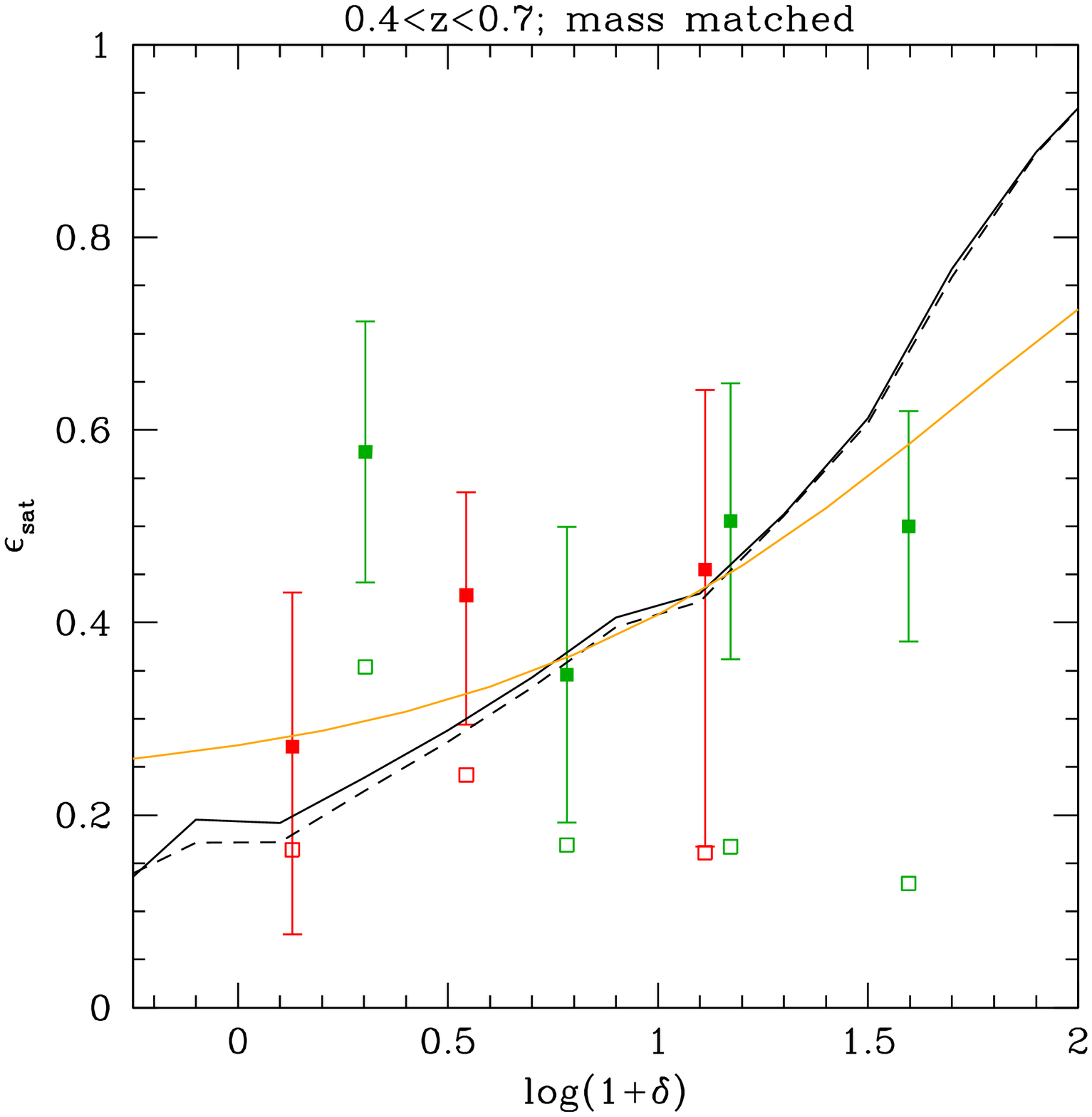}
\caption{\label{fig_epssatvsdelta}Satellite quenching efficiency $\epsilon_{sat}$ as a function of overdensity in $0.1<z<0.4$ (left) and $0.4<z<0.7$ (right). The green symbols are the $\epsilon_{sat}$ values measured from the data mass-matched in the quartiles of the satellite overdensity (with median mass $\log(M_*/$\Msol$)=$10.40 and 10.59 in the lower and higher redshift bins, respectively). The red symbols are the $\epsilon_{sat}$ values measured in the narrower bins of overdensity with at least 20 central and 20 satellite galaxies in each bin (with median masses $\log(M_*/$\Msol$)=$10.24, 10.32 and 10.38, and 10.49, 10.52, and 10.60 in the lower and higher redshift bins, respectively). The solid and open symbols are for the purity-corrected and -uncorrected quantities, respectively. For clarity, the error bars are shown only for the former, as an 16-84$\%$ uncertainty interval (see text for more details). The black lines correspond to the model prediction $\epsilon_{\rho} (\delta) / f_{sat}(\delta, M_*)$ valid in the case when the environmental quenching efficiency of centrals is negligible. The continuous and dashed lines are obtained for the fraction of satellites in the $\Delta \log(M_*/$\Msol$)=0.5$ bin centered at  $\log(M_*/$\Msol$)=10.40$ (left) and 10.59 (right), and 10.24 (left) and 10.49 (right), respectively. For comparison, the orange line is the $z \sim 0$ $\epsilon_{sat}(\delta)$ function averaged over all stellar masses from \citet{Peng.etal.2012}.}
\end{figure*}

Though the samples of centrals and satellites are now matched in mass, this still does not alleviate the fact that the red fractions of centrals and satellites are measured in systematically shifted ranges of overdensity. We deal with this by extrapolating a constant $f_{r,cen}$ to the highest over-densities that are probed by the satellites, by taking the average value of $f_{r,cen}$ that is measured for the centrals (the dashed lines in Figure~\ref{fig_redfraccensat}). The other possibility would have been to simply adopt the value expected from $\epsilon_m$, but we prefer to use the measured values from the data whenever possible. Using this adopted $f_{r,cen}$ for the whole range of overdensities, we can now also finally calculate the purity corrected $f_{r,sat}$, which we did not do in Section~\ref{sec_redfraccenandsat}, using equations (\ref{eq_impredfrsat}) and (\ref{eq_impredfrcen}). This is shown, for the two redshift bins, as the solid circles in Figure~\ref{fig_redfraccensat}.

The red fractions of satellites are, as expected, systematically above the red fractions of centrals up to $z=0.7$ at all overdensities, due to the additional quenching  process operating on the satellites. The excess of red satellites with respect to the centrals at a given overdensity $\delta$ and mass $M_*$ is quantified by {\it satellite quenching efficiency} $\epsilon_{sat}(\delta, M_*)$ defined as

\be 
\epsilon_{sat}(\delta, M_*) = \frac{f_{r,sat}(\delta, M_*) - f_{r, cen}(M_*)}{f_{b, cen}(M_*)}
\label{eq_epssat}
\ee

\noindent
where $f_{b, cen}(M_*)$ is the blue fraction of centrals \citep{Peng.etal.2012}. The defined $\epsilon_{sat}(\delta, M_*)$ quantity can then be interpreted as the excess fraction of satellite galaxies (in a given environment $\delta$) which are environmentally quenched during the accretion process with respect to the population of star-forming centrals of the same stellar mass.

It can be argued \citep[e.g.][]{vandenBosch.etal.2008, Wetzel.etal.2013} that one should use, in equation (\ref{eq_epssat}), the $f_{r, cen}(M_*)$ of the centrals at the (earlier) epoch at which the satellites first entered the halo. We digress for a moment to address this point, even though we will then argue that it is of no significance, because $f_{r, cen}(M_*)$ actually changes little with redshift (see Figure 3 in \citealt{Knobel.etal.2012b}). The choice of which $f_{r, cen}(M_*)$ to use in equation (\ref{eq_epssat}) depends on what is assumed for the mass quenching process. If mass quenching for some reason does not operate (at all) for satellites, then one should indeed use the $f_{r, cen}$ at the earlier epoch (and at the slightly reduced stellar mass at which the satellite entered the halo). Alternatively, if mass quenching operates in exactly the same way for centrals and for satellites, even if mass quenching changes with time, then one should use the $f_{r, cen}$ at the epoch of observation, simply because some of the satellites will have been mass quenched since they became satellites. We can not be sure which scenario holds, although \citet{Peng.etal.2012} argued that the mass quenching should operate identically for satellites and centrals because star-forming centrals and star-forming satellites exhibit the same characteristic Schechter M*. However, regardless of these points, we would argue that in practice  $f_{r, cen}(M_*)$ changes little with redshift over the redshifts of interest \citep[c.f.][]{Wetzel.etal.2013}. This is also evidenced by the constancy of $\epsilon_m(M_*)$ over $0<z<0.7$ shown in Figure~\ref{fig_epsmvar}.

We proceed with measuring $\epsilon_{sat}(\delta, M_*)$ from the zCOSMOS data using the red fractions of centrals and satellites computed (at the same epoch) in the mass-matched samples. The $\epsilon_{sat}(\delta, M_*)$ values, computed in the four quartiles of satellite overdensity, are shown in Figure~\ref{fig_epssatvsdelta} in green, using the solid and open symbols for the purity corrected and uncorrected quantities, respectively. The errors are shown only for the purity-corrected values and they are estimated by the propagation of errors on the individual fractions, using the symmetric $0.5(f_{r,cen}(84\%) - f_{r, cen}(16\%))$ and $0.5(f_{r,sat}(84\%) - f_{r, sat}(16\%))$ errors for the propagation. It should be noted that the $\epsilon_{sat}(\delta, M_*)$ efficiencies measured above are obtained for a range of overdensities, especially for the lowest and highest quartiles. This will tend to smooth the measured $\epsilon_{sat}(\delta, M_*)$ values in the extreme parts of the overdensity distribution.

As an independent estimate, we also measure $\epsilon_{sat}(\delta, M_*)$ within the narrower range of overdensity where both populations exist in significant numbers. We calculate the red fractions in the bins of overdensity that contain at least 20 centrals and 20 satellites, and still matching the mass distribution of centrals to the satellites in a given overdensity bin following the procedure described in Section~\ref{sec_redfraccenandsat}. The resulting  $\epsilon_{sat}(\delta, M_*)$ values are shown in red in Figure~\ref{fig_epssatvsdelta}, using the open and filled symbols for the purity-uncorrected and -corrected quantities. The median stellar masses in the overdensity bins, going from the lower to the higher values are $\log(M_*/$\Msol$)=$10.24, 10.32 and 10.38 in $0.1<z<0.4$, and $\log(M_*/$\Msol$)=$10.49, 10.52, and 10.60 in $0.4<z<0.7$. The error interval corresponds to the 16-84$\%$ range in the distribution of the $\epsilon_{sat}(\delta, M_*)$ values obtained from the 50 mass-matched samples of centrals.

Leaving aside the differences in the stellar mass (which we will address later), we find that the differently measured $\epsilon_{sat}(\delta, M_*)$ all paint a consistent picture on the dependence of satellite quenching efficiency on overdensity. The  $\epsilon_{sat}(\delta, M_*)$ measurements at the lower redshift show the satellite quenching efficiency increasing with the overdensity. We also show the value of $\epsilon_{sat}(\delta)$ at $z \sim 0$ from the group analysis of \citet[][orange lines in Figure~\ref{fig_epssatvsdelta}]{Peng.etal.2012}. We conclude that there is little evidence for evolution in $\epsilon_{sat}(\delta)$.

In the higher redshift bin, the $\epsilon_{sat}(\delta, M_*)$ estimated in the two ways at the lowest overdensities are just outside their respective $1\sigma$ error intervals. Considering the other points and their uncertainties our data do not, by themselves, show much evidence for a strong environmental dependence of $\epsilon_{sat}(\delta, M_*)$. However, considering the larger errors on the $\epsilon_{sat}(\delta, M_*)$ measurements and the smaller range in $\log(1+\delta)$ in the higher redshift bin, our higher redshift data also do not provide compelling evidence for any change of $\epsilon_{sat}(\delta, M_*)$ with redshift between $z \sim 0$ and $ z \sim 0.7$.

Turning back to the broader picture, in the case when the quenched fraction of centrals is produced solely through the mass quenching process $\epsilon_m$ (as strongly supported by the red fractions of zCOSMOS centrals, Figure~\ref{fig_redfraccen}), then the $\epsilon_{\rho}$ of the overall population should be due to environmental quenching of the satellite population alone. In that case, $\epsilon_{sat}(\delta, M_*)$ will be related to the environmental quenching through the fraction of satellites as $\epsilon_{\rho} / f_{sat}(\delta, M_*)$.

We check this using $f_{sat}(\delta, M_*)$ at the median stellar mass of satellites in the reference overdensity bin (i.e. $\log(M_*/$\Msol$)=$10.40 and 10.59 in the two redshift bins), and at the lowest median stellar mass of satellites in the narrower overdensity bins (i.e. $\log(M_*/$\Msol$)=$10.24 and 10.49 in the two redshift bins). These satellite fractions are calculated in $\log(M_*/$\Msol$)=0.5$ bins, centred at the quoted median mass. The resulting $\epsilon_{\rho}(\delta) / f_{sat}(\delta, M_*)$ functions are shown in Figure~\ref{fig_epssatvsdelta} as the black solid and dashed lines respectively. They differ only slightly, reflecting the weak dependence of $f_{sat}(\delta, M_*)$ on mass.

The $\epsilon_{sat}(\delta, M_*)$ measurements from the narrower overdensity bins show an excellent agreement with the expected $\epsilon_{\rho} (\delta) / f_{sat}(\delta, M_*)$ values in both redshifts. The $\epsilon_{sat}(\delta, M_*)$ measurements from the quartiles in the overdensity are also broadly consistent, and only the highest quartile point at the lower redshift bin and the lowest quartile point at the higher redshift bin are significantly discrepant. As we have already mentioned, these points cover the broadest ranges in overdensity and therefore must be rather smoothed. Taking this all together, still the agreement between the measured $\epsilon_{sat}(\delta, M_*)$ points and the value of $\epsilon_{\rho} (\delta)/ f_{sat}(\delta, M_*)$ is rather striking.

Our detailed analysis strongly suggests that the satellite galaxies must be the dominant population of galaxies that is driving the overall environmental trends at least up to $z=0.7$. The net effect in the overall population $\epsilon_{\rho}$ is the combination of two dependencies on overdensity: the $\epsilon_{sat}(\delta, M_*)$ and $f_{sat}(\delta, M_*)$ functions. Furthermore, the effect of the environment, as defined by our overdensity parameter, on the satellites, is consistent with being unchanged at $z = 0.7$ compared with $z \sim 0$.

It is clear that the constancies highlighted in this paper are unlikely to be exact.  First, the functions $\epsilon_{\rho}(\delta)$ and $\epsilon_{sat}(\delta, M_*)$ are simple representations of no doubt complex physical processes. Secondly, our chosen environment parameter is observationally convenient but is unlikely to correspond {\it exactly} to a physically relevant parametrisation of environment (we return to this point in the next section). Finally since $f_{sat}$ is unlikely to be completely independent of stellar mass and of epoch, it is clear that it is not possible for both (or possibly either) $\epsilon_{\rho}$ and $\epsilon_{sat}$ to be strictly independent of mass and time. It is also clear that, even with our very large sample, the statistical uncertainties become significant when the sample is split into centrals and satellites, into bins of overdensity and further selected via the mass-matching procedure. For all these reasons, our statements should be taken as `approximations to reality' rather than as a statement of physical exactitude.

\section[]{Which environment matters?}
\label{sec_whichenv}

One of the difficulties when interpreting the observed correlation between galaxy environment and other properties is the choice of the definition of environment. The quantification of environment by the number density of galaxies within some apertures (fixed or adaptive) is the most commonly used environmental indicator based on the observational data, as it is rather straightforward to calculate (but see the discussion on the reconstruction process and various uncertainties in e.g. \citealt{Kovac.etal.2010a}). However, it is questionable what is the physical meaning of such defined environment.

In most theoretical studies, dark matter haloes are considered to be the key drivers of galaxy evolution, and their properties have been most commonly used as an environment indicator. The number density of galaxies and their halo masses are certainly correlated, but the dispersion of the halo masses at a measured density can be 1 dex and more \citep[e.g.][]{Haas.etal.2012}. In order to reconcile the two indicators and constrain the physical process responsible for the observed environmental differences in the galaxy population, one ideally needs to constrain both environment estimators on the same data set. Assignment of the observed galaxies to common haloes (i.e. the group finding process) and measuring the properties of the haloes is observationally not a straightforward task as it usually requires for all except the richest structures some calibration against mock catalogues or other theoretical distributions. The reconstructed groups and their properties will suffer from a range of the uncertainties quantified commonly by the purity, completeness, and particularly by the error on the estimated halo properties, which increases for groups with only a few members.

 \begin{figure}
\includegraphics[width=0.45\textwidth]{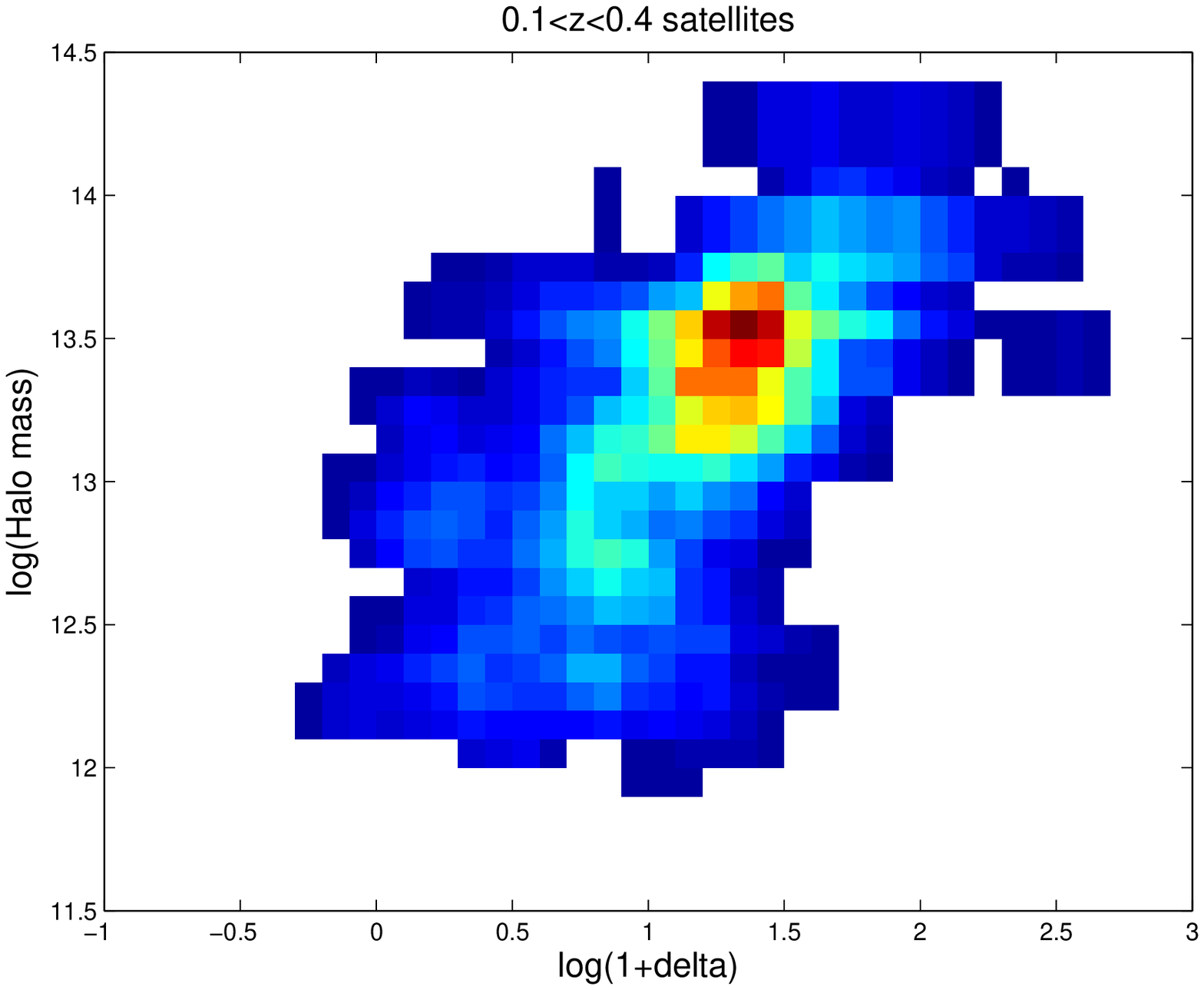}
\includegraphics[width=0.45\textwidth]{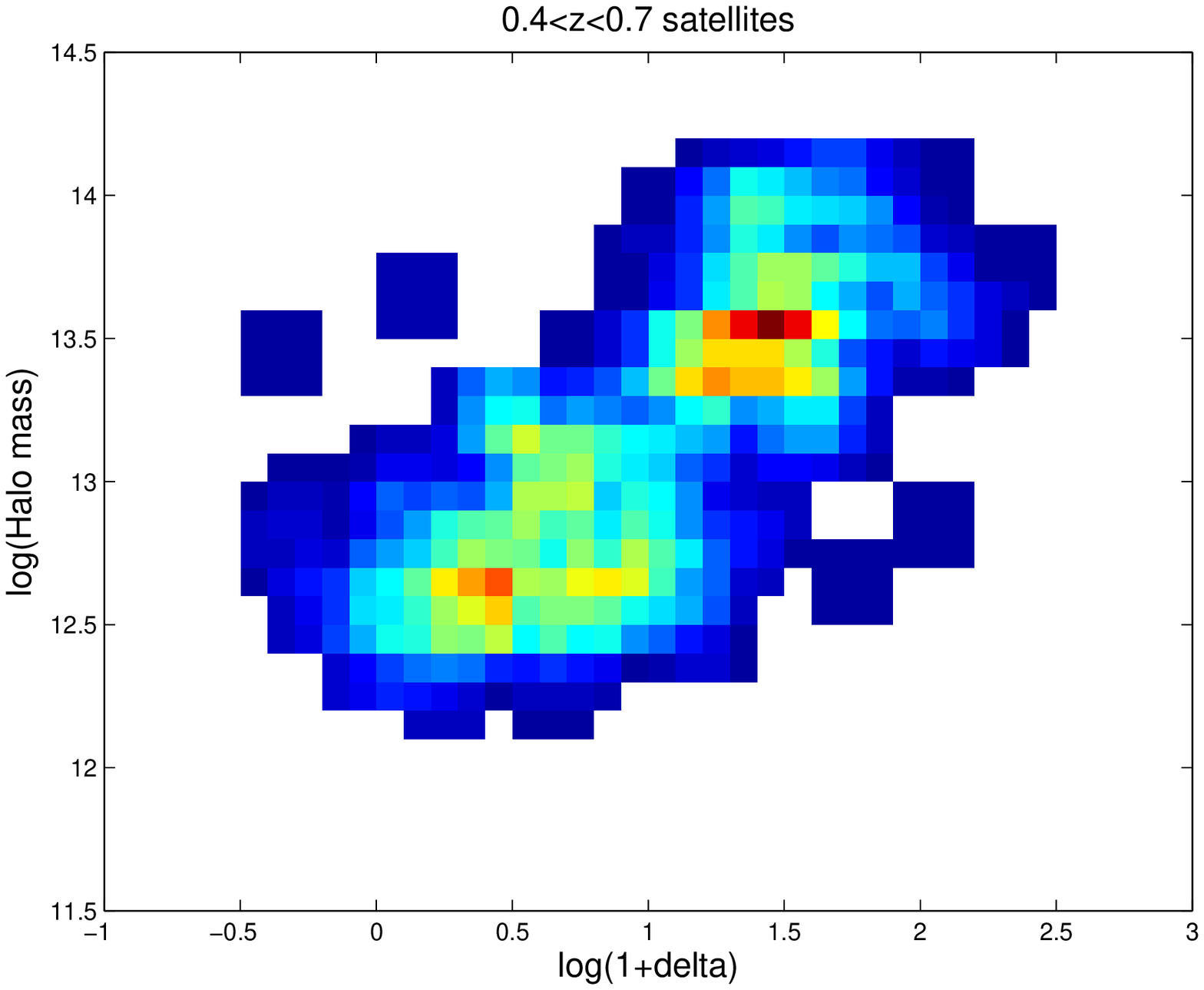}
\caption{\label{fig_halovsdensity}Relation between halo mass and overdensity for satellite galaxies in the mass complete samples. The top and bottom panels are for $0.1<z<0.4$ and $0.4<z<0.7$, respectively. The colour coding corresponds to the number density of galaxies in a given bin (going from blue to red as the number density increases).}
\end{figure}

It has been pointed out previously that the environment indicator based on the nearest N neighbour counts, as used in this paper, has a dual nature with respect to the parent galaxy halo. For the richest structures with more than N tracer galaxies, the reconstructed $\delta$ is a measure of the density within the halo, while for less rich structures the $\delta$ will be a measure of the environment surrounding the halo \citep[e.g.][]{Weinmann.etal.2006, Haas.etal.2012, Peng.etal.2012, Woo.etal.2012}.

The virial radius of a halo is often considered to be a physically motivated scale up to which the environment plays a role, as dynamically galaxies within the virial radius cannot be largely influenced by galaxies outside of it \citep{Weinmann.etal.2006}. The virial radius marks also a transition between infall and virial motions, the latter providing support for the strong shock fronts and hot, thermalised gas \citep[e.g.][]{Dekel&Birnboim.2006}. However, there is observational evidence that the environmental trends observed within the virial radius extend to the distance of a few virial radii \citep{Balogh.etal.1999, Hansen.etal.2009, vonderLinden.etal.2010, Lu.etal.2012}. Part of this can be explained through the non-sphericity of haloes \citep[e.g.][]{Weinmann.etal.2006}, or it can originate from galaxies already quenched  within the group infalling into more massive structure \citep[i.e. pre-processing;][]{Berrier.etal.2009, Li.etal.2009, McGee.etal.2009}, satellite galaxies outside of the virial radius after the first pericentric passage caused by the highly eccentric orbits \citep[i.e. overshooting;][]{Benson.2005, Ludlow.etal.2009, Wetzel.2011} or the recently proposed interaction of galaxies with the hot gas in filaments \citep{Bahe.etal.2012}. In addition, from the theoretical point of view  properties of dark matter haloes also depend on their environment through the so called `assembly bias' \citep{Gao.etal.2005, Gao&White.2006, Wechsler.etal.2006}. All this complicates the identification of a key environment (or a key environmental scale) in the evolution of galaxies.

\begin{figure}
\includegraphics[width=0.45\textwidth]{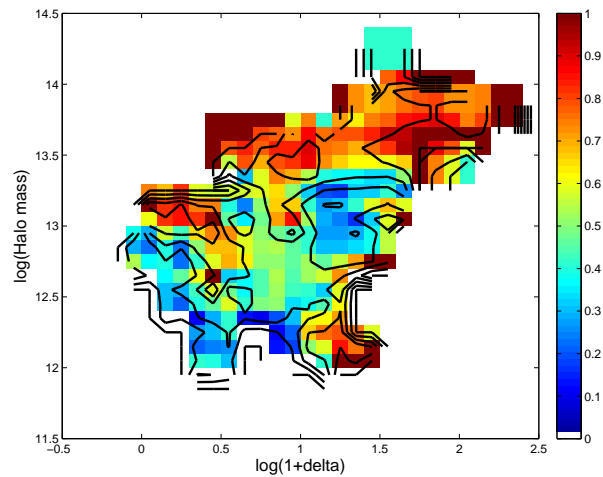}
\caption{\label{fig_redfrhalovsdensity}Red fraction of satellite galaxies in the halo mass-overdensity plane. Galaxies have stellar mass in the range $9.82 < \log(M_*/$\Msol$) < 10.32$ in $0.1<z<0.4$. The colour scale corresponds to the red fraction of satellites measured in a given halo mass-overdensity bin, as indicated by the bar on the right-hand side. The black contours follow the same colour scale and are separated by 0.2.}
\end{figure}

In the case of zCOSMOS satellites studied in the present paper, the projected distance to the 5th nearest neighbour in the $M_B<-19.3-z$ sample of tracer galaxies varies between $0.41-4.85$ \hh Mpc (mean 2.63 \hh Mpc) and $0.35-5.07$ \hh Mpc (mean 2.71 \hh Mpc) in $0.1<z<0.4$ and $0.4<z<0.7$, respectively. These values are much larger than the average virial radius of the reconstructed zCOSMOS groups in the \citet{Henriques.etal.2012} mocks, which is 0.33 and 0.35 physical Mpc in the lower and higher redshift bin, respectively. Furthermore, the $\pm 1000$ \kms\ interval over which we project galaxies in the radial dimension corresponds to  21.1, 23.4, and 24.2 \hh Mpc  at 0.1, 0.4, and 0.7, respectively. Even though at a first instance this interval seems large, it is similar to the  line-of-sight length of a cylinder used for the Voronoi-Delaunay method for the group reconstruction in both the zCOSMOS \citep{Knobel.etal.2009, Knobel.etal.2012a} and the DEEP2 and AEGIS \citep{Gerke.etal.2007, Gerke.etal.2012}, optimised to account for the peculiar velocities of group galaxies.

As described in Section~\ref{sec_groups}, \citet{Knobel.etal.2012a} estimate the halo masses for our zCOSMOS groups. This is based on calibration against mock catalogues, coupled with a study of the cross-correlation function of groups and galaxies \citep{Knobel.etal.2012c}. The relation between the halo mass and the overdensity of the zCOSMOS20k satellites used in this work is shown in Figure~\ref{fig_halovsdensity}. There is a broad correlation  between the two environment indicators: the satellites in the most massive haloes simultaneously reside in the most overdense environments, and vice versa, though the correlation has a lot of scatter. In addition, one needs to keep in mind that the reconstructed halo masses have substantial uncertainties \citep{Knobel.etal.2009, Knobel.etal.2012a}.

Are the observed trends with overdensity discussed above due to a primary correlation with halo mass? Figure~\ref{fig_redfrhalovsdensity} shows the fraction of red satellite galaxies in the halo mass-overdensity plane for a $9.82 < \log(M_*/$\Msol$) < 10.32$ subset in $0.1<z<0.4$. It should be noted that we cannot correct the obtained fractions for the purities of central and satellite samples since we do not know (at this point) how these depend on halo mass. The red fraction does not show a stronger correlation with either of the two environmental indicators (black contours in Figure~\ref{fig_redfrhalovsdensity}). We reach the same conclusion also for the satellites of higher stellar mass at both redshift bins. Within our own data set, it is hard to distinguish whether the halo mass or the overdensity of the zCOSMOS satellite galaxies at $0.1<z<0.7$ has a more important role in building the red fraction of satellite galaxies.

The situation at low redshift is better: in SDSS, \citet{Peng.etal.2012} concluded that halo mass was a relatively minor component and that local overdensity was the dominant driver of the relation between the red fraction of satellites and their environment. \citet{Woo.etal.2012} presented a more complex picture in which a halo mass dependence was present for only the highest density environments.

\citet{Peng.etal.2012} used the $9.5<\log(M_*/$\Msol$)<10.0$ subsample of satellites to reach the above conclusion. These are amongst the most populous satellites and they dominate the outskirts of the groups (see the white contours in the right hand panel in Figure 10 in \citealt{Woo.etal.2012}), for which Woo et al. also concluded that the halo quenching is weakest, bringing these two works into agreement on the roles of halo mass and local overdensity in quenching of satellites. However, using also SDSS, \citet{Tinker.etal.2011} found that the increase in the quenched fraction of satellites at a given luminosity (or stellar mass) with the large scale environment is due to variations of the halo mass with overdensity, i.e. that at a given large-scale overdensity and a luminosity bin, the fraction of quenched satellites increases with the halo mass. It still needs to be understood what is the reason for the different dependence of the quenched fraction of satellites on halo mass and $\delta$ measured by Peng et al. and Woo et al. on one side, and Tinker et al. on the other. It is possible that for satellites the small scale environment measured by the nearest neighbour technique (used by Peng et al. and Woo et al.) correlates stronger with halo mass than with the overdensity measured on 10 \hh Mpc scale (used by Tinker et al.).

We also have too few centrals with measurable halo masses to meaningfully examine any residual dependence of the red fraction of centrals on halo mass, although noting that this could not be large because of the apparent independence of this quantity with local overdensity. In SDSS, there is some evidence that, at fixed stellar mass, central galaxies in higher mass haloes have a higher red fraction \citep{Woo.etal.2012}. On the other hand, given that the properties of central galaxies in the relaxed and unrelaxed $z \sim 0$ ZENS groups do not change much, \citet{Carollo.etal.2012} concluded that the properties of central galaxies must be driven by stellar mass.

\citet{Woo.etal.2012} also explored the relations between the quenched fractions of galaxies and their stellar and halo mass and overdensity in $0.75<z<1$ using the AEGIS data. Due to the low number of satellites, they obtained results only for the central galaxies at these redshifts, finding that for the majority of centrals their quenched fraction correlates better with stellar than halo mass. Moreover, Woo et al. found no correlation between the halo mass and the overdensity for the central galaxies in the AEGIS sample. The quenched fraction of AEGIS centrals clearly increases with stellar mass, in agreement with the zCOSMOS results in \citet{Knobel.etal.2012b} and the $\epsilon_m$-model for the red fraction of centrals presented in this paper. In addition, the quenched fraction of the AEGIS centrals does not show any correlation with the local overdensity, echoing the results presented in this work.

\section{Conclusions}
\label{sec_concl}

In this paper we have studied the origin of environmental effects in the star-formation properties of galaxies at significant look-back times. Our analysis is based on the final zCOSMOS-bright data set, from which we select a set of some 17,000 galaxies with reliable spectroscopic redshifts. These same galaxies, complemented by some information on the galaxies from the parent catalogue that do not have reliable spectroscopic redshift, are also used to reconstruct the overdensities based on the nearest neighbour technique (which we use as a primary environmental indicator in this paper). Separation of the galaxies into centrals and satellites is done using the group catalogue of \citet{Knobel.etal.2012a}. Finally, we carefully considered, in deriving our results, the effects of incompleteness in stellar mass and the impurity of the central-satellite classification.

We first studied the colour-mass-density relation by measuring the red fraction of galaxies $f_{red}(M_*, \delta)$ in bins of stellar mass and overdensity $\delta$ in two redshift bins, $0.1<z<0.4$ and $0.4<z<0.7$.

1. As previously observed, the red fraction of galaxies increases with the environmental overdensity, for galaxies of a given stellar mass, and with mass at a given overdensity, out to $z=0.7$, the highest redshift up to which we carried out the analysis.

2. As in the local galaxy population \citep{Baldry.etal.2006, Peng.etal.2010}, such dependence of $f_{red}$ on mass and environment can be approximated by the functional form $1 - f_{red} (M_*, \delta) = (1 - \epsilon_m(M_*))(1 - \epsilon_{\rho}(\delta))$ in which the dependence on mass and environment is separable. The measured $f_{red}$ can be well represented by this function in both our redshift bins, within 2$\sigma$ and without systematic deviations in either mass or overdensity. Given the limited size of the current data set, we cannot exclude the existence of a cross term in both mass and environment, but this must be within the estimated statistical uncertainties. If exists, such effects would represent a small addition to the provided description, rather than changing the model completely.

3.  Following our previous paper we interpret this separability as indicating that two independent processes operate that dominate the quenching of galaxies, mass quenching and environmental quenching. The parameters that describe these two processes, i.e. $\epsilon_m(M_*)$ and $\epsilon_{\rho}(\delta)$ do not, within the uncertainties, change with redshift, and are consistent with their respective equivalents measured at $z \sim 0$ from SDSS \citep{Peng.etal.2010}. Lack of change in $\epsilon_m(M_*)$ is expected from constant $M^*$ of star-forming galaxies, lack of change in $\epsilon_{\rho}(\delta)$ is less obvious.

Using samples of centrals and satellites that are carefully matched in stellar mass, we investigated further the colour-density relation by measuring red fractions at different overdensities for the centrals and satellites separately.

4. We find that the red fraction of centrals $f_{r,cen}$ is primarily a function of mass and is almost independent of overdensity at both epochs studied. The red fraction of centrals is therefore consistent with being produced almost entirely by the mass quenching process alone, with any environmental quenching increasing the red fraction by at most a few percent (less than 5$\%$).

5. In contrast to this, the red fraction of satellite galaxies $f_{r,sat}$ requires additional environmental quenching in order to explain the observed red fractions at a given stellar mass and overdensity. This is particularly obvious in $0.1<z<0.4$, where we find that $f_{r,sat}$ clearly increases with the overdensity.

6. The overall satellite fraction $f_{sat}$ is a strong function of overdensity, increasing from practically zero at the lowest overdensities and reaching more than 0.9 at the highest overdensities. There is also a weak dependence of $f_{sat}$ on stellar mass. For satellites with $\log(M_*/$\Msol$) \sim 10.5$ (for which we are statistically complete at both redshifts) there is no evidence for a redshift evolution in the satellite fraction at a given overdensity.

7. At the same stellar mass, the red fraction of satellites is higher than the red fraction of centrals over the whole range of overdensities. This excess of red satellites with respect to the centrals at a given overdensity $\delta$ and mass $M_*$, normalised by the blue fraction of centrals, is used to measure the satellite quenching efficiency, $\epsilon_{sat}(\delta, M_*)$. Satellite quenching efficiency at $0.1<z<0.4$ increases with overdensity and is noticeably consistent with the equivalent $z \sim 0$ measurements in the SDSS \citep{Peng.etal.2012}. In $0.4<z<0.7$, our $\epsilon_{sat}(\delta, M_*)$ measurements do not show much evidence for a strong environmental dependence. However, taking into account the statistical uncertainties as well as the smaller overdensity range covered in the higher redshift bin, the overall conclusion is that $\epsilon_{sat}(\delta, M_*)$ stays relatively unchanged from $z \sim 0$ to $ z \sim 0.7$.

8. Moreover, our results demonstrate an overall agreement between the measured $\epsilon_{sat}(\delta, M_*)$ points and the value of $\epsilon_{\rho}(\delta) / f_{sat}(\delta, M_*)$, where the equality between the last two expressions will hold when the environmental quenching in the overall population is entirely produced through the satellite quenching process.

The main conclusion from our analysis is that satellite galaxies are the main drivers of the observed trends with the overdensity at least up to $z=0.7$. The relative role of environment, quantified by either $\epsilon_{\rho}$ or $\epsilon_{sat}$, does not change significantly with redshift, and is consistent with $z \sim 0$ measurements. However, for the satellite population, the dependence of the red fraction on the environmental overdensity is very similar to its dependence on the halo mass, making these two environment indicators statistically equivalent within our data set of satellites. Finally, given the limited size of our sample when split in bins of redshift, mass, overdensity, and populations of centrals and satellites, the above relations should be regarded as approximations to a complex reality, rather than as mathematical identities.

\section*{Acknowledgments}

This research was supported by the Swiss National Science Foundation, and it is based on observations undertaken at the European Southern Observatory (ESO) Very Large Telescope (VLT) under the Large Program 175.A-0839. The Millennium Simulation databases used in this paper and the web application providing online access to them were constructed as part of the activities of the German Astrophysical Virtual Observatory.

\bibliography{refs}

\begin{thebibliography}{}

\bibitem[\protect\citeauthoryear{{Abadi}, {Moore} \& {Bower}}{{Abadi}
  et~al.}{1999}]{Abadi.etal.1999}
{Abadi} M.~G.,  {Moore} B.,    {Bower} R.~G.,  1999, \mnras, 308, 947

\bibitem[\protect\citeauthoryear{{Bah{\'e}}, {McCarthy}, {Crain} \&
  {Theuns}}{{Bah{\'e}} et~al.}{2012}]{Bahe.etal.2012}
{Bah{\'e}} Y.~M.,  {McCarthy} I.~G.,  {Crain} R.~A.,    {Theuns} T.,  2012,
  \mnras, 424, 1179

\bibitem[\protect\citeauthoryear{{Baldry}, {Balogh}, {Bower}, {Glazebrook},
  {Nichol}, {Bamford} \& {Budavari}}{{Baldry} et~al.}{2006}]{Baldry.etal.2006}
{Baldry} I.~K.,  {Balogh} M.~L.,  {Bower} R.~G.,  {Glazebrook} K.,  {Nichol}
  R.~C.,  {Bamford} S.~P.,    {Budavari} T.,  2006, \mnras, 373, 469

\bibitem[\protect\citeauthoryear{{Balogh} et~al.,}{{Balogh}
  et~al.}{2004}]{Balogh.etal.2004}
{Balogh} M.  et~al., 2004, \mnras, 348, 1355

\bibitem[\protect\citeauthoryear{{Balogh} \& {Morris}}{{Balogh} \&
  {Morris}}{2000}]{Balogh&Morris.2000}
{Balogh} M.~L.,  {Morris} S.~L.,  2000, \mnras, 318, 703

\bibitem[\protect\citeauthoryear{{Balogh}, {Morris}, {Yee}, {Carlberg} \&
  {Ellingson}}{{Balogh} et~al.}{1999}]{Balogh.etal.1999}
{Balogh} M.~L.,  {Morris} S.~L.,  {Yee} H.~K.~C.,  {Carlberg} R.~G.,
  {Ellingson} E.,  1999, \apj, 527, 54

\bibitem[\protect\citeauthoryear{{Balogh}, {Navarro} \& {Morris}}{{Balogh}
  et~al.}{2000}]{Balogh.etal.2000}
{Balogh} M.~L.,  {Navarro} J.~F.,    {Morris} S.~L.,  2000, \apj, 540, 113

\bibitem[\protect\citeauthoryear{{Bardelli} et~al.,}{{Bardelli}
  et~al.}{2010}]{Bardelli.etal.2010}
{Bardelli} S.  et~al., 2010, \aap, 511, A1

\bibitem[\protect\citeauthoryear{{Benson}}{{Benson}}{2005}]{Benson.2005}
{Benson} A.~J.,  2005, \mnras, 358, 551

\bibitem[\protect\citeauthoryear{{Berrier}, {Stewart}, {Bullock}, {Purcell},
  {Barton} \& {Wechsler}}{{Berrier} et~al.}{2009}]{Berrier.etal.2009}
{Berrier} J.~C.,  {Stewart} K.~R.,  {Bullock} J.~S.,  {Purcell} C.~W.,
  {Barton} E.~J.,    {Wechsler} R.~H.,  2009, \apj, 690, 1292

\bibitem[\protect\citeauthoryear{{Blanton}, {Eisenstein}, {Hogg}, {Schlegel} \&
  {Brinkmann}}{{Blanton} et~al.}{2005}]{Blanton.etal.2005}
{Blanton} M.~R.,  {Eisenstein} D.,  {Hogg} D.~W.,  {Schlegel} D.~J.,
  {Brinkmann} J.,  2005, \apj, 629, 143

\bibitem[\protect\citeauthoryear{{Bolzonella} et~al.,}{{Bolzonella}
  et~al.}{2010}]{Bolzonella.etal.2009}
{Bolzonella} M.  et~al., 2010, \aap, 524, A76

\bibitem[\protect\citeauthoryear{{Boselli} \& {Gavazzi}}{{Boselli} \&
  {Gavazzi}}{2006}]{Boselli&Gavazzi.2006}
{Boselli} A.,  {Gavazzi} G.,  2006, \pasp, 118, 517

\bibitem[\protect\citeauthoryear{{Bruzual} \& {Charlot}}{{Bruzual} \&
  {Charlot}}{2003}]{Bruzual&Charlot.2003}
{Bruzual} G.,  {Charlot} S.,  2003, \mnras, 344, 1000

\bibitem[\protect\citeauthoryear{{Calzetti}, {Armus}, {Bohlin}, {Kinney},
  {Koornneef} \& {Storchi-Bergmann}}{{Calzetti}
  et~al.}{2000}]{Calzetti.etal.2000}
{Calzetti} D.,  {Armus} L.,  {Bohlin} R.~C.,  {Kinney} A.~L.,  {Koornneef} J.,
    {Storchi-Bergmann} T.,  2000, \apj, 533, 682

\bibitem[\protect\citeauthoryear{{Capak} et~al.,}{{Capak}
  et~al.}{2007}]{Capak.etal.2007}
{Capak} P.  et~al., 2007, \apjs, 172, 99

\bibitem[\protect\citeauthoryear{{Caputi} et~al.,}{{Caputi}
  et~al.}{2009}]{Caputi.etal.2009}
{Caputi} K.~I.  et~al., 2009, \apj, 691, 91

\bibitem[\protect\citeauthoryear{{Carollo} et~al.,}{{Carollo}
  et~al.}{2012}]{Carollo.etal.2012}
{Carollo} C.~M.  et~al., 2012, ArXiv e-prints

\bibitem[\protect\citeauthoryear{{Chabrier}}{{Chabrier}}{2003}]{Chabrier.2003}
{Chabrier} G.,  2003, \pasp, 115, 763

\bibitem[\protect\citeauthoryear{{Chuter} et~al.,}{{Chuter}
  et~al.}{2011}]{Chuter.etal.2011}
{Chuter} R.~W.  et~al., 2011, \mnras, 413, 1678

\bibitem[\protect\citeauthoryear{{Cooper} et~al.,}{{Cooper}
  et~al.}{2007}]{Cooper.etal.2007}
{Cooper} M.~C.  et~al., 2007, \mnras, 376, 1445

\bibitem[\protect\citeauthoryear{{Cooper} et~al.,}{{Cooper}
  et~al.}{2010}]{Cooper.etal.2010}
{Cooper} M.~C.  et~al., 2010, \mnras, 409, 337

\bibitem[\protect\citeauthoryear{{Cucciati} et~al.,}{{Cucciati}
  et~al.}{2006}]{Cucciati.etal.2006}
{Cucciati} O.  et~al., 2006, \aap, 458, 39

\bibitem[\protect\citeauthoryear{{Cucciati} et~al.,}{{Cucciati}
  et~al.}{2010}]{Cucciati.etal.2010}
{Cucciati} O.  et~al., 2010, \aap, 524, A2

\bibitem[\protect\citeauthoryear{{Davis} \& {Geller}}{{Davis} \&
  {Geller}}{1976}]{Davis&Geller.1976}
{Davis} M.,  {Geller} M.~J.,  1976, \apj, 208, 13

\bibitem[\protect\citeauthoryear{{Dekel} \& {Birnboim}}{{Dekel} \&
  {Birnboim}}{2006}]{Dekel&Birnboim.2006}
{Dekel} A.,  {Birnboim} Y.,  2006, \mnras, 368, 2

\bibitem[\protect\citeauthoryear{{Dressler}}{{Dressler}}{1980}]{Dressler.1980}
{Dressler} A.,  1980, \apj, 236, 351

\bibitem[\protect\citeauthoryear{{Farouki} \& {Shapiro}}{{Farouki} \&
  {Shapiro}}{1981}]{Farouki&Shapiro.1981}
{Farouki} R.,  {Shapiro} S.~L.,  1981, \apj, 243, 32

\bibitem[\protect\citeauthoryear{{Feldmann} et~al.,}{{Feldmann}
  et~al.}{2006}]{Feldmann.etal.2006}
{Feldmann} R.  et~al., 2006, \mnras, 372, 565

\bibitem[\protect\citeauthoryear{{Gao} \& {White}}{{Gao} \&
  {White}}{2006}]{Gao&White.2006}
{Gao} L.,  {White} S.~D.~M.,  2006, \mnras, 373, 65

\bibitem[\protect\citeauthoryear{{Gao}, {Springel} \& {White}}{{Gao}
  et~al.}{2005}]{Gao.etal.2005}
{Gao} L.,  {Springel} V.,    {White} S.~D.~M.,  2005, \mnras, 363, L66

\bibitem[\protect\citeauthoryear{{Gerke} et~al.,}{{Gerke}
  et~al.}{2007}]{Gerke.etal.2007}
{Gerke} B.~F.  et~al., 2007, \mnras, 376, 1425

\bibitem[\protect\citeauthoryear{{Gerke} et~al.,}{{Gerke}
  et~al.}{2012}]{Gerke.etal.2012}
{Gerke} B.~F.  et~al., 2012, \apj, 751, 50

\bibitem[\protect\citeauthoryear{{Gunn} \& {Gott}}{{Gunn} \&
  {Gott}}{1972}]{Gunn&Gott.1972}
{Gunn} J.~E.,  {Gott} J.~R.~I.,  1972, \apj, 176, 1

\bibitem[\protect\citeauthoryear{{Guo} et~al.,}{{Guo}
  et~al.}{2011}]{Guo.etal.2011}
{Guo} Q.  et~al., 2011, \mnras, 413, 101

\bibitem[\protect\citeauthoryear{{Haas}, {Schaye} \& {Jeeson-Daniel}}{{Haas}
  et~al.}{2012}]{Haas.etal.2012}
{Haas} M.~R.,  {Schaye} J.,    {Jeeson-Daniel} A.,  2012, \mnras, 419, 2133

\bibitem[\protect\citeauthoryear{{Hansen}, {Sheldon}, {Wechsler} \&
  {Koester}}{{Hansen} et~al.}{2009}]{Hansen.etal.2009}
{Hansen} S.~M.,  {Sheldon} E.~S.,  {Wechsler} R.~H.,    {Koester} B.~P.,  2009,
  \apj, 699, 1333

\bibitem[\protect\citeauthoryear{{Henriques}, {White}, {Lemson}, {Thomas},
  {Guo}, {Marleau} \& {Overzier}}{{Henriques}
  et~al.}{2012}]{Henriques.etal.2012}
{Henriques} B.~M.~B.,  {White} S.~D.~M.,  {Lemson} G.,  {Thomas} P.~A.,  {Guo}
  Q.,  {Marleau} G.-D.,    {Overzier} R.~A.,  2012, \mnras, 421, 2904

\bibitem[\protect\citeauthoryear{{Hogg} et~al.,}{{Hogg}
  et~al.}{2004}]{Hogg.etal.2004}
{Hogg} D.~W.  et~al., 2004, \apjl, 601, L29

\bibitem[\protect\citeauthoryear{{Hubble}}{{Hubble}}{1939}]{Hubble.1939}
{Hubble} E.,  1939, in Publications of the American Astronomical Society.
  p.~249

\bibitem[\protect\citeauthoryear{{Iovino} et~al.,}{{Iovino}
  et~al.}{2010}]{Iovino.etal.2010}
{Iovino} A.  et~al., 2010, \aap, 509, A40

\bibitem[\protect\citeauthoryear{{Jian}, {Lin} \& {Chiueh}}{{Jian}
  et~al.}{2012}]{Jian.etal.2012}
{Jian} H.-Y.,  {Lin} L.,    {Chiueh} T.,  2012, \apj, 754, 26

\bibitem[\protect\citeauthoryear{{Kampczyk} et~al.,}{{Kampczyk}
  et~al.}{2013}]{Kampczyk.etal.2013}
{Kampczyk} P.  et~al., 2013, \apj, 762, 43

\bibitem[\protect\citeauthoryear{{Kauffmann} et~al.,}{{Kauffmann}
  et~al.}{2003}]{Kauffmann.etal.2003b}
{Kauffmann} G.  et~al., 2003, \mnras, 341, 54

\bibitem[\protect\citeauthoryear{{Kauffmann}, {White}, {Heckman}, {M{\'e}nard},
  {Brinchmann}, {Charlot}, {Tremonti} \& {Brinkmann}}{{Kauffmann}
  et~al.}{2004}]{Kauffmann.etal.2004}
{Kauffmann} G.,  {White} S.~D.~M.,  {Heckman} T.~M.,  {M{\'e}nard} B.,
  {Brinchmann} J.,  {Charlot} S.,  {Tremonti} C.,    {Brinkmann} J.,  2004,
  \mnras, 353, 713

\bibitem[\protect\citeauthoryear{{Kennicutt}
  Jr.}{{Kennicutt}}{1998}]{Kennicutt.1998}
{Kennicutt} Jr. R.~C.,  1998, \araa, 36, 189

\bibitem[\protect\citeauthoryear{{Kitzbichler} \& {White}}{{Kitzbichler} \&
  {White}}{2007}]{Kitzbichler&White.2007}
{Kitzbichler} M.~G.,  {White} S.~D.~M.,  2007, \mnras, 376, 2

\bibitem[\protect\citeauthoryear{{Knobel} et~al.,}{{Knobel}
  et~al.}{2009}]{Knobel.etal.2009}
{Knobel} C.  et~al., 2009, \apj, 697, 1842

\bibitem[\protect\citeauthoryear{{Knobel} et~al.,}{{Knobel}
  et~al.}{2012a}]{Knobel.etal.2012a}
{Knobel} C.  et~al., 2012a, \apj, 753, 121

\bibitem[\protect\citeauthoryear{{Knobel} et~al.,}{{Knobel}
  et~al.}{2012b}]{Knobel.etal.2012c}
{Knobel} C.  et~al., 2012b, \apj, 755, 48

\bibitem[\protect\citeauthoryear{{Knobel} et~al.,}{{Knobel}
  et~al.}{2013}]{Knobel.etal.2012b}
{Knobel} C.  et~al., 2013, \apj, 769, 24

\bibitem[\protect\citeauthoryear{{Kova{\v c}} et~al.,}{{Kova{\v c}}
  et~al.}{2010a}]{Kovac.etal.2010a}
{Kova{\v c}} K.  et~al., 2010a, \apj, 708, 505

\bibitem[\protect\citeauthoryear{{Kova{\v c}} et~al.,}{{Kova{\v c}}
  et~al.}{2010b}]{Kovac.etal.2010b}
{Kova{\v c}} K.  et~al., 2010b, \apj, 718, 86

\bibitem[\protect\citeauthoryear{{Larson}, {Tinsley} \& {Caldwell}}{{Larson}
  et~al.}{1980}]{Larson.etal.1980}
{Larson} R.~B.,  {Tinsley} B.~M.,    {Caldwell} C.~N.,  1980, \apj, 237, 692

\bibitem[\protect\citeauthoryear{{Le F{\`e}vre} et~al.,}{{Le F{\`e}vre}
  et~al.}{2003}]{LeFevre.etal.2003}
{Le F{\`e}vre} O.  et~al., 2003, in {Iye} M.,  {Moorwood} A.~F.~M.,  eds,
  Society of Photo-Optical Instrumentation Engineers (SPIE) Conference Series
  Vol. 4841, Society of Photo-Optical Instrumentation Engineers (SPIE)
  Conference Series. pp 1670--1681

\bibitem[\protect\citeauthoryear{{Le Floc'h} et~al.,}{{Le Floc'h}
  et~al.}{2009}]{LeFloch.etal.2009}
{Le Floc'h} E.  et~al., 2009, \apj, 703, 222

\bibitem[\protect\citeauthoryear{{Li}, {Yee} \& {Ellingson}}{{Li}
  et~al.}{2009}]{Li.etal.2009}
{Li} I.~H.,  {Yee} H.~K.~C.,    {Ellingson} E.,  2009, \apj, 698, 83

\bibitem[\protect\citeauthoryear{{Lilly} et~al.,}{{Lilly}
  et~al.}{2007}]{Lilly.etal.2007}
{Lilly} S.~J.  et~al., 2007, \apjs, 172, 70

\bibitem[\protect\citeauthoryear{{Lilly} et~al.,}{{Lilly}
  et~al.}{2009}]{Lilly.etal.2009}
{Lilly} S.~J.  et~al., 2009, \apjs, 184, 218

\bibitem[\protect\citeauthoryear{{Lilly}, {Peng}, {Carollo} \&
  {Renzini}}{{Lilly} et~al.}{2013}]{Lilly.etal.2013}
{Lilly} S.,  {Peng} Y.-j.,  {Carollo} M.,    {Renzini} A.,  2013, ArXiv
  e-prints

\bibitem[\protect\citeauthoryear{{Lu}, {Gilbank}, {McGee}, {Balogh} \&
  {Gallagher}}{{Lu} et~al.}{2012}]{Lu.etal.2012}
{Lu} T.,  {Gilbank} D.~G.,  {McGee} S.~L.,  {Balogh} M.~L.,    {Gallagher} S.,
  2012, \mnras, 420, 126

\bibitem[\protect\citeauthoryear{{Ludlow}, {Navarro}, {Springel}, {Jenkins},
  {Frenk} \& {Helmi}}{{Ludlow} et~al.}{2009}]{Ludlow.etal.2009}
{Ludlow} A.~D.,  {Navarro} J.~F.,  {Springel} V.,  {Jenkins} A.,  {Frenk}
  C.~S.,    {Helmi} A.,  2009, \apj, 692, 931

\bibitem[\protect\citeauthoryear{{McCarthy}, {Frenk}, {Font}, {Lacey}, {Bower},
  {Mitchell}, {Balogh} \& {Theuns}}{{McCarthy}
  et~al.}{2008}]{McCarthy.etal.2008}
{McCarthy} I.~G.,  {Frenk} C.~S.,  {Font} A.~S.,  {Lacey} C.~G.,  {Bower}
  R.~G.,  {Mitchell} N.~L.,  {Balogh} M.~L.,    {Theuns} T.,  2008, \mnras,
  383, 593

\bibitem[\protect\citeauthoryear{{McDonald}, {Veilleux} \&
  {Mushotzky}}{{McDonald} et~al.}{2011}]{McDonald.etal.2011}
{McDonald} M.,  {Veilleux} S.,    {Mushotzky} R.,  2011, \apj, 731, 33

\bibitem[\protect\citeauthoryear{{McGee}, {Balogh}, {Bower}, {Font} \&
  {McCarthy}}{{McGee} et~al.}{2009}]{McGee.etal.2009}
{McGee} S.~L.,  {Balogh} M.~L.,  {Bower} R.~G.,  {Font} A.~S.,    {McCarthy}
  I.~G.,  2009, \mnras, 400, 937

\bibitem[\protect\citeauthoryear{{Moore}, {Lake} \& {Katz}}{{Moore}
  et~al.}{1998}]{Moore.etal.1998}
{Moore} B.,  {Lake} G.,    {Katz} N.,  1998, \apj, 495, 139

\bibitem[\protect\citeauthoryear{{Moresco} et~al.,}{{Moresco}
  et~al.}{2013}]{Moresco.etal.2013}
{Moresco} M.  et~al., 2013, ArXiv e-prints

\bibitem[\protect\citeauthoryear{{Oemler} Jr.}{{Oemler}}{1974}]{Oemler.1974}
{Oemler} Jr. A.,  1974, \apj, 194, 1

\bibitem[\protect\citeauthoryear{{Oesch} et~al.,}{{Oesch}
  et~al.}{2010}]{Oesch.etal.2010}
{Oesch} P.~A.  et~al., 2010, \apjl, 714, L47

\bibitem[\protect\citeauthoryear{{Peng} et~al.,}{{Peng}
  et~al.}{2010}]{Peng.etal.2010}
{Peng} Y.-j.  et~al., 2010, \apj, 721, 193

\bibitem[\protect\citeauthoryear{{Peng}, {Lilly}, {Renzini} \&
  {Carollo}}{{Peng} et~al.}{2012}]{Peng.etal.2012}
{Peng} Y.-j.,  {Lilly} S.~J.,  {Renzini} A.,    {Carollo} M.,  2012, \apj, 757,
  4

\bibitem[\protect\citeauthoryear{{Quadri}, {Williams}, {Franx} \&
  {Hildebrandt}}{{Quadri} et~al.}{2012}]{Quadri.etal.2012}
{Quadri} R.~F.,  {Williams} R.~J.,  {Franx} M.,    {Hildebrandt} H.,  2012,
  \apj, 744, 88

\bibitem[\protect\citeauthoryear{{Quilis}, {Moore} \& {Bower}}{{Quilis}
  et~al.}{2000}]{Quilis.etal.2000}
{Quilis} V.,  {Moore} B.,    {Bower} R.,  2000, Science, 288, 1617

\bibitem[\protect\citeauthoryear{{Rasmussen}, {Ponman} \&
  {Mulchaey}}{{Rasmussen} et~al.}{2006}]{Rasmussen.etal.2006}
{Rasmussen} J.,  {Ponman} T.~J.,    {Mulchaey} J.~S.,  2006, \mnras, 370, 453

\bibitem[\protect\citeauthoryear{{Read}, {Wilkinson}, {Evans}, {Gilmore} \&
  {Kleyna}}{{Read} et~al.}{2006}]{Read.etal.2006}
{Read} J.~I.,  {Wilkinson} M.~I.,  {Evans} N.~W.,  {Gilmore} G.,    {Kleyna}
  J.~T.,  2006, \mnras, 366, 429

\bibitem[\protect\citeauthoryear{{Sanders} et~al.,}{{Sanders}
  et~al.}{2007}]{Sanders.etal.2007}
{Sanders} D.~B.  et~al., 2007, \apjs, 172, 86

\bibitem[\protect\citeauthoryear{{Scoville} et~al.,}{{Scoville}
  et~al.}{2007a}]{Scoville.etal.2007a}
{Scoville} N.  et~al., 2007a, \apjs, 172, 38

\bibitem[\protect\citeauthoryear{{Scoville} et~al.,}{{Scoville}
  et~al.}{2007b}]{Scoville.etal.2007b}
{Scoville} N.  et~al., 2007b, \apjs, 172, 150

\bibitem[\protect\citeauthoryear{{Scoville} et~al.,}{{Scoville}
  et~al.}{2013}]{Scoville.etal.2013}
{Scoville} N.  et~al., 2013, \apjs, 206, 3

\bibitem[\protect\citeauthoryear{{Silverman} et~al.,}{{Silverman}
  et~al.}{2009}]{Silverman.etal.2009}
{Silverman} J.~D.  et~al., 2009, \apj, 695, 171

\bibitem[\protect\citeauthoryear{{Tasca} et~al.,}{{Tasca}
  et~al.}{2009}]{Tasca.etal.2009}
{Tasca} L.~A.~M.  et~al., 2009, \aap, 503, 379

\bibitem[\protect\citeauthoryear{{Tinker}, {Wetzel} \& {Conroy}}{{Tinker}
  et~al.}{2011}]{Tinker.etal.2011}
{Tinker} J.,  {Wetzel} A.,    {Conroy} C.,  2011, ArXiv e-prints

\bibitem[\protect\citeauthoryear{{van den Bosch}, {Aquino}, {Yang}, {Mo},
  {Pasquali}, {McIntosh}, {Weinmann} \& {Kang}}{{van den Bosch}
  et~al.}{2008}]{vandenBosch.etal.2008}
{van den Bosch} F.~C.,  {Aquino} D.,  {Yang} X.,  {Mo} H.~J.,  {Pasquali} A.,
  {McIntosh} D.~H.,  {Weinmann} S.~M.,    {Kang} X.,  2008, \mnras, 387, 79

\bibitem[\protect\citeauthoryear{{van der Burg} et~al.,}{{van der Burg}
  et~al.}{2013}]{vanderBurg.etal.2013}
{van der Burg} R.~F.~J.  et~al., 2013, ArXiv e-prints

\bibitem[\protect\citeauthoryear{{Vergani} et~al.,}{{Vergani}
  et~al.}{2010}]{Vergani.etal.2010}
{Vergani} D.  et~al., 2010, \aap, 509, A42

\bibitem[\protect\citeauthoryear{{von der Linden}, {Wild}, {Kauffmann}, {White}
  \& {Weinmann}}{{von der Linden} et~al.}{2010}]{vonderLinden.etal.2010}
{von der Linden} A.,  {Wild} V.,  {Kauffmann} G.,  {White} S.~D.~M.,
  {Weinmann} S.,  2010, \mnras, 404, 1231

\bibitem[\protect\citeauthoryear{{Wechsler}, {Zentner}, {Bullock}, {Kravtsov}
  \& {Allgood}}{{Wechsler} et~al.}{2006}]{Wechsler.etal.2006}
{Wechsler} R.~H.,  {Zentner} A.~R.,  {Bullock} J.~S.,  {Kravtsov} A.~V.,
  {Allgood} B.,  2006, \apj, 652, 71

\bibitem[\protect\citeauthoryear{{Weinmann}, {van den Bosch}, {Yang} \&
  {Mo}}{{Weinmann} et~al.}{2006}]{Weinmann.etal.2006}
{Weinmann} S.~M.,  {van den Bosch} F.~C.,  {Yang} X.,    {Mo} H.~J.,  2006,
  \mnras, 366, 2

\bibitem[\protect\citeauthoryear{{Weinmann}, {Kauffmann}, {van den Bosch},
  {Pasquali}, {McIntosh}, {Mo}, {Yang} \& {Guo}}{{Weinmann}
  et~al.}{2009}]{Weinmann.etal.2009}
{Weinmann} S.~M.,  {Kauffmann} G.,  {van den Bosch} F.~C.,  {Pasquali} A.,
  {McIntosh} D.~H.,  {Mo} H.,  {Yang} X.,    {Guo} Y.,  2009, \mnras, 394, 1213

\bibitem[\protect\citeauthoryear{{Wetzel}}{{Wetzel}}{2011}]{Wetzel.2011}
{Wetzel} A.~R.,  2011, \mnras, 412, 49

\bibitem[\protect\citeauthoryear{{Wetzel}, {Tinker} \& {Conroy}}{{Wetzel}
  et~al.}{2012}]{Wetzel.etal.2012}
{Wetzel} A.~R.,  {Tinker} J.~L.,    {Conroy} C.,  2012, \mnras, 424, 232

\bibitem[\protect\citeauthoryear{{Wetzel}, {Tinker}, {Conroy} \& {van den
  Bosch}}{{Wetzel} et~al.}{2013}]{Wetzel.etal.2013}
{Wetzel} A.~R.,  {Tinker} J.~L.,  {Conroy} C.,    {van den Bosch} F.~C.,  2013,
  \mnras, 432, 336

\bibitem[\protect\citeauthoryear{{White}, {Jones} \& {Forman}}{{White}
  et~al.}{1997}]{White.etal.1997}
{White} D.~A.,  {Jones} C.,    {Forman} W.,  1997, \mnras, 292, 419

\bibitem[\protect\citeauthoryear{{Woo} et~al.,}{{Woo}
  et~al.}{2013}]{Woo.etal.2012}
{Woo} J.  et~al., 2013, \mnras, 428, 3306

\bibitem[\protect\citeauthoryear{{Wuyts} et~al.,}{{Wuyts}
  et~al.}{2011}]{Wuyts.etal.2011}
{Wuyts} S.  et~al., 2011, \apj, 738, 106

\bibitem[\protect\citeauthoryear{{York} et~al.,}{{York}
  et~al.}{2000}]{York.etal.2000}
{York} D.~G.  et~al., 2000, \aj, 120, 1579

\end{thebibliography}

\appendix

\section{Difference between the modelled and measured red fractions of zCOSMOS galaxies}
\label{appendix_fiterr}

We showed, in Figure~\ref{fig_thfit} (Section~\ref{sec_redallmodel}), the red fraction of galaxies both expected from the model and measured from the zCOSMOS data in the $\log(1+\delta)-\log(M_*/$\Msol$)$ plane. Here, we show in Figure~\ref{fig_fiterr} the differences between the two, normalised by the adopted 1$\sigma$ uncertainties at each $\Delta \log(1+\delta) = \Delta \log(M_*/$\Msol$) = 0.3$ bin. We show the results only for the range of stellar mass and overdensity which were used to fit equation (\ref{eq_redfrmodel}). The key conclusion is that there are no systematic trends in the normalised differences between the model and the data across the overdensity-mass range probed by our data.

\begin{figure}
\includegraphics[width=0.45\textwidth]{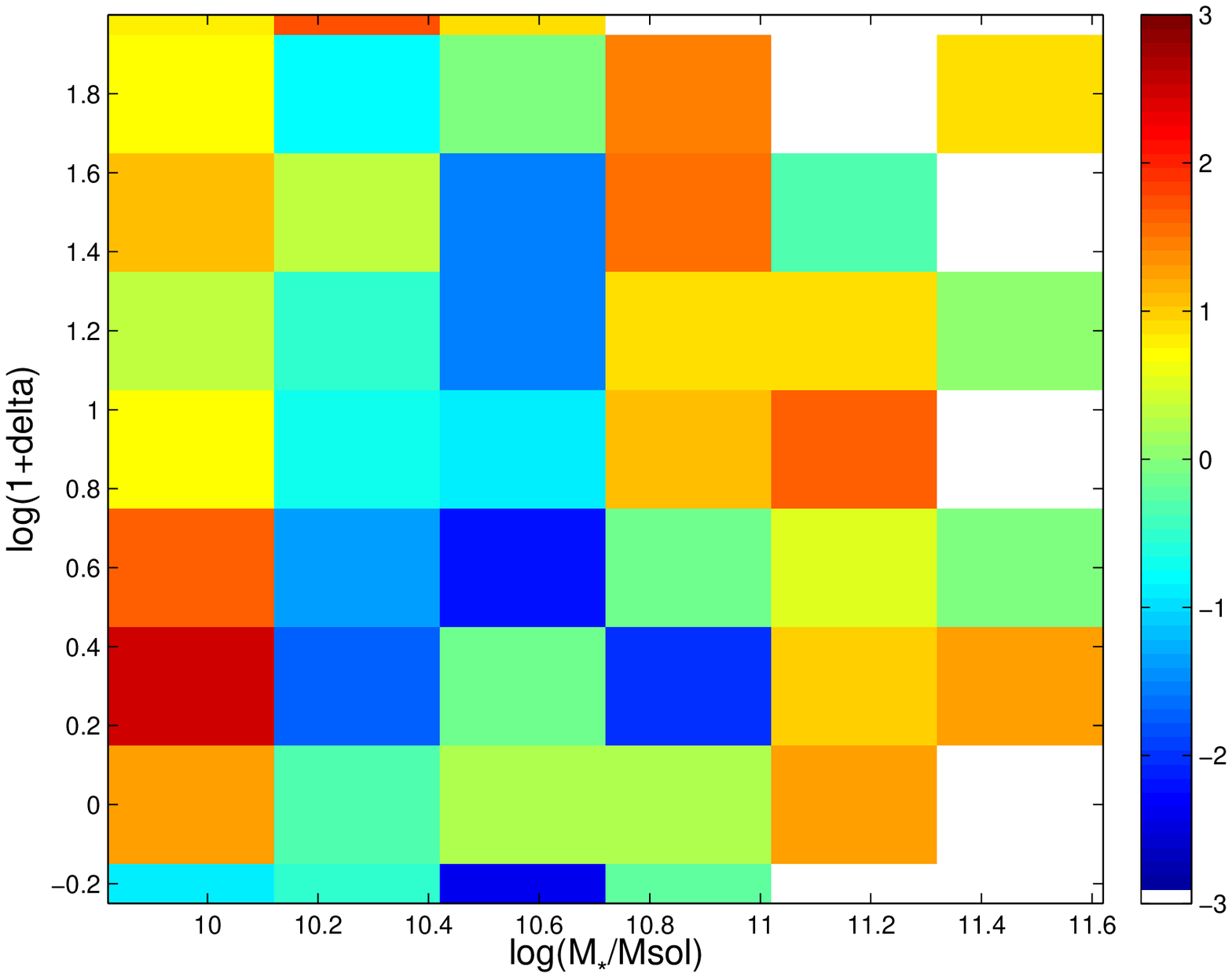}
\includegraphics[width=0.45\textwidth]{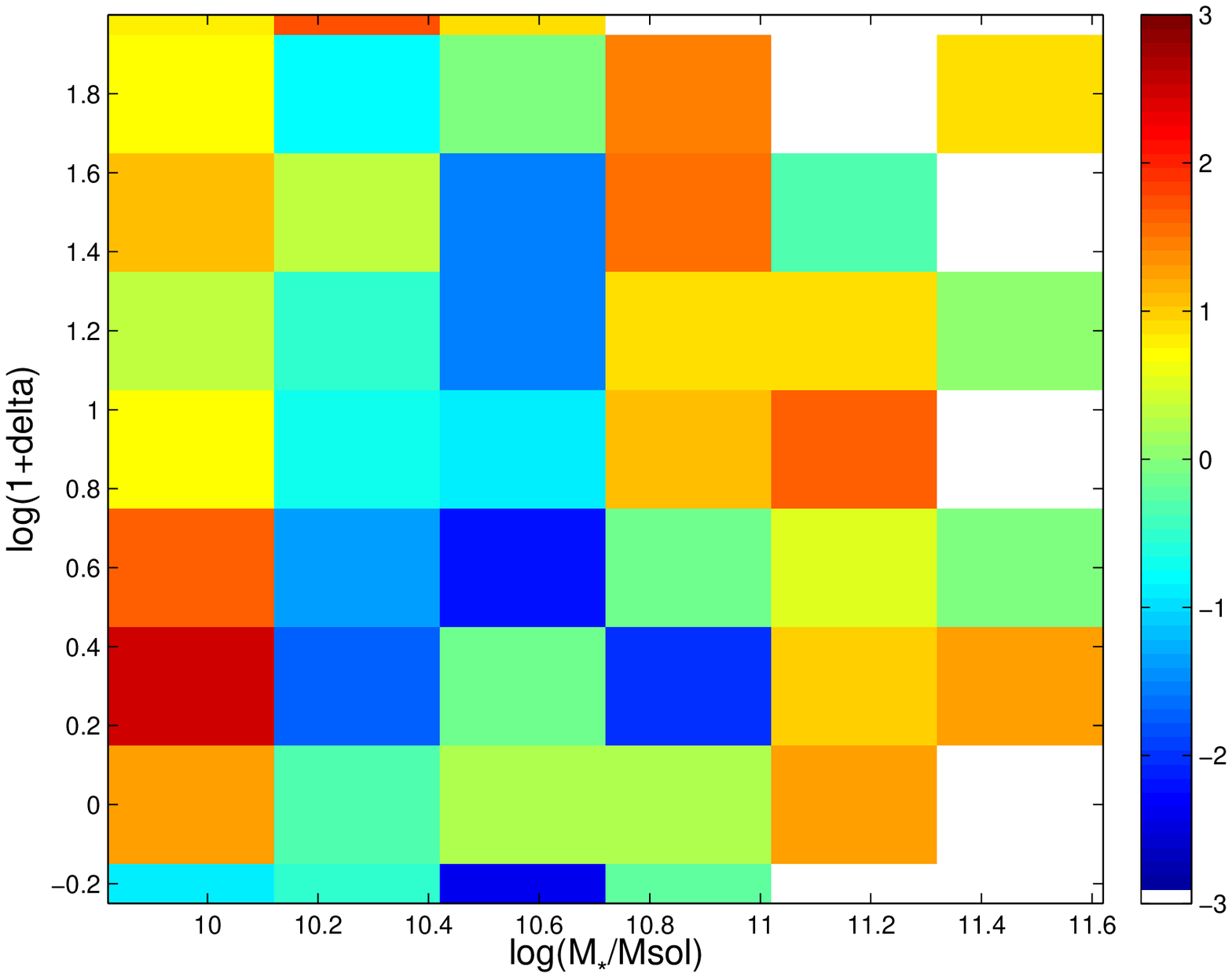}
\caption{\label{fig_fiterr}Difference between the modelled and measured red fraction normalised by the adopted error. The obtained values are colour coded according to the bar shown on the right hand side. The bins without data are artificially set to have the low negative values (white fields). The top and bottom panels are for the two redshift bins: $0.1<z<0.4$ and $0.4<z<0.7$, respectively.}
\end{figure}

\begin{figure}
\includegraphics[width=0.395\textwidth]{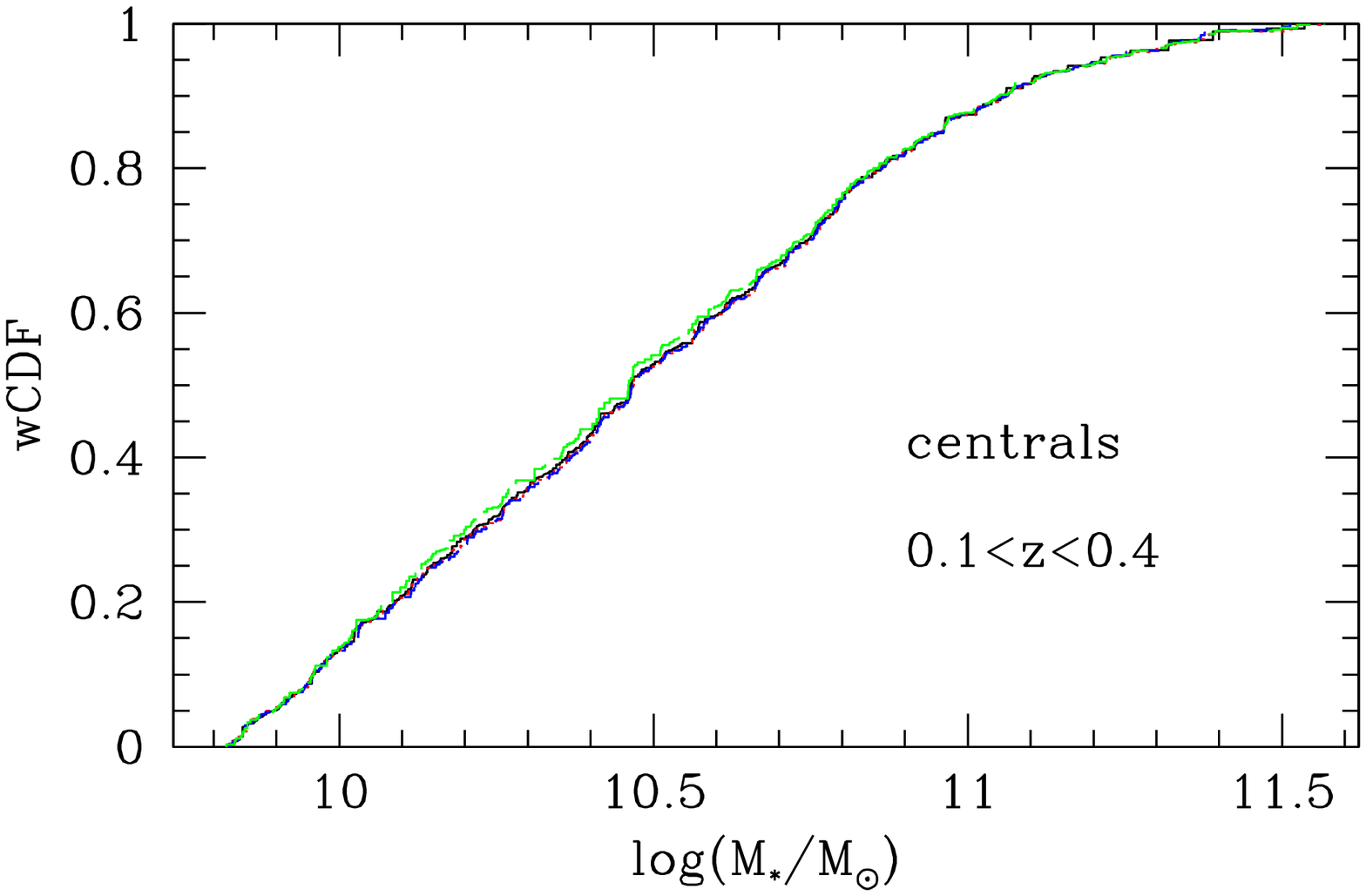}
\includegraphics[width=0.395\textwidth]{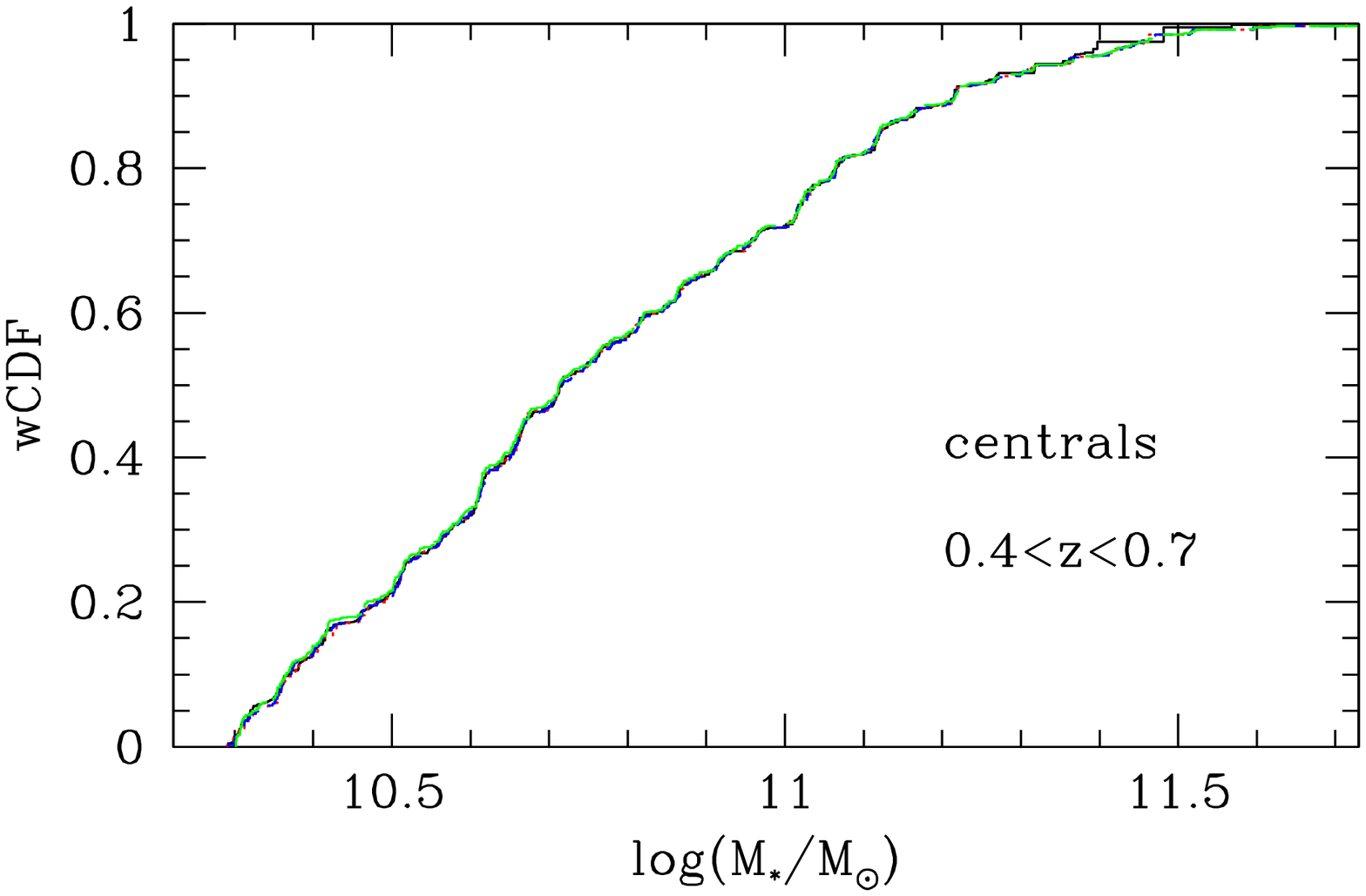}
\caption{\label{fig_masscenmatch}Weighted cumulative distribution functions of stellar mass in the samples of mass-matched centrals. The top and bottom panels are for $0.1<z<0.4$ and $0.4<z<0.7$, respectively. Different curves are constructed from the samples of centrals in the four quartiles of overdensity distribution defined by the central galaxies in a given redshift bin.}
\end{figure}

\begin{figure}
\includegraphics[width=0.395\textwidth]{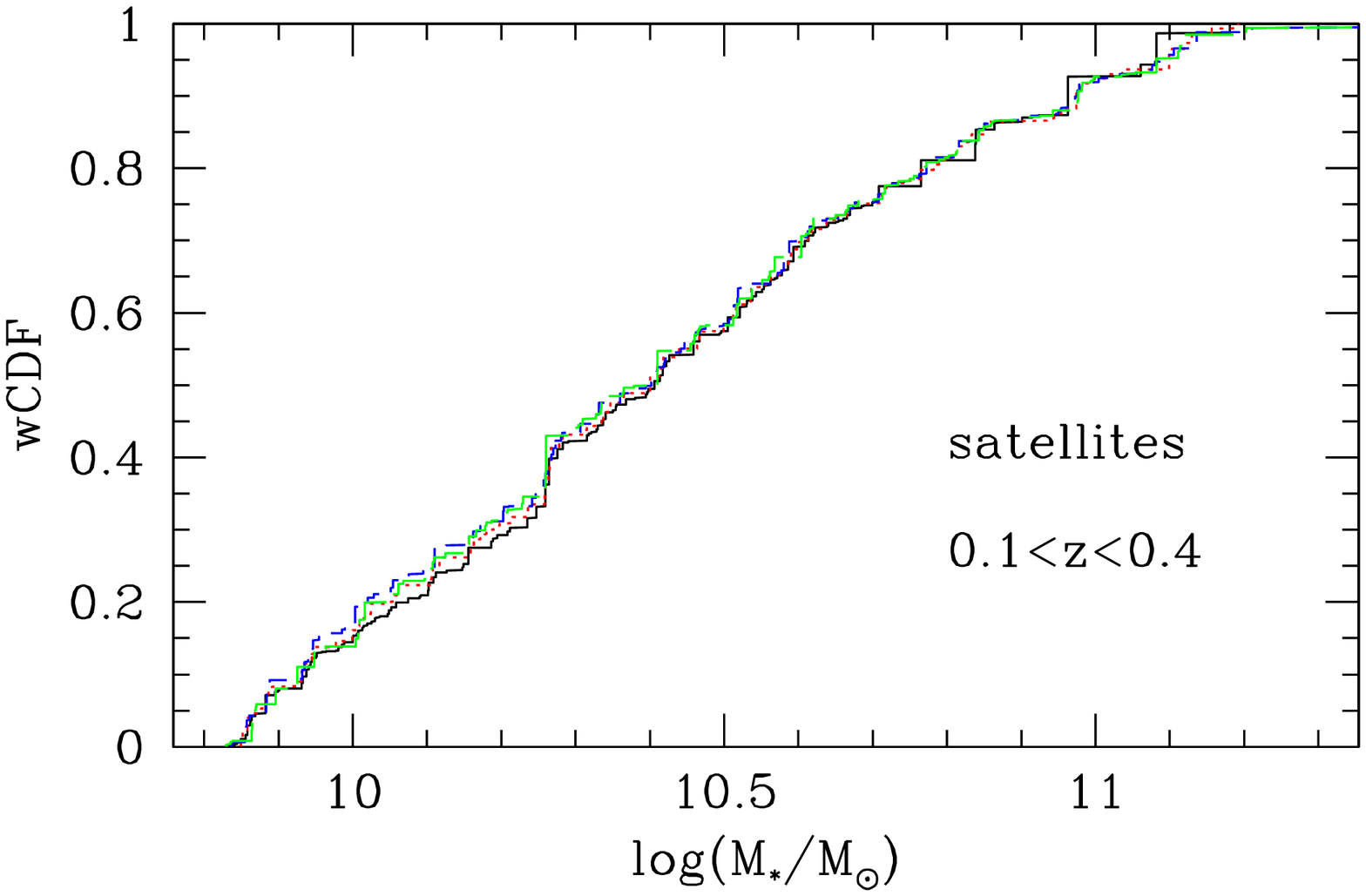}
\includegraphics[width=0.395\textwidth]{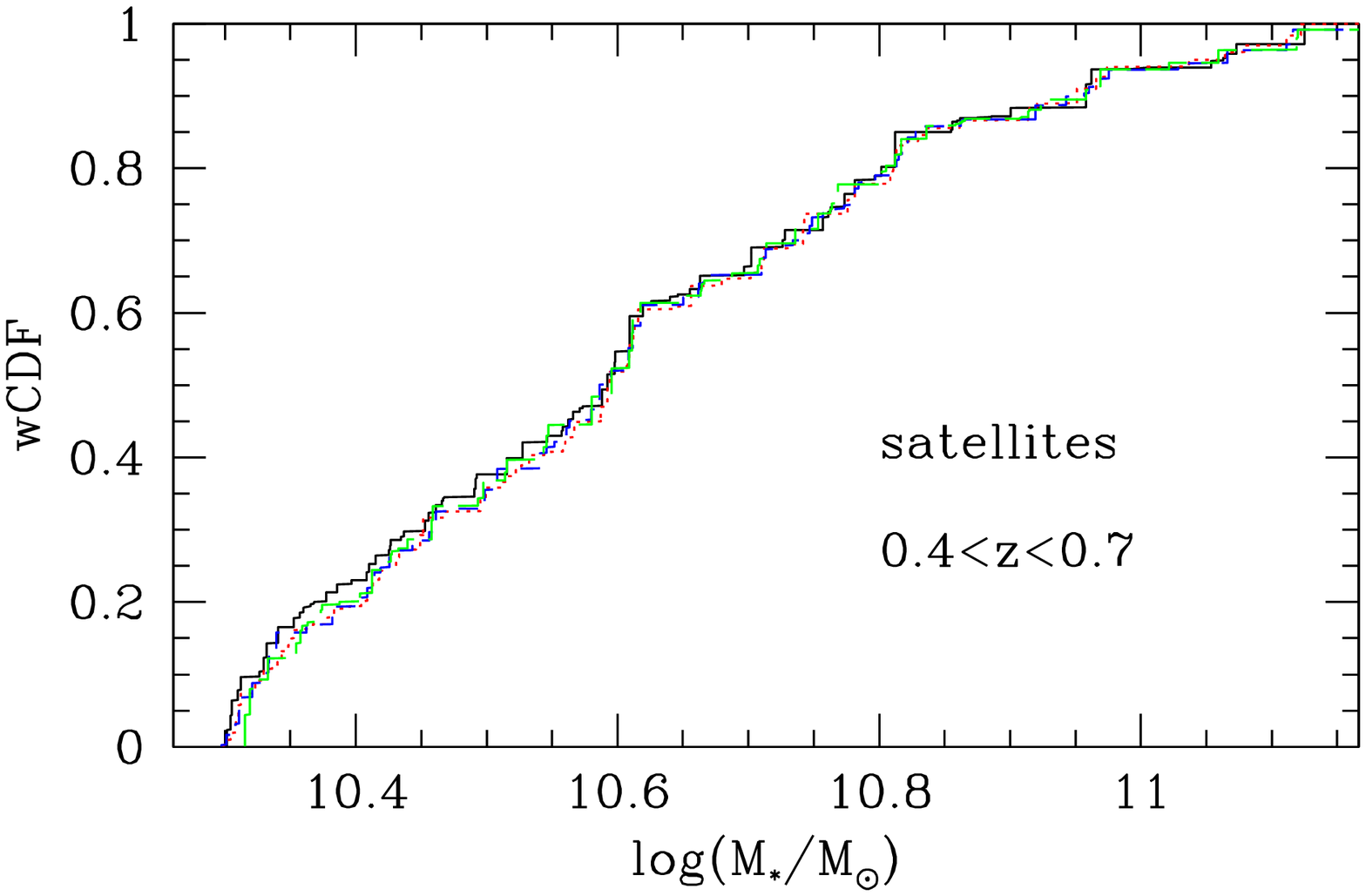}
\caption{\label{fig_masssatmatch}Weighted cumulative distribution functions of stellar mass in the samples of mass-matched satellites. The top and bottom panels are for $0.1<z<0.4$ and $0.4<z<0.7$, respectively. Different curves are constructed from the samples of satellites in the four quartiles of overdensity distribution defined by the satellite galaxies in a given redshift bin.}
\end{figure}

\section{Mass distribution of galaxies in the matched samples}
\label{appendix_massdist}

We described in Section~\ref{sec_redfraccenandsat} the procedure to match in stellar mass different samples of galaxies residing in different bins of overdensities. In this appendix, we provide some additional, more quantitative details of the matching procedure and show the resulting weighted stellar mass distributions of the matched samples of centrals and satellites.  

The mass-matching process consists of randomly choosing a galaxy in each of the overdensity quartiles with stellar mass close to a galaxy in the reference sample, for each galaxy in the reference sample. The nominal absolute difference in the $\log(M_*/$\Msol$)$ values which we use to pair galaxies is 0.01, though this difference could be enlarged if no suitable galaxy is found. The created samples of central or satellite galaxies in a given quartile of overdensity are then compared to their respective reference sample by comparing their wCDFs of stellar mass. We consider the wCDFs of stellar masses in the two samples to be matched if the absolute normalised difference between the two wCDFs at any percentile is less than 0.01 (see \citealt{Kauffmann.etal.2004} for a similar approach in comparing distributions). We use for normalisation the range in mass corresponding to the 1-99$\%$ interval of the wCDF in the reference sample. In practice, we measure the differences at 99 points equally distributed between 1 and 99$\%$ of the wCDF in the reference sample. If the distributions are not matched according to our criteria we discard the matched sample and redo the matching. We repeat the process until we have produced 20 samples matched according to the outlined criteria in each overdensity and redshift bin. The resulting wCDFs of stellar mass of the mass-matched central and satellite galaxies are shown in Figures~\ref{fig_masscenmatch} and~\ref{fig_masssatmatch}, respectively.

\end{document}